\documentclass[fleqn,usenatbib]{mnras}

\usepackage{newtxtext,newtxmath}

\usepackage[T1]{fontenc}
\usepackage{ae,aecompl}
\DeclareRobustCommand{\VAN}[3]{#2}
\let\VANthebibliography\thebibliography
\def\thebibliography{\DeclareRobustCommand{\VAN}[3]{##3}\VANthebibliography}

\usepackage{graphicx}	
\usepackage{amsmath}	
\usepackage{gensymb}
\usepackage{siunitx}
\usepackage{array}
\usepackage{float}
\usepackage{physics}
\usepackage{siunitx}
\usepackage{mathtools}
\usepackage{mhchem}
\usepackage{color}
\usepackage{xcolor}
\usepackage{listings}
\usepackage{booktabs}
\usepackage{makecell}
\usepackage{subcaption}
\usepackage{tabularx}
\usepackage[T1]{fontenc}
\usepackage{longtable}
\usepackage{afterpage}
\usepackage{tablefootnote}
\usepackage[flushleft]{threeparttable}
\usepackage{soul}
\usepackage{enumitem}

\setstcolor{red}
\DeclareSIUnit \parsec {pc}
\DeclareSIUnit \year {yr}
\DeclareSIUnit \erg {erg}
\DeclareSIUnit \gauss {g}

\newcommand{\ergspersecond}{\si{\erg\per\second}}
\newcommand{\pc}{\si{\parsec}}
\newcommand{\kpc}{\si{\kilo\parsec}}

\newcommand{\GeV}{\si{\giga\electronvolt}}
\newcommand{\TeV}{\si{\tera\electronvolt}}
\newcommand{\PeV}{\si{\peta\electronvolt}}
\newcommand{\centimeterminusthree}{\si{\per\centi\meter\cubed}}
\newcommand{\MeV}{\si{\mega\electronvolt}}
\newcommand{\keV}{\si{\kilo\electronvolt}}

\newcommand{\kmpersec}{\si{\kilo\meter\per\second}}
\newcommand{\kiloyear}{\si{\kilo\year}}

\newcommand{\fermi}{\mbox{\emph{Fermi}}}

\newcommand{\HESSmain}{\mbox{\mbox{HESS\,J1825-137}} }
\newcommand{\HESSminor}{\mbox{\mbox{HESS\,J1826-130}} }

\title[Model of e$^-$ transport towards \mbox{HESS\,J1825-137}]{A 3D Diffusive and Advective Model of Electron Transport Applied to the Pulsar Wind Nebula \mbox{HESS\,J1825-137}}

\author[T. Collins et al.]{
T. Collins,$^{1}$\thanks{E-mail: tiffany.collins@adelaide.edu.au}, G. Rowell$^{1}$, S. Einecke${^1}$,  F. Voisin$^{1}$, Y. Fukui$^{2}$ and H. Sano$^{3}$ 
 \\
$^{1}$School of Physical Sciences, University of Adelaide, Adelaide 5005, Australia\\
$^{2}$Department of Physics, University of Nagoya, Furo-cho, Chikusa-ku, Nagoya, 464-8601, Japan \\
$^{3}$Faculty of Engineering, Gifu University, Yanagido 1-1, Gifu, 501-1193, Japan\\
}

\date{Accepted 2024 January 11. Received 2024 January 11; in original form 2023 September 5}

\pubyear{2023}

\begin{document}
\label{firstpage}
\pagerange{\pageref{firstpage}--\pageref{lastpage}}
\maketitle

\begin{abstract}
\HESSmain is one of the most powerful and luminous TeV gamma-ray pulsar wind nebulae (PWNe), making it an excellent laboratory to study particle transportation around pulsars. We present a model of the (diffusive and advective) transport and radiative losses of electrons from the pulsar \mbox{PSR\,J1826-1334} powering \mbox{HESS\,J1825-137} using interstellar medium gas (ISM) data, soft photon fields and a spatially varying magnetic field. We find that for the characteristic age of $21\,\kiloyear$, \mbox{PSR\,J1826-1334} is unable to meet the energy requirements to match the observed X-ray and gamma-ray emission. An older age of $40\,\kiloyear$, together with an electron conversion efficiency of $0.14$ and advective flow of $v=0.002c$, can reproduce the observed multi-wavelengh emission towards \mbox{HESS\,J1825-137}. A turbulent ISM with magnetic field of $B=\SIrange{20}{60}{\micro G}$ to the north of {\mbox{HESS\,J1825-137}} (as suggested by ISM observations) is required to prevent significant gamma-ray contamination towards the northern $\TeV$ source \mbox{HESS\,J1826-130}.
\end{abstract}



\section{Introduction}
\HESSmain is a luminous pulsar wind nebula (PWN) powered by pulsar \mbox{PSR\,J1826-1334} with spin-down power $\dot{E}=2.8\times 10^{36}\,\ergspersecond$ and characteristic age $\tau_c=P/2\dot{P}=21.4\,\kiloyear$  \citep{2005AJ....129.1993M}. The distance to \mbox{PSR\,J11826-1334} has been estimated to lie at $3.6\,\kpc$ based on dispersion measurements \citep{1993ApJ...411..674T,2002astro.ph..7156C}, however we will use a distance of $4\,\kpc$ in line with \cite{2011ApJ...742...62V} and \cite{2019A&A...621A.116H}. The $\TeV$ gamma-ray emission from \mbox{HESS\,J1825-137} has a characteristic ($1/e$) radius of $\ang{0.66}\pm\ang{0.03}_\text{stat}\pm\ang{0.04}_\text{sys}$, implying a radius of $\approx 46\,\pc$ based on a distance of $4\,\kpc$ \citep{2019A&A...621A.116H}.  Owing to its brightness in TeV gamma rays, \HESSmain is an ideal laboratory to study relativistic particle transport in and around middle-aged PWNe. 
Several studies (e.g. \cite{2016MNRAS.460.4135P}, \cite{2020A&A...636A.113G}) suggest that both diffusive and advective transport mechanisms are required to explain the extended gamma-ray morphology towards PWNe.
\par
\begin{figure}
    \centering
    \includegraphics[width=\columnwidth]{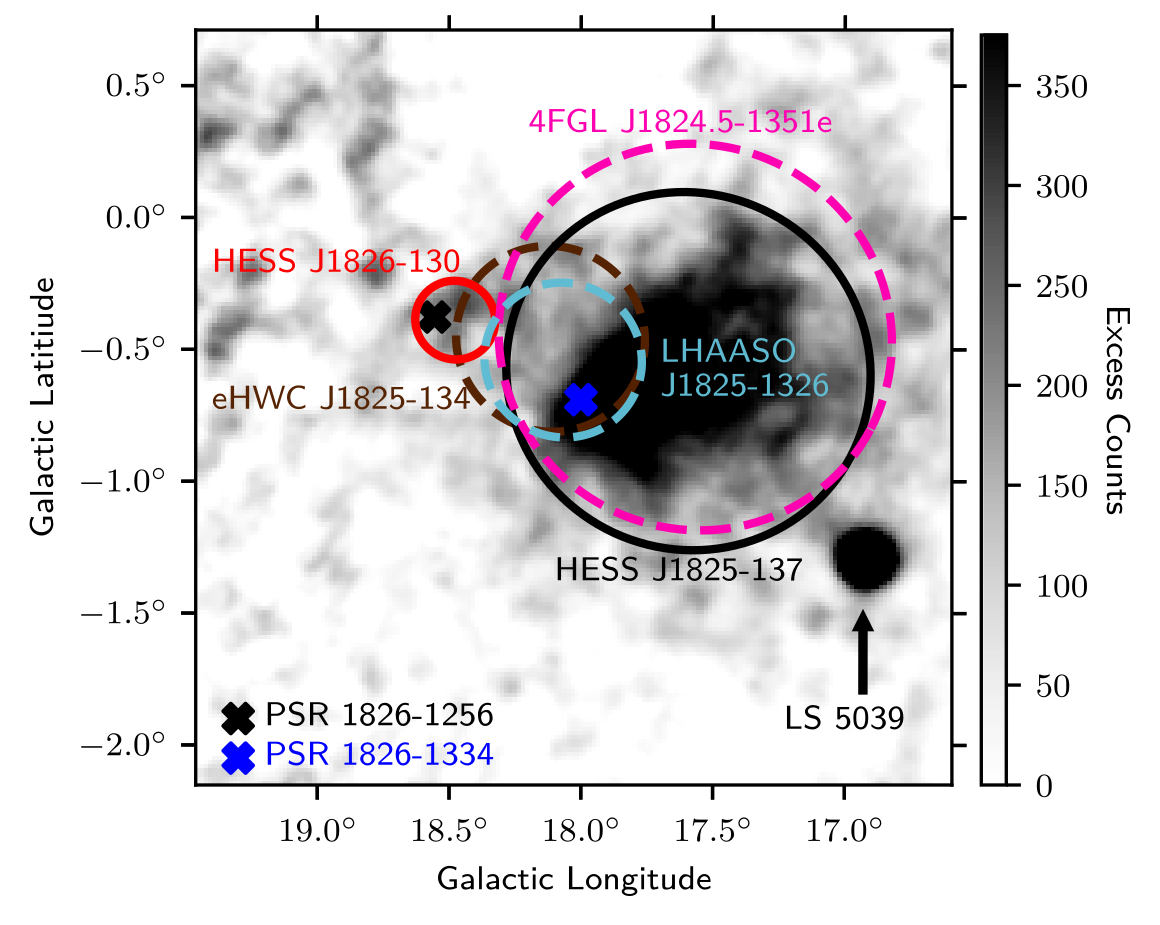}
    \caption{HESS excess counts towards \mbox{HESS\,J1825-137} \protect\citep{2019A&A...621A.116H} overlayed by the regions used to extract the gamma-ray spectra towards \mbox{HESS\,J1825-137} (black) and \mbox{HESS\,J1826-130} (red). \mbox{4FGL\,J1824.5-1351e}, \mbox{eHWC\,J1825-134} and \mbox{LHAASO\,J1825-1326} are shown by the purple, brown and cyan dashed circles respectively with the positions of \mbox{PSR\,J11826-1334} (blue) and \mbox{PSR\,J11826-1256} (black). The position of the nearby binary system \mbox{LS\,5039} is indicated by the black arrow.}
    \label{fig:1825_sources}
\end{figure}

Situated $\ang{0.7}$ north of \mbox{HESS\,J1825-137} (see \autoref{fig:1825_sources}), \HESSminor is a TeV gamma-ray source and possible accelerator of cosmic rays up to \PeV energies \citep{2020PhRvL.124b1102A, 2021Natur.594...33C}. Due to its close proximity to \mbox{HESS\,J1825-137}, \HESSminor was originally considered an extension of \mbox{HESS\,J1825-137} until it was revealed to be a separate source of gamma rays \citep{2018A&A...612A...1H,}. The two nearby supernova remnants (SNRs) \mbox{SNR\,G018.1-0.1} and \mbox{SNR\,G018.6-0.2} \citep{1986AJ.....92.1372O,2006ApJ...639L..25B} were deemed to be unlikely to be associated with \mbox{HESS\,J1826-137} due to their offset positions and small angular diameters \citep{2020A&A...644A.112H}. Instead, the Eel PWN (\mbox{PWN\,G18.5-0.4}) and \mbox{PSR\,J11826-1256} are associated with \mbox{HESS\,J1826-130} based on spatial coincidence \citep{2018A&A...612A...1H}. \mbox{PSR\,J11826-1256} has a spin-down power of $3.6\times 10^{36}\,\ergspersecond$ and characteristic age of $14\,\kiloyear$, well within the range of pulsar properties associated with $\TeV$ PWNe \citep{2005AJ....129.1993M,2018A&A...612A...2H}.
\par
\cite{2019MNRAS.485.1001A} revealed $\GeV$ gamma-ray emission $\ang{\sim2.5}$ to the Galactic south of \mbox{HESS\,J1825-137}. The same study postulated that the $\GeV$ emission from this region originates from cosmic rays accelerated by the SNR or PWN associated with \mbox{HESS\,J1825-137} or a star-forming region such as the Cygnus Cocoon. Comprehensive modelling of the spectral energy distribution (SED) towards the $\GeV$ region suggests that the emission may be reflective of an earlier epoch of the PWN or a combination of \mbox{HESS\,J1825-137} and nearby compact object \mbox{LS\,5039} \citep{2021MNRAS.tmp..976C}.
\par
The PWN associated with HESS\,J1825-137 must be expanding within the progenitor SNR. A large H$\alpha$ rim-like structure discovered by \cite{2008MNRAS.390.1037S} is present towards the south of \mbox{HESS\,J1825-137}. \cite{2016MNRAS.458.2813V} postulated a connection between this rim and another southern H$\alpha$ rim and the progenitor SNR of \mbox{HESS\,J1825-137}. Both structures lie $\approx\!\!\ang{1.7}$ away from \mbox{PSR\,J1826-1334} ($\approx 120\,\pc$ for a distance of $4\,\kpc$), which is consistent with the predicted SNR radius of $130\,\pc$ as suggested by \cite{2009ASSL..357..451D}.
\par 
Electrons released by a pulsar are subject to varying transport processes such as diffusion and/or advection. It has been proposed that advection dominates the particle transport close to the pulsar while diffusion dominates the outer reaches of the nebula \citep{2012ApJ...752...83T,2016MNRAS.460.4135P}. TeV halos around PWN have been suggested to form when electrons escape the PWN into the surrounding ISM where diffusion dominates particle transport \citep{2020A&A...636A.113G,2021PhRvD.104l3017R}. \cite{2019A&A...621A.116H} found that the energy-dependent radial extent of the $\TeV$ PWN associated with \mbox{HESS\,J1825-137} is unlikely to be explained with a diffusion-only scenario and requires an overall bulk flow towards lower Galactic longitudes.
\par
\cite{1984ApJ...283..694K,1984ApJ...283..710K} developed the first 1D magnetohydrodynamic model of the Crab Nebula as an extension to the model first developed by \cite{1974MNRAS.167....1R} by considering the PWNe evolving in a slowly expanding SNR shell. \cite{2018ApJ...860...59K} applied this approach to \mbox{HESS\,J1825-137} and was able to reproduce the size of the PWN and the position of the termination shock ($r_\text{ts}\approx0.03\,\pc$) by assuming a short initial period of \mbox{PSR\,J11826-1334} ($P\approx 1\,\si{\milli\second}$), small braking index ($n\leq 2$), birth spin-down power $\geq 10^{41}\,\ergspersecond$ and evolution in dense environment ($n_\text{ISM}\geq 1\,\centimeterminusthree$). This is supported by the presence of dense molecular clouds towards \mbox{HESS\,J1825-137} as described by \cite{2016MNRAS.458.2813V}. Alternatively,\cite{2011ApJ...742...62V} treated the transport of electrons from \mbox{PSR\,J11826-1334} as a series of uniform, spherical `bubbles' to study the inverse Compton and Synchrotron emission from PWN (e.g. \citep{1997MNRAS.291..162A}. They found that a combined diffusive and advective model was able to predict the multi-wavelengh SED and radial profile towards \mbox{HESS\,J1825-137}. Recently, \cite{2023MNRAS.518.3949L} investigated the gamma-ray emission towards \mbox{HESS\,J1825-137} by combining a 1D diffusion-advection particle transport model with Markov chain Monte Carlo techniques to obtain the best-fitting parameters. However, the observed asymmetric gamma-ray morphology observed towards HESS J1825-137 suggests a similarly asymmetric electron density and/or magnetic field. Moreover, dense molecular clouds would prohibit the escape of electrons out of the PWN, leading to an irregular electron number density distribution around the pulsar, which cannot be predicted by a 1D model and therefore requires a more complex model.
\par
The evolution of the cosmic-ray number density distribution can be described by the Fokker-Planck equation for particle transport (e.g. \cite{1975MNRAS.172..557S,1978A&A....70..367C}). Henceforth, this equation will be called the transport equation. Analytical solutions of the transport equation can be found for specific cases, e.g. isotropic diffusion in a homogeneous environment \citep{1970RvMP...42..237B,1995PhRvD..52.3265A,1996A&A...309..917A}. However, it can only be solved numerically for more complex systems, e.g.\,anisotropic diffusion where the diffusion coefficient varies with position.
\par
In this paper, we present a model that incorporates 3D distributions of the ISM hydrogen number density and magnetic field and solves the transport equation numerically. This model assumes \mbox{PSR\,J11826-1334} to be a source of high-energy electrons and aims at reproducing the X-ray and gamma-ray morphology, spectrum and surface brightness radial profiles towards \mbox{HESS\,J1825-137}.
\section{Particle Transport and Multi-wavelength Emission} \label{sec:particle_transport}

Upon the release from an accelerator, such as a SNR or PWN, cosmic rays are transported through the ISM and experience radiative losses. The evolution of the number density distribution of cosmic rays, $n\equiv n\qty(\gamma,t,\vec{r})$, with Lorentz factor $\gamma\equiv \gamma\qty(\vec{r})$, at position $\vec{r}\equiv (x,y,z)$ and time $t$ after the birth of the accelerator, can be described by (e.g. \cite{1975MNRAS.172..557S,1978A&A....70..367C}):

\begin{equation}
    \begin{aligned}
        \pdv{n}{t}=&\pdv{ }{\gamma}\qty(\dot{\gamma}n)+\nabla\qty(\bar{\bar{D}}\cdot\nabla n)-\nabla\cdot\qty(n\vec{v}_A)-\frac{1}{3}\pdv{ }{\gamma}\qty(\gamma\qty(\nabla\cdot\vec{v}_A))n \\
        &+\pdv{ }{\gamma}\qty(\gamma^2 D_{\gamma\gamma}\pdv{ }{\gamma}\qty(\frac{n}{\gamma^2})) + S\qty(\gamma,t,\vec{r})  \text{ .}     
    \end{aligned} \label{eq:Diffusion_Equation}
\end{equation} 
The first term in \autoref{eq:Diffusion_Equation} gives the evolution of cosmic-ray density due to radiative losses. The second term considers the spatial diffusion of cosmic rays as a second-rank tensor ($\bar{\bar{D}}\equiv\bar{\bar{D}}\qty(\gamma,t,\vec{r})$), allowing preferential direction of transport. The third term describes the evolution of cosmic-ray density due to advection as a co-moving fluid with velocity $\vec{v}_A\equiv\vec{v}_A\qty(\gamma,t,\vec{r})$. The fourth term considers losses due to adiabatic expansion. The fifth term represents the re-acceleration of cosmic rays due to stochastic processes with $D_{\gamma\gamma}$ being the acceleration rate. Finally, $S\qty(\gamma,t,\vec{r})$ is the cosmic-ray source/injection function.
\par
To numerically solve \autoref{eq:Diffusion_Equation}, explicit finite difference techniques forward in time can be used after discretising a region of interest into a grid of voxels with dimension $\Delta x\Delta y \Delta z$ and time step $\Delta t$:
\begin{equation}
    \begin{aligned}
        \frac{n\qty(\gamma,t+\Delta t, \vec{r})-n\qty(\gamma+\Delta\gamma,t, \vec{r})}{\Delta t}&=\qty(\pdv{n}{t})'_\text{diff}+\qty(\pdv{n}{t})'_\text{adv}+\qty(\pdv{n}{t})'_\text{adb} \\
        &+\qty(\pdv{n}{t})'_\text{re-acc}+S\qty(\gamma,t,\vec{r})\text{ ,}
    \end{aligned} \label{eq:Diffusion_Equation_numerical}
\end{equation}
\noindent where prime represents the evolution of the number density distribution \textit{after} radiative losses. The following discussion describes how the implemented model treats individual terms in \autoref{eq:Diffusion_Equation_numerical}.

\subsection{Radiation Losses}

High-energy electrons interact with the ISM via inverse Compton interactions on ambient photons, via Bremsstrahlung with interstellar gas and via synchrotron interactions against magnetic fields (see Appendix\,\ref{sec:non_thermal_em}). The evolution of the electron number density with Lorentz factor $\gamma$ due to radiative losses is given by:

\begin{align}
    \pdv{n}{t}=\pdv{ }{\gamma}\qty(\dot{\gamma}n)\text{ ,} \label{eq:cooling_term}
\end{align}

\noindent where $\dot{\gamma}$ is the cooling rate as given by \cite{2007A&A...474..689M}:

\begin{equation}
	    \begin{aligned}
    		 \dot{\gamma} = b_s\gamma^2+b_c\qty(3\ln\gamma+18.8)+5.3b_b + \sum\limits_{j} b_{\text{IC}}^j\gamma^2 F_{\text{KN}}^j\qty(\gamma)
	     \end{aligned}\text{ ,} \label{eq:cooling_rate}
	\end{equation}

\noindent for the case of ionisation or Bremsstrahlung losses in neutral hydrogen. Here, $j$ sums over all radiation fields (CMB, infrared and optical photons), $b_s$, $b_c$, $b_b$ and $b_{IC}$ are the coefficients for synchrotron losses, Coulomb losses, Bremsstrahlung losses and inverse Compton losses respectively and $F_\text{KN}$ is the Klein-Nishina cross section (see \autoref{eq:klein_nishina_cross_section}). The photon fields were assumed to be constant across the 3D grid. The general solution to \autoref{eq:cooling_term} is:

\begin{align}
    n\qty(\gamma,t+\Delta t)&=\frac{\dot{\gamma}_0}{\dot{\gamma}}n\qty(\gamma_0,t)\text{ ,}
\end{align}

\noindent where $\gamma_0\equiv\gamma_0\qty(x,y,z) = \gamma+\Delta\gamma$ is the Lorentz factor at time $t$ before electrons cool to Lorentz factor $\gamma$ at time $t+\Delta t$.

\subsection{Diffusion}\label{sec:diffusion}

Over distances smaller than the gyro-radius, $r_g$, electrons propagate through the ISM via ballistic motion. In a medium with randomised magnetic turbulence ($\delta B$), electrons scatter and the motion switches to a diffusive regime for distances larger than the gyro-radius (e.g. \cite{2015PhRvD..92h3003P}). For a simple case of isotropic diffusion in magnetic field $B\qty(\vec{r})$, the diffusion tensor in \autoref{eq:Diffusion_Equation} becomes a scalar; $\overline{\overline{D}}\rightarrow D\qty(E,\vec{r})$, where $E$ is the energy of the cosmic ray. 
\par
Suppression of cosmic-ray diffusion (compared to the Galactic average) is to be expected towards PWNe and SNRs where magnetic field turbulence is enhanced and the diffusion coefficient, $D\qty(E,\vec{r})$, can be parameterised by (e.g. \cite{2007Ap&SS.309..365G}):

\begin{align}
    D\qty(E,\vec{r})&=\chi D_0\qty(\frac{E/\GeV}{B\qty(\vec{r})/3\,\si{\micro G}})^{\delta}\text{ ,}\label{eq:diffusion}
\end{align}

\noindent where $D_0=3\times 10^{27}\,\si{\centi\meter\squared\per\second}$ is the average Galactic diffusion coefficient at $1\,\GeV$, $\delta=0.5$ following cosmic-ray observations (e.g. see \cite{2007ARNPS..57..285S}) and the diffusion suppression factor, $\chi$, takes values\,$\leq1$ depending on the environment \citep{1990acr..book.....B}. For example, \cite{2007Ap&SS.309..365G} found that highly suppressed diffusion ($\chi\sim 0.01$) in molecular clouds can significantly affect the shape of the observed gamma-ray spectrum. However, the diffusion suppression factor is not well constrained and a variety of $\chi$ have been found, e.g. \cite{2010MNRAS.409L..35L}, \cite{2010A&A...516L..11G} and \cite{2010sf2a.conf..313G} found values of $\chi=0.1$, $0.01$ and $0.06$ towards \mbox{SNR\,W28}, respectively. Similarly, \cite{2008MNRAS.390..683P} showed that the suppression factor towards the star-forming region Sgr\,B2 takes values $<0.02$ based on the radio synchotron flux. \cite{2023MNRAS.518.3949L} found a diffusion coefficient of $1.4\times 10^{26}\,\si{\centi\meter\squared\per\second}$ at $1\,\GeV$ towards \mbox{HESS\,J1825-137}. 
\par
Assuming isotropic inhomogeneous diffusion, the diffusive component of \autoref{eq:Diffusion_Equation_numerical} is given by:

\renewcommand{\arraystretch}{0.8}
\begin{equation}
    \begin{aligned}
        \qty(\pdv{n}{t})_\text{diff}
        &= \frac{1}{\Delta i^2}\sum_{i=x,y,z} \\
        &\qty[\frac{D\qty(\gamma,i+\Delta i)+D\qty(\gamma,i)}{2}]\cdot 
        \qty[n\qty(\gamma,t, i+\Delta i)-n\qty(\gamma,t,i)] \\
        +&\qty[\frac{D\qty(\gamma,i-\Delta i)+D\qty(\gamma,i)}{2}]\cdot
        \qty[n\qty(\gamma,t, i-\Delta i)-n\qty(\gamma,t,i)]\text{ ,}
    \end{aligned} \label{eq:diff_num}
\end{equation} 
\noindent where $D\qty(\gamma,i)$ is the diffusion coefficient from \autoref{eq:diffusion}. The central finite difference technique used in \autoref{eq:diff_num} only considers the transport of electrons to/from the surrounding voxels. If the time step is too large, electrons travel across more than one voxel and are lost from the system. The finite difference technique is then said to be numerically `unstable'. Using Von Neuman stability analysis (e.g. see \cite{alma9929637001811}), \autoref{eq:diff_num} is stable when:

\begin{equation}
    \begin{aligned}
        \Delta t&\leq\frac{\Delta i^2}{2D\qty(i)}\bigg |_\text{min}\text{ .}
    \end{aligned}\label{eq:time_step_diffusion}
\end{equation}

\subsection{Advection} \label{sec:advection_theory}

For simplicity, the velocity due to the bulk flow of electrons ($\vec{v}_A\equiv [v_\text{A,x},v_\text{A,y},v_\text{A,z}]$) was assumed to be spatially-independent and energy-independent across the region of interest. This assumption is reasonable for a model describing HESS J1825-137 (see \autoref{sec:multizone}) as \citep{2019A&A...621A.116H} implied an overall bulk motion towards lower Galactic longitudes. Using explicit finite difference techniques, the advective component of \autoref{eq:Diffusion_Equation_numerical} is given by \autoref{eq:advec_eq_num}:

\renewcommand{\arraystretch}{0.8}
\begin{equation}
    \begin{aligned}
        \qty(\pdv{n}{t})_\text{adv}&=-\sum_{i=x,y,z} v_{\text{A,i}}\frac{1}{\Delta i}
        \begin{cases}
            n\qty(\gamma,t,i+\Delta i)-n\qty(\gamma,t,i)\text{,}& v_\text{A,i}<0 \\
            n\qty(\gamma,t,i)-n\qty(\gamma,t,i-\Delta i)\text{,}& v_\text{A,i}>0
        \end{cases}\text{,}
    \end{aligned} \label{eq:advec_eq_num}
\end{equation} 
\par
\noindent where $v_\text{A,i}$ is the component of advective velocity in the ith direction. \autoref{eq:advec_eq_num} uses the forward difference method to approximate the derivative in \autoref{eq:Diffusion_Equation_numerical} when $v_\text{A,i}<0$ and the backward difference method when $v_\text{A,i}>0$. 
\par
For \autoref{eq:advec_eq_num} to be numerically stable, the time step must be chosen so that an electron does not travel across more than one voxel in time $\Delta t$:
\begin{equation}
    \begin{aligned}
        \Delta t \leq \frac{\Delta i}{\abs{v_\text{A,i}}}\bigg|_\text{min} \text{ .}
    \end{aligned} \label{eq:time_step_advection}
\end{equation}
\noindent The time step must satisfy both \autoref{eq:time_step_diffusion} and \autoref{eq:time_step_advection} when modelling a scenario including both diffusion and advection.

\subsection{Adiabatic Expansion and Re-acceleration of Electrons} \label{sec:adiabatic_reacc}

The spatially-independent advective velocity assumed in our model results in zero adiabatic losses in \autoref{eq:Diffusion_Equation} ($\nabla\cdot \vec{v}_A=0$). Moreover, studies such as \cite{2010ApJ...715.1248T} and \cite{2016MNRAS.460.4135P} who considered spherically symmetric advection concluded that adiabatic losses are dominant over radiative losses for electrons $< 1\,\TeV$ (equivalent to gamma-ray emission $<20\,\GeV$). As we are interested in the VHE gamma-ray range which is not dominated by adiabatic losses, adiabatic expansion is not considered here but is left for future work.
\par
The termination shock (TS) of pulsar wind has been proposed as a site for the re-acceleration of electrons through diffusive shock acceleration (DSA). By ensuring the voxel width ($\Delta x$,$\Delta y$,$\Delta z$) is larger than the diameter of the TS ($0.2\,\pc$, \citep{2006ARA&A..44...17G}), electrons are both injected and re-accelerated within the same voxel. Therefore, the source term in \autoref{eq:Diffusion_Equation_numerical} treats the injected electron spectra as the spectra obtained after re-acceleration due to the TS. Furthermore, magnetohydrodynamic models (e.g. \cite{2010MNRAS.402..321L,2015SSRv..191..519S}) suggest that DSA at the TS is too suppressed for electron acceleration up to energies responsible for the $\TeV$ emission seen towards PWNe. Hence, the re-acceleration of electrons is left for future work.

\subsection{Multi-wavelength Photon Production}

The final electron number density distribution was obtained by solving \autoref{eq:Diffusion_Equation_numerical} in discrete time steps $\Delta t$ until the desired age was reached. Based on the obtained electron number densities, the multi-wavelength photon emission was derived for each voxel and summed along the line of sight, $z$, to obtain the 2D photon distribution. Equations\,\ref{eq:synchrotron}, \ref{eq:IC_emission} and \ref{eq:bremsstrahlung_flux} gives the flux from synchrotron, inverse Compton and Bremsstrahlung interactions respectively.

\section{Application to HESS\,J1825-137} \label{sec:multizone}

The modelling described in \autoref{sec:particle_transport} was applied to the PWN \mbox{HESS\,J1825-137} with the pulsar \mbox{PSR\,J11826-1334} being the accelerator of high-energy electrons. \mbox{PSR\,J11826-1334} is located at $\ell=\ang{18}$ $b=-\ang{0.69}$ and has a proper motion of $\approx 440\,\si{\kilo\meter\per\second}$ (assuming a distance of $4\,\kpc$) approximately perpendicular to the extended $\TeV$ emission (see \autoref{fig:nanten_data}) \citep{2005AJ....129.1993M}. Hence, the proper motion of the pulsar is unlikely to be related to the asymmetric gamma-ray emission and our model assumed that electrons are injected at the current position of the pulsar for simplicity. Two different ages of \mbox{PSR\,J11826-1334} were considered, the characteristic age of $21.4\,\kiloyear$ and the older age of $40\,\kiloyear$ suggested by \cite{2011ApJ...742...62V}. The presence of the TeV halo toward \mbox{HESS\,J1825-137} indicates that the system is a middle aged PWN where diffusive particle transport dominates the outer reaches of the Nebula \citep{2012ApJ...752...83T,2016MNRAS.460.4135P,2020A&A...636A.113G,2021PhRvD.104l3017R}. However, \cite{2019A&A...621A.116H} suggested that both diffusive and advective transport mechanisms are present in \mbox{HESS\,J1825-137}.
\par
Each voxel in the 3D grid had a volume of $\Delta x \Delta y  \Delta z$, where $\Delta z$ is the voxel length in the line of sight and $\Delta x$ and $\Delta y$ are the voxel length along Galactic longitude and latitude respectively. For the purposes of this study, we utilised a $200\,\pc \times 200\,\pc \times 200\,\pc$ grid consisting of voxels of size $2\,\pc \times 2\,\pc \times 2\,\pc$ ($\approx \ang{0.03}\times\ang{0.03}\times 2\,\pc$). The pulsar was located in the centre of the grid with the central $z$ slice lying at distance $4\,\kpc$. The time step used for the finite difference technique was $\approx 8\,\text{yr}$.

\subsection{Electron Injection}\label{ref:particle_spectra_evolution}

High-energy electrons were injected into the 3D grid by \mbox{PSR\,J1826-1334} and follow an exponential cutoff power-law:

\begin{equation}
    \begin{aligned}
        S\qty(E,t)&=A\cdot\qty(\frac{E}{1\,\TeV})^{-\Gamma}\exp(-\frac{E}{E_c})\text{ ,}
    \end{aligned}
\end{equation}
\noindent following the observed $\TeV$ gamma-ray emission (e.g. see \cite{1970RvMP...42..237B}) as observed by H.E.S.S. \citep{2019A&A...621A.116H}, where $E_c$ is the cutoff energy and $A$ is the normalisation factor such that:

\begin{equation}
    \begin{aligned}
        L_\text{inj}\qty(t)&=\int_{E_\text{min}}^{E_\text{max}} S\qty(E,t) \dd{E}\text{ ,}
    \end{aligned}
\end{equation}
\noindent with $L_\text{inj}\equiv\eta\dot{E}$ being the electron injection luminosity, $\eta<1$ is the conversion efficiency of the pulsar spin-down power, $E_\text{min}=1\,\MeV$ and $E_\text{max}=500\,\TeV$. The spin-down power, $\dot{E}\qty(t)$ at time~$t$ is given by \citep{neutron_star}:

\begin{equation}
    \begin{aligned}
        \dot{E}\qty(t)&=
        \dot{E}\qty(t=t_\text{age})\qty[1+\qty(n-1)\frac{\dot{P}\qty(t-t_\text{age})}{P}]^{-\Gamma_{\mathscr{n}}} \text{ ,}
    \end{aligned} \label{eq:Edot_evo}
\end{equation}
\noindent where $n$ is the braking index of the pulsar, $\Gamma_{n}\equiv \qty(n+1)/\qty(n-1)$ and $\dot{E}\qty(t=t_\text{age})$, $P$ and $\dot{P}$ are the spin-down power, period and spin-down period of the pulsar at the current age $t_\text{age}$.

\subsection{The Environment Towards HESS\,J1825-137}

\subsubsection{Magnetic Field} \label{sec:B_field}

\begin{figure}
    \centering
    \includegraphics[width=\columnwidth]{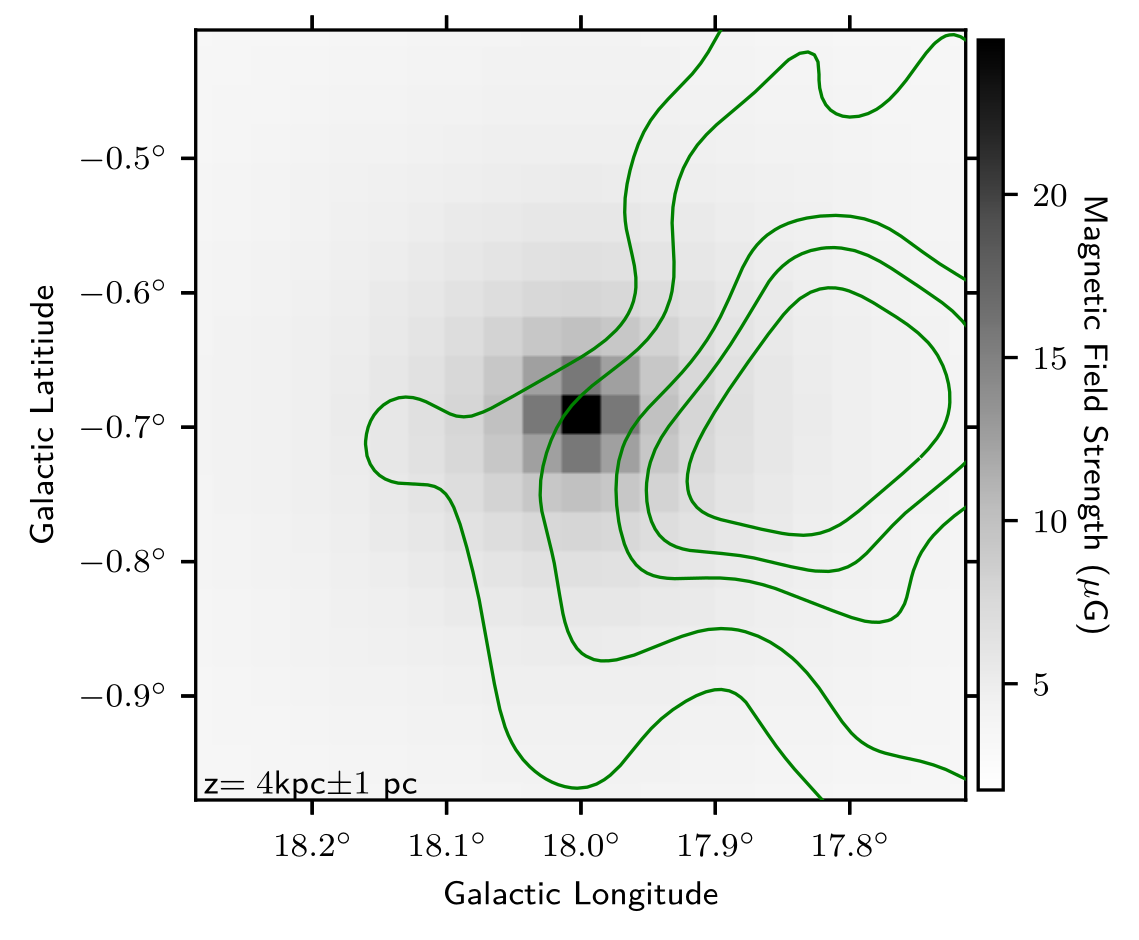}
    \caption{The radially symmetric magnetic field due to the pulsar at slice $z=4\,\kpc\pm 1\,\pc$. Overlaid are the green H.E.S.S. contours at 4, 5, 6, 7, 8 and 9$\sigma$ \protect\citep{2019A&A...621A.116H}.}
    \label{fig:Bfield}
\end{figure}

Following \cite{2011ApJ...742...62V}, the magnetic field due to the PWN was assumed to follow a time-independent power-law with a decreasing magnetic field strength varying with distance $r$ from the pulsar:

\begin{equation}
    \begin{aligned}
        B_\text{PWN}\qty(r) &=B_0\qty(\frac{r}{r_\text{ts}})^{-\beta}\text{ ,}  \label{eq:PWN_Bfield}
    \end{aligned}
\end{equation}

\noindent where $r_\text{ts}=0.03\,\pc$ is the radius of the termination shock, and $B_0$ and $\beta$ are free parameters optimised to match the multi-wavelength SED of \mbox{HESS\,J1825-137}. \cite{2011ApJ...742...62V} suggested $\beta=-0.69$ and $B_0=400\,\si{\micro G}$ for an age of $40\,\kiloyear$. Note that \cite{2011ApJ...742...62V} considered an additional dependence on the spin-down energy of the pulsar which was not considered in this study.

\subsubsection{Interstellar Medium} \label{sec:morphology}

\begin{figure}
    \centering
    \includegraphics[width=\columnwidth]{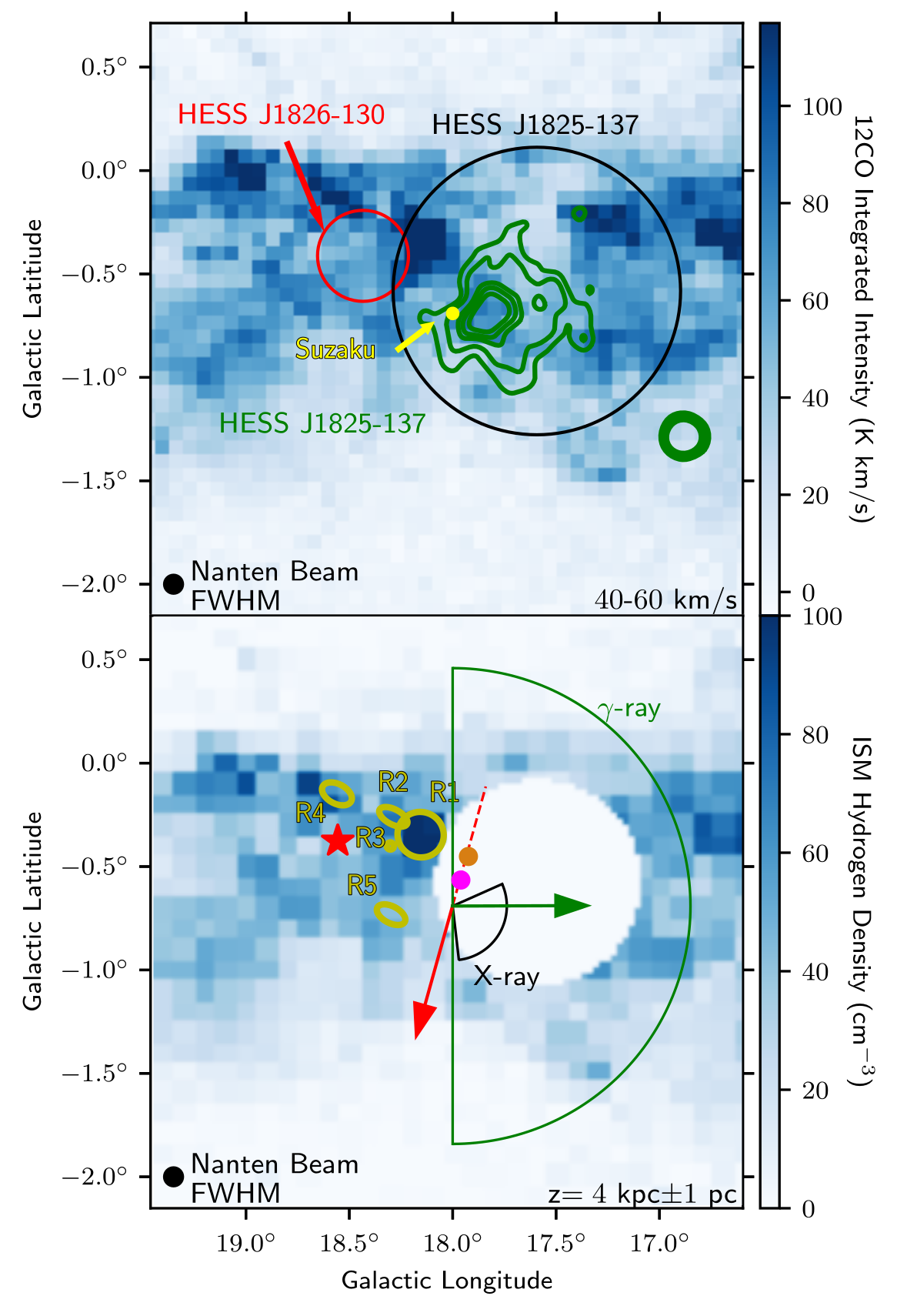}
    \caption{(\textit{top}) Nanten $^{12}$CO(1-0) integrated intensity in the velocity range $40-60\,\kmpersec$ corresponding to $3.5-4.5\,\si{\kilo\parsec}$ overlaid by green H.E.S.S. contours (at 4, 5, 6, 7, 8 and 9$\sigma$). The yellow dot represents the Suzaku region~A as defined in \protect\cite{2009PASJ...61S.189U} and is used to extract the X-ray SED. The region used to obtain the gamma-ray spectra towards \mbox{HESS\,J1825-137} and \mbox{HESS\,J1826-130} are shown in black and red respectively. (\textit{bottom}) Calculated ISM number density across the 3D grid at $4\,\kpc\pm 1\,\pc$ where the voxels within the PWN extent ($R<\ang{0.5}$) are set to a density of $0.5\,\centimeterminusthree$ to represent the bubble that has been swept out by the stellar wind from the progenitor star. The proper motion of the pulsar is shown by the red arrow with the projected birthplaces indicated by the red-dashed line. The projected birthplaces for ages $21\,\kiloyear$ and $40\,\kiloyear$ are indicated by the magenta and brown dot respectively. The direction of advective particle transport suggested by \protect\cite{2019A&A...621A.116H} is shown by the green arrow. The black and green segments represent the regions used to extract the X-ray and gamma-ray surface brightness radial profile respectively. Molecular clouds R1-R5 from \protect\cite{2016MNRAS.458.2813V} are shown in yellow with the position of \mbox{PSR\,J11826-1256} indicated by the red star. The width and height of the voxel in the 2D slice is $2\,\pc$ compared to the minimum Nanten resolution of $5\,\pc$ (assuming a distance of $4\,\kpc$).}
    \label{fig:nanten_data}
\end{figure}

The Nanten $^{12}$CO(1-0) survey \citep{2004ASPC..317...59M} was used to trace the column density of molecular hydrogen towards \mbox{HESS\,J1825-137}:
\begin{equation}
    \begin{aligned}
        N_{H_2}=X_\mathrm{12CO}W_\mathrm{12CO}
    \end{aligned}
\end{equation}
\noindent where $W_\mathrm{12CO}$ is the integrated intensity of the gas. The scaling factor $X_\mathrm{12CO}=1.5\times 10^{20}\,\si{\per\centi\meter\squared\per\kelvin\per\kilo\meter\second}$ is assumed to be constant over the Galactic plane but may vary with galactocentric radius ($1.3-1.5$ per $\kpc$) \citep{2004A&A...422L..47S}. The length of the 3D grid ($<1\,\kpc$) allows the assumption of a constant $X_\mathrm{12CO}$ towards the region of interest. PSR\,J1826-1334 has a dispersion measure distance of $4\,\kpc$, corresponding to a velocity of $50\,\kmpersec$ using the Galactic Rotation model \citep{1993A&A...275...67B}. As there may be local motion of the gas unrelated to Galactic rotation, we considered a velocity range of $40-60\,\kmpersec$ ($\SIrange{3.5}{4.5}{\kilo\parsec}$) consistent with \cite{2016MNRAS.458.2813V}.  Atomic hydrogen in the same velocity range contributes less than $1\%$ to the total column density towards \mbox{HESS\,J1825-137} and thus was not considered \citep{2016MNRAS.458.2813V, 2021MNRAS.tmp..976C}. Assuming that all the gas in the $40-60\,\kmpersec$ velocity range lies within the 3D grid and the density along the line of sight is constant, the number density of a voxel with column density $N_{H_2}$ is given by:

\begin{equation}
    \begin{aligned}
    n_H&= \frac{N_H}{200\,\pc}\text{ ,}
    \end{aligned} \label{eq:density_form}
\end{equation}

\noindent where $N_{H}\equiv 2.8N_{H_2}$ considers a $20\%$ He component.
\par
Stellar winds from the progenitor star of \mbox{PSR\,J11826-1334} pushes out gas in the nearby vicinity \citep{1975ApJ...200L.107C}. The subsequent supernova explosion creates a `bubble' of hot dense gas around a low-density interior. A region of low-density gas in the $40-60\,\kmpersec$ velocity range can be seen towards the centre of the $\TeV$ emission in \autoref{fig:nanten_data}. To include this, any voxels lying within the extent of the PWN volume (a sphere centered on the pulsar with radius $\ang{0.5}\approx 35\,\pc$) was set to a density of $0.5\,\centimeterminusthree$ based on the average densities expected within massive stellar wind bubbles \citep{1977ApJ...218..377W}. The Nanten $^{12}$CO(1-0) integrated intensity between $40-60\,\kmpersec$ and calculated ISM number density for the central slice lying at distance $4\,\kpc$ can be seen in \autoref{fig:nanten_data}. Any difference between the bottom and top panel of \autoref{fig:nanten_data} was due to the different resolutions of the 3D grid and Nanten.
\par
Turbulent motion in the ISM results in an amplification of the magnetic field, suppressing the diffusion of electrons as they travel through the ISM as given by \autoref{eq:diffusion}. Figure\,6 of \cite{2016MNRAS.458.2813V} shows a three-coloured image of the CS(1-0) and $\text{NH}_3$ integrated intensity between $40-60\,\kmpersec$ and the $\text{H62}\alpha$ integrated intensity between $45-65\,\kmpersec$ towards the cloud defined as R1 (see \autoref{fig:nanten_data} for the position of clouds R1-R5 from \cite{2016MNRAS.458.2813V}). This suggests that cloud R1 is highly turbulent with a minimum magnetic field strength of $21\,\si{\micro G}$ based on the density of $600\,\centimeterminusthree$ calculated by \cite{2016MNRAS.458.2813V} (see \autoref{eq:crutchers_relation}). The amplification of the magnetic field towards cloud R1 was considered in \autoref{sec:result_model3}. Given the likely physical proximity to \mbox{HESS\,J1825-137}, cloud R1 may act as a barrier for electrons escaping into \mbox{HESS\,J1826-130} from PSR\,J1826-1334 \citep{2016MNRAS.458.2813V}.

\subsubsection{Soft Photon Fields}
The photon fields around \HESSmain was estimated utilising the radiation field model described by \cite{2017MNRAS.470.2539P}; the far-infrared field (FIR) with temperature $T=40\si{\kelvin}$ and energy density $U=1\,\si{\electronvolt\per\centi\meter\cubed}$, near infrared field (NIR) with temperature $T=500\,\si{\kelvin}$ and energy density $0.4\,\si{\electronvolt\per\centi\meter\cubed}$ and optical light with temperature $T=3500\si{\kelvin}$ and energy density of $U=1.9\,\si{\electronvolt\per\centi\meter\cubed}$.

\subsection{Multi-wavelength Observations} \label{sec:observations}

The modelled gamma-ray SED of \mbox{HESS\,J1825-137} was optimised to the $\TeV$ gamma-ray energy flux presented by \cite{2019A&A...621A.116H} and the $\GeV$ spectrum from the 4FGL catalogue \citep{2020ApJS..247...33A}. To compare the modelled surface brightness radial profile to \mbox{Figure\,6} from \cite{2019A&A...621A.116H}, a collection area of  $0.25\,\si{\kilo\meter\squared}$ \citep{2005AIPC..745..611B} and observation time of $387\,\si{hr}$ was used. The X-ray SED and surface brightness radial profile was optimised to the results presented by \cite{2009PASJ...61S.189U} using a collection area of $0.029\,\si{\meter\squared}$.
\par
To investigate the gamma-ray contamination of \mbox{HESS\,J18260-130}, by \mbox{HESS\,J1825-137}, we utilised the gamma-ray SED presented by  \cite{2020A&A...644A.112H} and the spectrum from the 4FGL catalogue \citep{2020ApJS..247...33A}. \cite{2020A&A...644A.112H} estimated that the gamma-ray contamination to be $40\%$ for photon energies below $1.5\,\TeV$ and $20\%$ above $1.5\,\TeV$. The modelled X-ray emission towards \mbox{HESS\,J1826-130} was constrained by the ROSAT X-ray upper limit calculated using the ROSAT X-ray background tool \citep{2019ascl.soft04001S}. The regions used to extract the X-ray and gamma-ray SED towards \mbox{HESS\,J1825-137} and \mbox{HESS\,J1826-130} are shown in \autoref{fig:nanten_data}.
\subsection{Results} \label{sec:results}
\begin{table}
    \centering
    \begin{threeparttable}
        \renewcommand{\thefootnote}{\fnsymbol{footnote}}
        \caption{Model parameters used for the application towards \mbox{HESS\,J1825-137}. Fixed parameters refer to those constrained by measurements and non-fixed refers to those that are optimised to observations discussed in \autoref{sec:observations}.}
        \centering
        \begin{tabular}{lll}
    		\toprule
    		Fixed Parameters & Value & Reference  \\
    		\midrule
            $t$ & $21\,\kiloyear$ \& $40\,\kiloyear$ & a, b  \\
            $d$ & $4\,\kpc$ & a \\ 
            $P$ & $101\,\si{\milli\second}$  & a  \\
            $\dot{P}$ & $7.5\times 10^{-14}\,\si{\second\per\second}$ & a  \\
            $\dot{E}$ & $2.8\times 10^{36}\,\ergspersecond$ & a \\
            $\Delta x$, $\Delta y$, $\Delta z$ & $2\,\pc$ & \\
            $\Delta t$ & $8\,\si{yr}$ & \\
            $E_\text{min}$ & $1\,\MeV$ & \\ 
            $E_\text{max}$ & $500\,\TeV$ & \\ 
            $D_0$ & $3\times 10^{27}\,\si{\centi\meter\squared\per\second}$ & c \\
            $r_\text{ts}$ & $0.03\,\pc$ & \\
            $U_\text{CMB}$, $T_\text{CMB}$ & $0.26\,\si{\electronvolt\per\centi\meter\cubed}$, $2.72\,\si{K}$ & e \\
            $U_\text{NIR}$, $T_\text{NIR}$ & $1\,\si{\electronvolt\per\centi\meter\cubed}$, $500\,\si{K}$ & e \\
            $U_\text{FIR}$, $T_\text{FIR}$ & $0.4\,\si{\electronvolt\per\centi\meter\cubed}$, $40\,\si{K}$ & e \\
            $U_\text{Opt}$, $T_\text{Opt}$ & $1.9\,\si{\electronvolt\per\centi\meter\cubed}$, $3500\,\si{K}$ & e \\
            \midrule
            Non-fixed Parameters & Value & Reference  \\
    		\midrule
            $\eta$ & $<1$ & \\ 
            $\chi$ & $<1$ & c \\
            $\Gamma$ & - & \\
            $E_c$ & - & \\
            $B_0$ &  - & \\
            $\beta$ & - & \\
            $n$ & 2-3 &  \\
            $\vec{v}_A(\ell,b,z)$ & ($<0.01c,0,0$) & d \\ 
            $B_{1826}$ & $\star$ & \\
    		\bottomrule
    	\end{tabular}
        \begin{tablenotes}
        \item $\star$ See \autoref{sec:result_model3}
        \item a. \cite{2005AJ....129.1993M}
        \item b. \cite{2011ApJ...742...62V}
        \item c. \cite{1990acr..book.....B}
        \item d. \cite{2019A&A...621A.116H}
        \item e. \cite{2017MNRAS.470.2539P}
        \end{tablenotes}
        \label{tab:paramter_range}
    \end{threeparttable}
\end{table}

\begin{table*}
    \begin{threeparttable}
        \renewcommand{\thefootnote}{\fnsymbol{footnote}}
        \caption{Model parameters that match the multi-wavelength SED and gamma-ray morphology towards \mbox{HESS\,J1825-137}. See \autoref{sec:systematic_variation} for the $10\%$ and $20\%$ systematic variation of parameters.}
        \centering
        \begin{tabular}{l|lllll}
        \toprule
        Parameter & Model 1 ($21\,\kiloyear$) & Model 1 ($40\,\kiloyear$) & Model 2 ($0.002c$) & Model 3$^*$ ($60\,\si{\micro G}$) & Model 3 ($60\,\si{\micro G}$)  \\
        \midrule
        $\eta$ & $10.7$ & $0.14$ & $0.14$ & $0.14$ & $0.14$ \\
        $\chi$ & $0.25$ & $0.1$ & $0.1$ & $0.1$ & $0.1$ \\
        $\Gamma$ & $2.0$ & $1.9$ & $1.9$ & $1.9$ & $1.9$ \\
        $E_c$ ($\TeV$) & $40$ & $500$ & $500$ & $500$ & $500$ \\
        $B_0$ ($\si{\micro G}$) & $70$ & $450$ & $450$ & $450$ & $450$ \\
        $\beta$ & $-0.9$ & $-0.7$ & $-0.7$ & $-0.7$ & $-0.7$ \\
        $n$ & $2$ & $2$ & $2$ & $2$ & $2$ \\
        $v_{A}$ & - & - & $0.002c$ & - & $0.002c$ \\
        $B_\text{J1826}$ ($\si{\micro G}$) & - & - & - & $60$ & $60$\footnote{1}  \\
        $\dot{E}_\text{birth}$ ($\ergspersecond$) & $2.1\times 10^{37}$ & $1.1\times 10^{40}$ & $1.1\times 10^{40}$ & $1.1\times 10^{40}$ & $1.1\times 10^{40}$ \\
        \bottomrule
    \end{tabular}
    \begin{tablenotes}
        \item \footnotemark[1]{See \autoref{sec:discussion_model3}}
    \end{tablenotes}
    \label{tab:matched_parameters}
    \end{threeparttable}
\end{table*}

The full list of model parameters is summarised in \autoref{tab:paramter_range}, including any constraints based on measurements. A computationally quick single-zone model, where the electron number density is derived using a uniform sphere, was utilised to investigate a large range of parameters to gain insight into \mbox{HESS\,J1825-137}. The results of the single-zone modelling are summarised in Appendix.\,\ref{sec:singlezone_mod}. For an age of $21\,\kiloyear$, the single-zone model required electrons to follow an exponential cutoff power-law with spectral index $\Gamma=2.1$ and cutoff $E_c=40\,\TeV$ to match the observed gamma-ray SED while an older age of $40\,\kiloyear$ required an index of $\Gamma=2.1$ and cutoff of $E_c=50\,\TeV$.
\par
In the following, we present three applications of our model towards \mbox{HESS\,J1825-137}. All models incorporated a simple assumption of isotropic diffusion and radiative losses as described in \autoref{sec:diffusion}. Model~1 considered both ages of \mbox{PSR\,J11862-1334}, $21\,\kiloyear$ and $40\,\kiloyear$. Model~2 introduced an additional advective component to Model~1 with velocity $\Vec{v}_A=[v_A,0,0]$ as suggested by \cite{2019A&A...621A.116H} to explain the asymmetric gamma-ray morphology. \cite{2019A&A...621A.116H} constrained the total flow velocity to be $<0.01c$. Model~3 expanded on Model~2 by including turbulent ISM towards \mbox{HESS\,J1826-130} (see \autoref{sec:morphology}) to reduce the contamination by \mbox{HESS\,J1825-137}. The model parameters were chosen based on the observations discussed in \autoref{sec:observations} with the parameter list shown in \autoref{tab:paramter_range}. The parameters we found to match the multi-wavelength SED and morphology are shown in \autoref{tab:matched_parameters}.

\subsubsection{Model~1 ($21$ \& $40\,\kiloyear$) - Isotropic Diffusion}

\begin{figure*}
    \centering

    \includegraphics{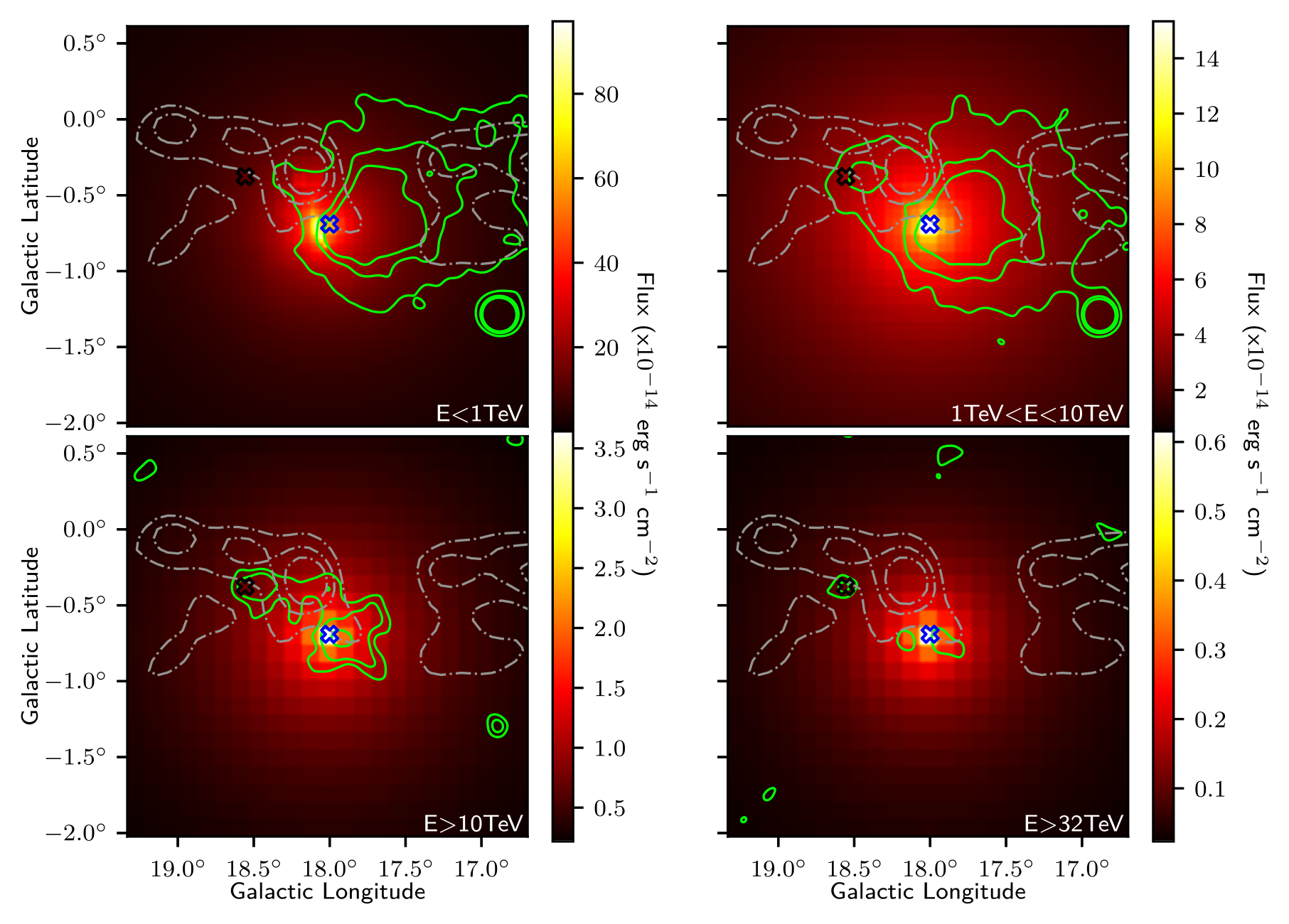}    
    
    \begin{subfigure}[c]{\columnwidth}
       \includegraphics{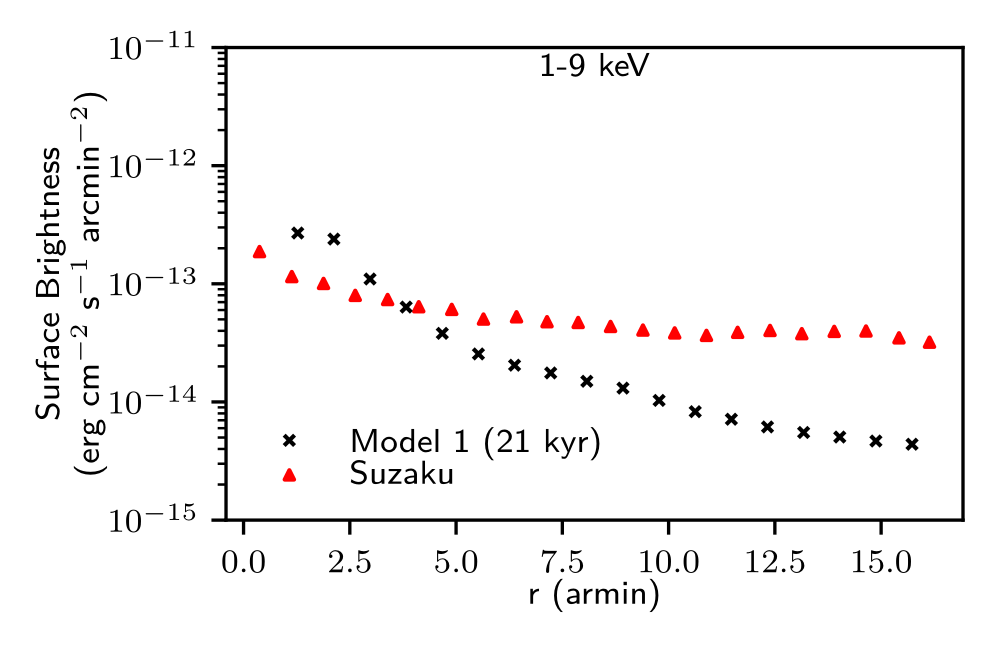}
    \end{subfigure}
    ~ 
    \begin{subfigure}[c]{\columnwidth}
       \includegraphics{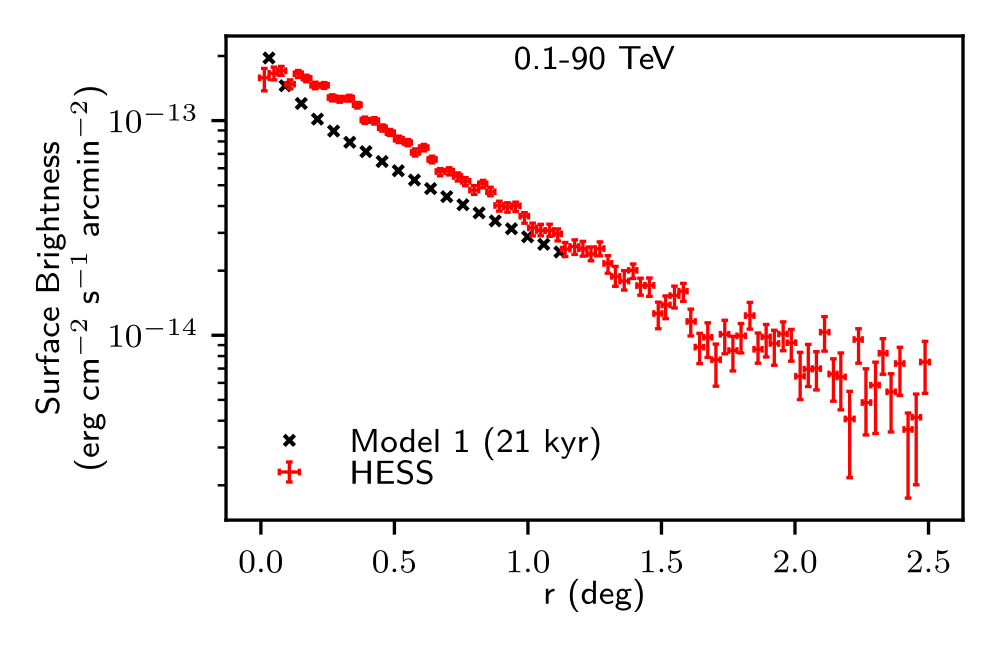}
    \end{subfigure}

    \begin{subfigure}[c]{\columnwidth}
       \includegraphics{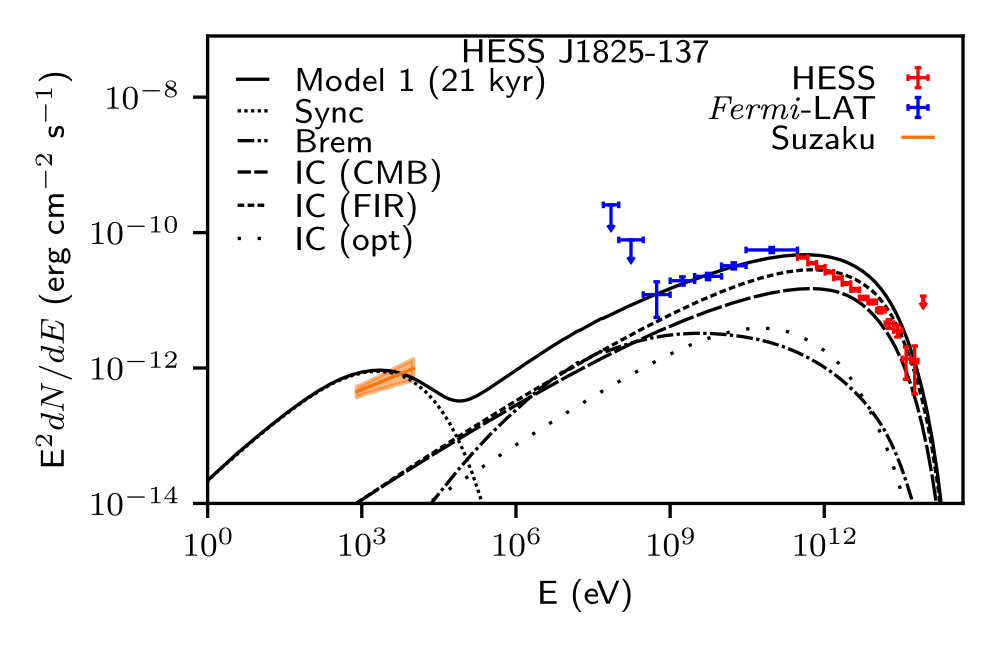}
    \end{subfigure}
    \caption{Model~1 ($21\,\kiloyear$), see \autoref{tab:matched_parameters} for model parameters. (\textit{top \& top-middle}) Modelled gamma-ray morphology towards HESS\,J1825-137 in different energy bands overlaid by green HESS significance contours ($5$, $10$ and $15\sigma$ for $E<10\,\TeV$ and $3$, $5$ and $10\sigma$ for $E>10\,\TeV$) and grey $40$, $50$ and $60\sigma$ Nanten $^{12}\text{CO}$ integrated intensity contours. The positions of \mbox{PSR\,J11826-1334} and \mbox{PSR\,J11826-1256} are indicated by the empty blue and black crosses respectively. (\textit{bottom-middle} )$1-9\,\keV$ X-ray (\textit{left}) and $1-91\,\TeV$ gamma-ray (\textit{right}) surface brightness radial profiles in comparison to Suzaku \protect\citep{2009PASJ...61S.189U} and HESS observations \protect\citep{2019A&A...621A.116H} respectively. (\textit{bottom}) SED towards HESS\,J1825-137 with the orange Suzaku X-ray spectral fit, blue \mbox{4FGL\,J1824.5-135e} flux observations \protect\citep{2020ApJS..247...33A} and red \mbox{HESS\,J1825-137} flux observations.} 
    \label{fig:multizone_21}
\end{figure*}

\begin{figure*}
    \centering
    \includegraphics{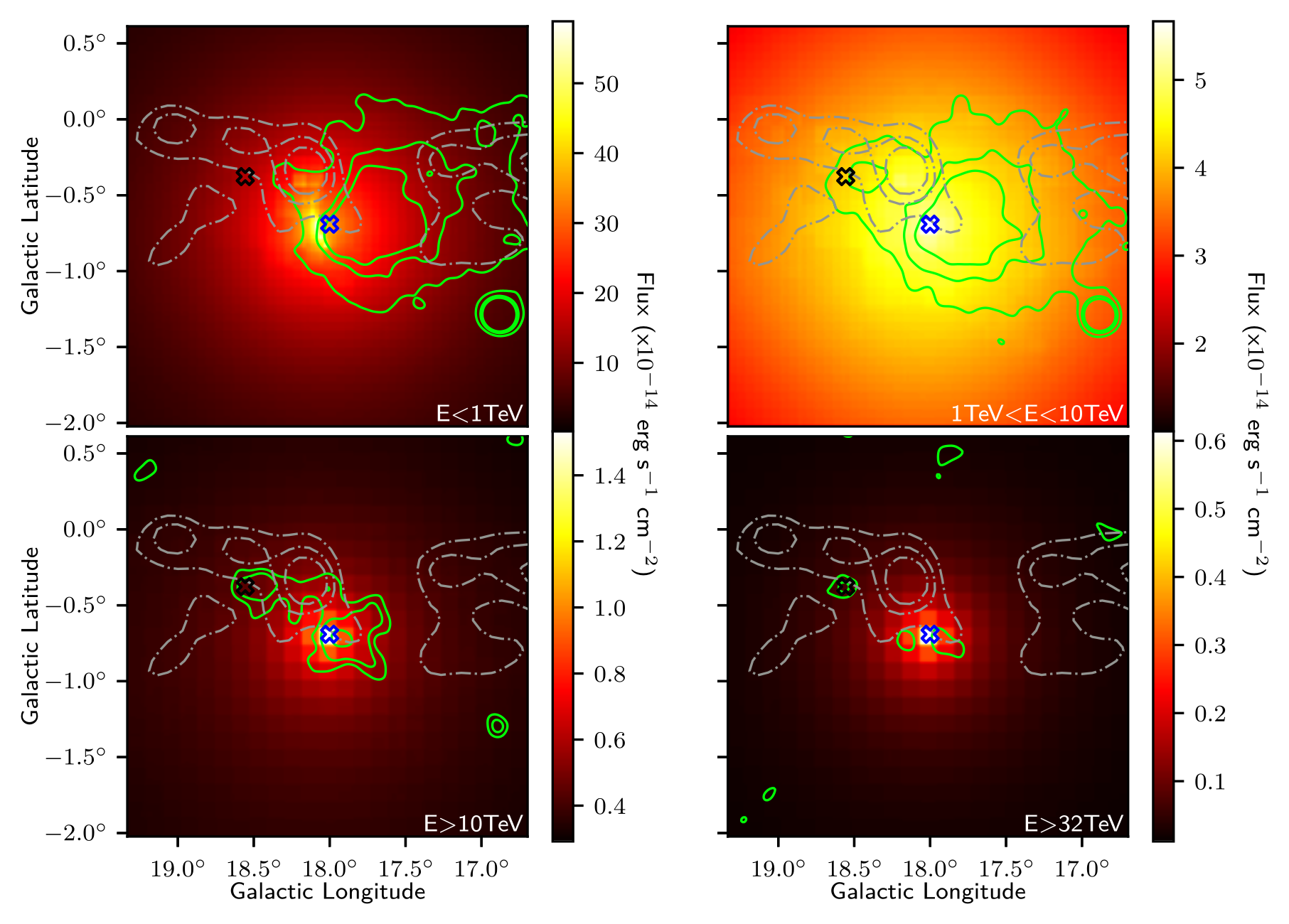}    
    
    \begin{subfigure}[c]{\columnwidth}
       \includegraphics{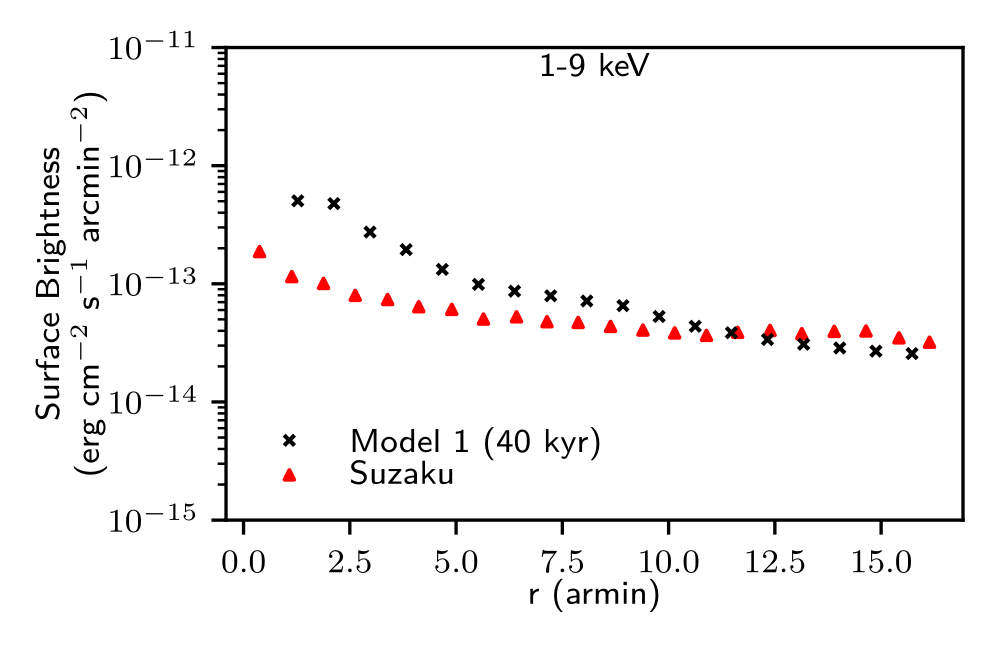}
    \end{subfigure}
    ~ 
    \begin{subfigure}[c]{\columnwidth}
       \includegraphics{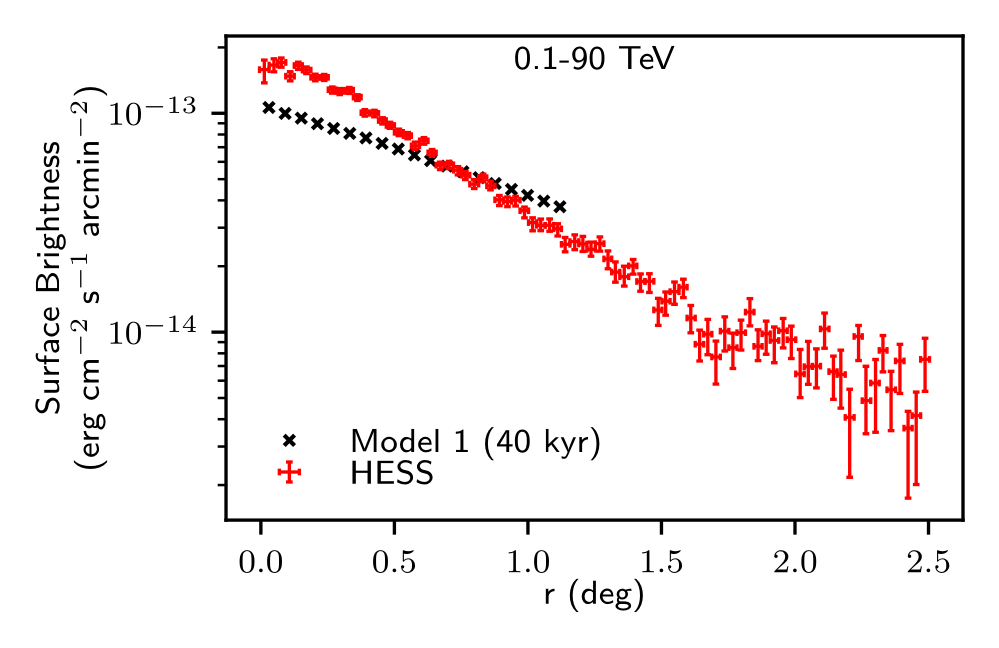}
    \end{subfigure}

    \begin{subfigure}[c]{\columnwidth}
       \includegraphics{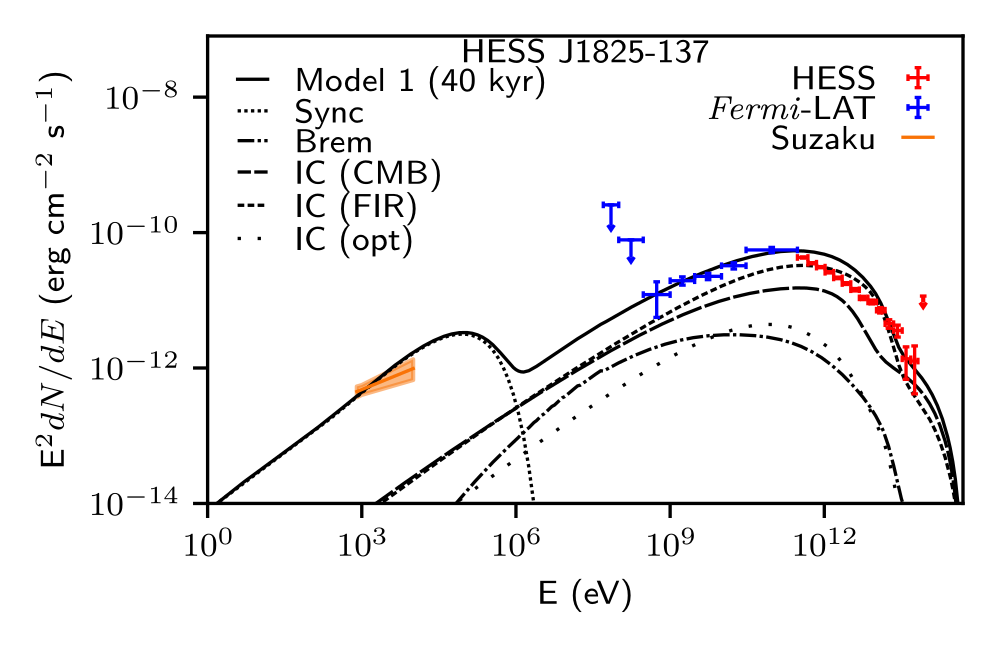}
    \end{subfigure}

    \caption{Model~1 ($40\,\kiloyear$), see \autoref{tab:matched_parameters} for model parameters. Same panel layout as in \autoref{fig:multizone_21}.}
    \label{fig:multizone_40}
\end{figure*}

\autoref{fig:multizone_21} and \ref{fig:multizone_40} show the modelled gamma-ray morphology in different energy bands, the multi-wavelength SED and the $1-9\,\keV$ X-ray and $0.1-91\,\TeV$ gamma-ray surface brightness radial profiles for the $21$ and $40\,\kiloyear$ models respectively.
\begin{figure}
    \centering
    \includegraphics[width=\columnwidth]{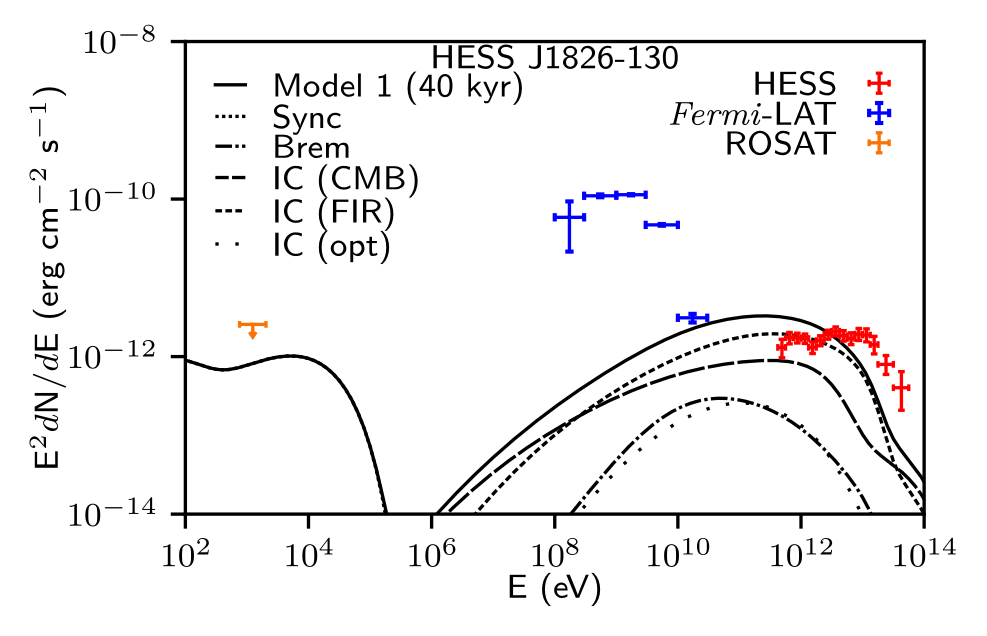}
    \caption{SED towards \mbox{HESS\,J1826-130} from electrons accelerated by PSR\,J1826-1256 by Model~1 ($40\,\kiloyear$) against the observed flux of \mbox{HESS\,J1826-130} \protect\citep{2020A&A...644A.112H} and \mbox{4FGL\,J1826.1-1256} \protect\citep{2020ApJS..247...33A}. The $\GeV$ and $\TeV$ gamma-ray flux observations towards \mbox{HESS\,J1826-130} are represented by blue and red respectively.
    }
    \label{fig:multizone_40_1826}
\end{figure}
Both models predicted that the gamma-ray morphology towards \mbox{HESS\,J1825-137} is symmetric around the powering pulsar with some gamma-ray contribution $<1\,\TeV$ via Bremsstrahlung radiation toward the region between \mbox{HESS\,J1825-137} and \mbox{HESS\,J1826-130} (see \autoref{fig:multizone_21} and \ref{fig:multizone_40}). The $40\,\kiloyear$ gamma-ray emission between $1-10\,\TeV$ extends further from the pulsar than the $21\,\kiloyear$ emission. Both models predicted a steep surface brightness radial profile for X-rays between $1-9\,\keV$ (see the bottom-middle panels of \autoref{fig:multizone_21} and \ref{fig:multizone_40}). The $21\,\kiloyear$ model was able to replicate the HESS surface brightness radial profile for gamma rays between $0.133-91\,\TeV$ (see the bottom-middle right panel of \autoref{fig:multizone_21}) while the $40\,\kiloyear$ model over-predicted the gamma-ray emission for distances $>\ang{0.5}$ from the pulsar (see the bottom-middle right panel of \autoref{fig:multizone_40}).
\par
The $21\,\kiloyear$ modelled gamma-ray SED predicted by the multi-zone model was able to match observations with a slight over-prediction ($\approx 94\%$) of the HESS data between $1-10\,\TeV$. While able to predict the normalisation of X-rays produced by synchrotron emission, the model was unable to replicate the slope of the observed Suzaku SED. The multi-zone $40\,\kiloyear$ SED was able to predict both the X-ray and gamma-ray SED with a similar over-prediction in $1-10\,\TeV$ photons as seen in the $21\,\kiloyear$ model. A slight 'bump' is present in the SED for photons around $50-100\,\TeV$ for both ages. 
\par 
The $21\,\kiloyear$ model required electrons with spectral index $\Gamma=2.0$ and cutoff $40\,\TeV$ to be injected into the ISM with a spin-down conversion factor of $10.4$ to match the multi-wavelength SED.  The $40\,\kiloyear$ required a conversion factor of $0.14$ with a spectral index and cutoff of $1.9$ and $500\,\TeV$ respectively. As $\eta<1$, Models 2 and 3 only considered an age of $40\,\kiloyear$. The initial birth spin-down power, $\dot{E}_\text{birth}$,  of the \mbox{PSR\,J11826-1334} predicted by the $21\,\kiloyear$ and $40\,\kiloyear$ models are $2.1\times 10^{37}\,\ergspersecond$ and $1.1\times 10^{40}\,\ergspersecond$ respectively.
\par
\autoref{fig:multizone_40_1826} shows the modelled SED towards \mbox{HESS\,J1826-130} due to electrons escaping from \mbox{HESS\,J1825-137} for the $40\,\kiloyear$ model. The SED towards \mbox{HESS\,J1826-130} as a result of \mbox{HESS\,J1825-137} exceeds observations for photons below $2\,\TeV$. In the model, too many low-energy electrons have escaped into the region towards \mbox{HESS\,J1826-130} before losing their energy to radiative losses. It is clear that further refinement of the model is required to accurately describe the region surrounding \mbox{HESS\,J1825-137}.

\subsubsection{Model~2 - Isotropic Diffusion + Advection}

\begin{figure*}
    \centering

    \includegraphics{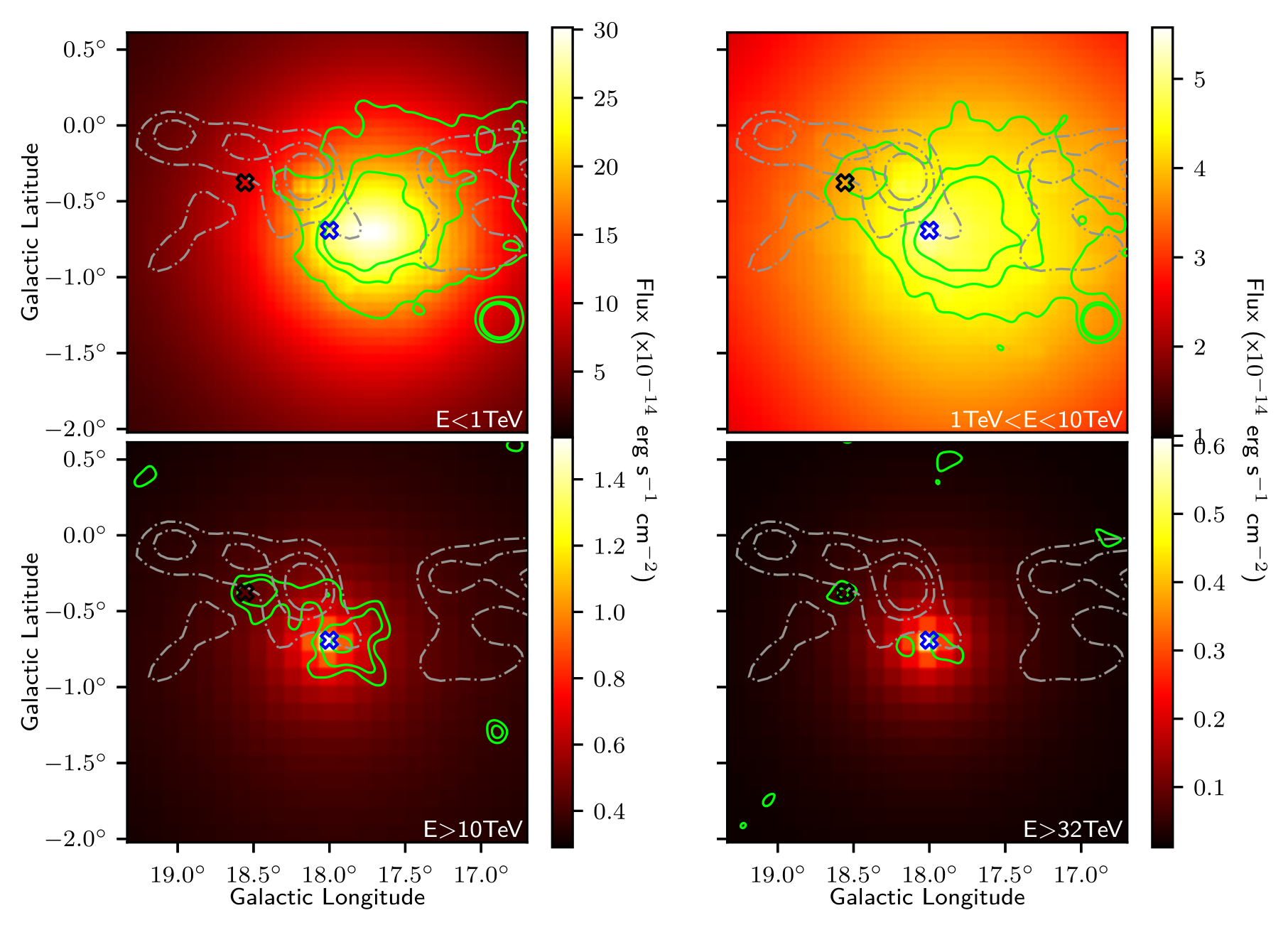}    
    
    \begin{subfigure}[c]{\columnwidth}
       \includegraphics[width=\textwidth]{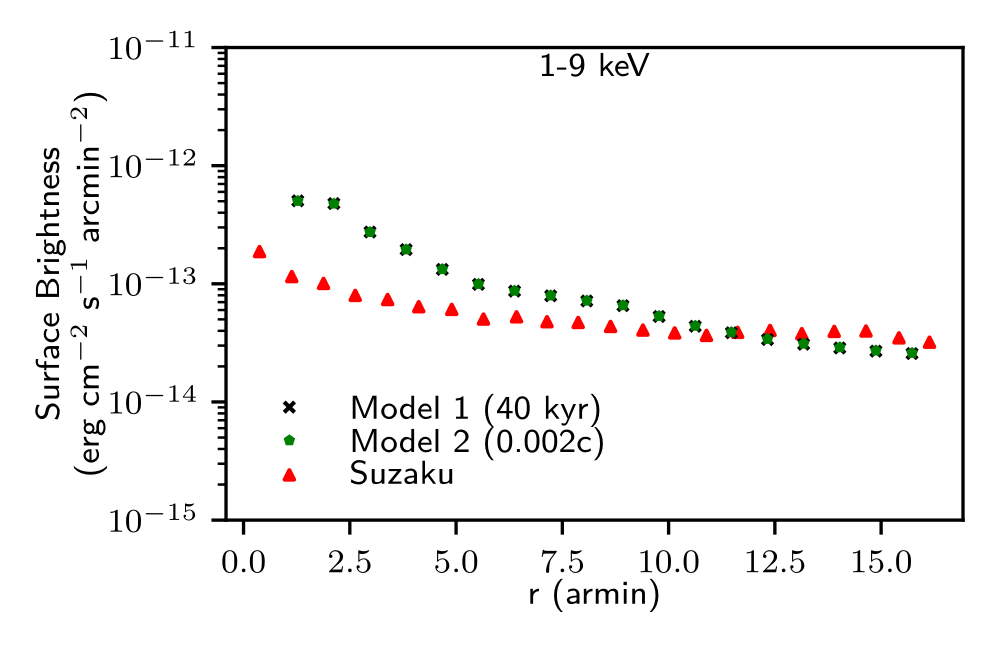}
    \end{subfigure}
    ~ 
    \begin{subfigure}[c]{\columnwidth}
       \includegraphics[width=\textwidth]{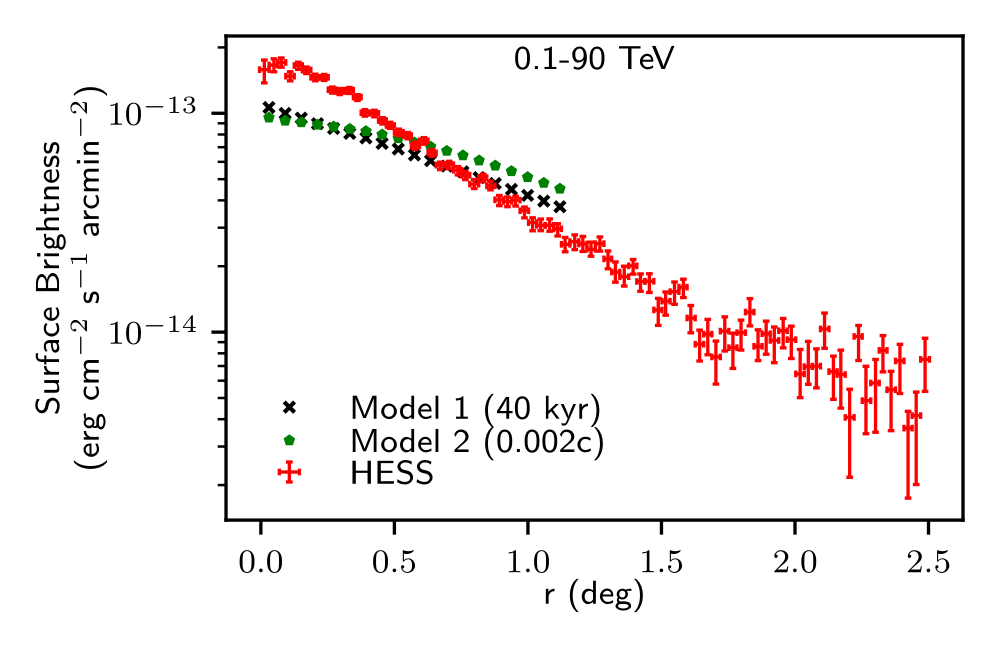}
    \end{subfigure}

    \begin{subfigure}[c]{\columnwidth}
       \includegraphics[width=\textwidth]{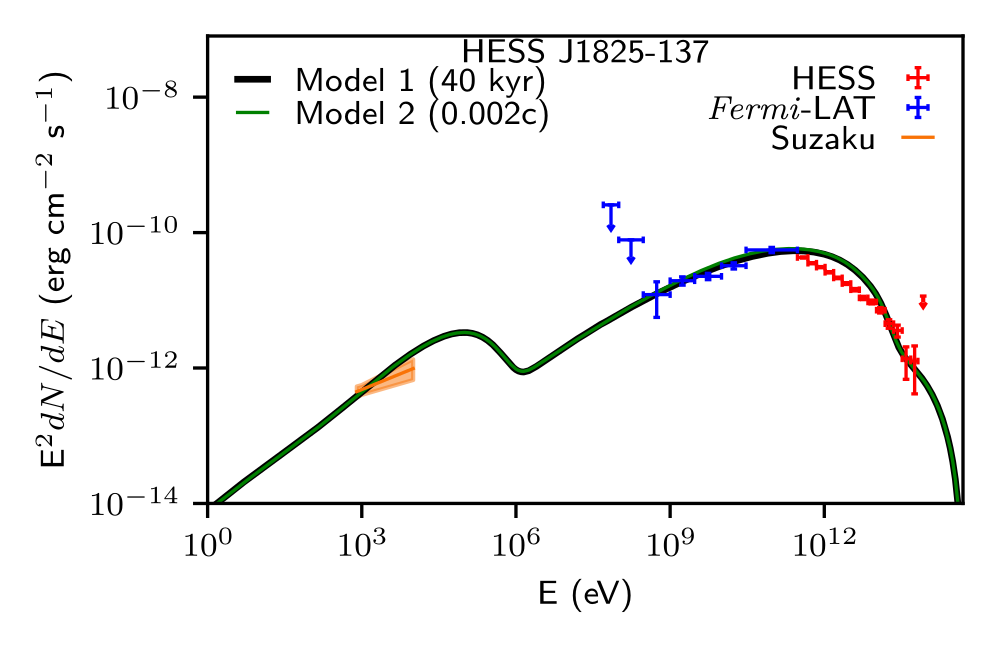}
    \end{subfigure}
    ~
    \begin{subfigure}[c]{\columnwidth}
       \includegraphics[width=\textwidth]{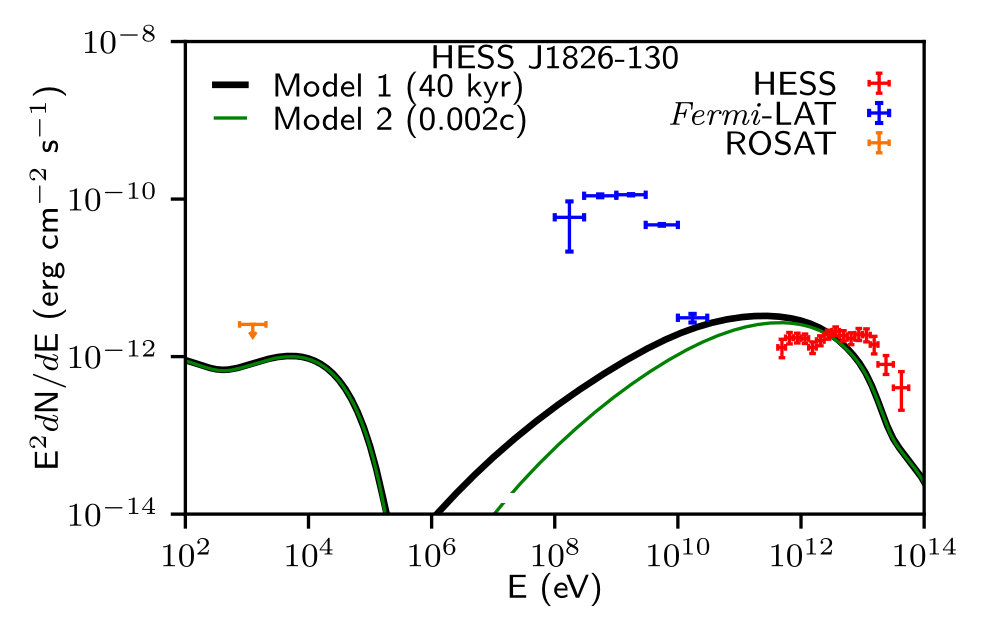}
    \end{subfigure}
    \caption{Model~2 ($0.002c$) vs Model~1 ($40\,\kiloyear$), see \autoref{tab:matched_parameters} for model parameters. (\textit{top and top-middle}) panels show the gamma-ray morphology for Model~2 ($0.002c$). (\textit{bottom-middle}) $1-9\,\keV$ X-ray (\textit{left}) and $1-91\,\TeV$ gamma-ray (\textit{right}) surface brightness radial profiles. (\textit{bottom-left}) SED towards \mbox{HESS\,J1825-137}. (\textit{bottom-right}) SED towards \mbox{HESS\,J1826-130}.}
    \label{fig:multizone_40_advection}
\end{figure*}

The gamma-ray morphology in \autoref{fig:multizone_40} shows that Model~1 ($40\,\kiloyear$) did not reproduce the extended $\TeV$ gamma-ray morphology towards \mbox{HESS\,J1825-137} at lower Galactic longitudes (see the top-middle left panel of \autoref{fig:multizone_40}). Thus, Model~2 introduced an additional advective component as suggested by \cite{2019A&A...621A.116H} towards lower Galactic longitudes. The modelled flux, surface brightness radial profiles and gamma-ray morphology for Model~2 ($40\,\kiloyear$) with an advective flow of $v=0.002$ are shown in \autoref{fig:multizone_40_advection}. A comparison between different advection speeds ($v=0.001c$, $v=0.002c$ and $v=0.003c$) is shown in \autoref{fig:multizone_40_advection_comparison}. All models otherwise have the same parameters as Model 1 ($40\,\kiloyear$) (see \autoref{tab:matched_parameters}).

\begin{figure*}
    \centering

    \includegraphics[width=\columnwidth]{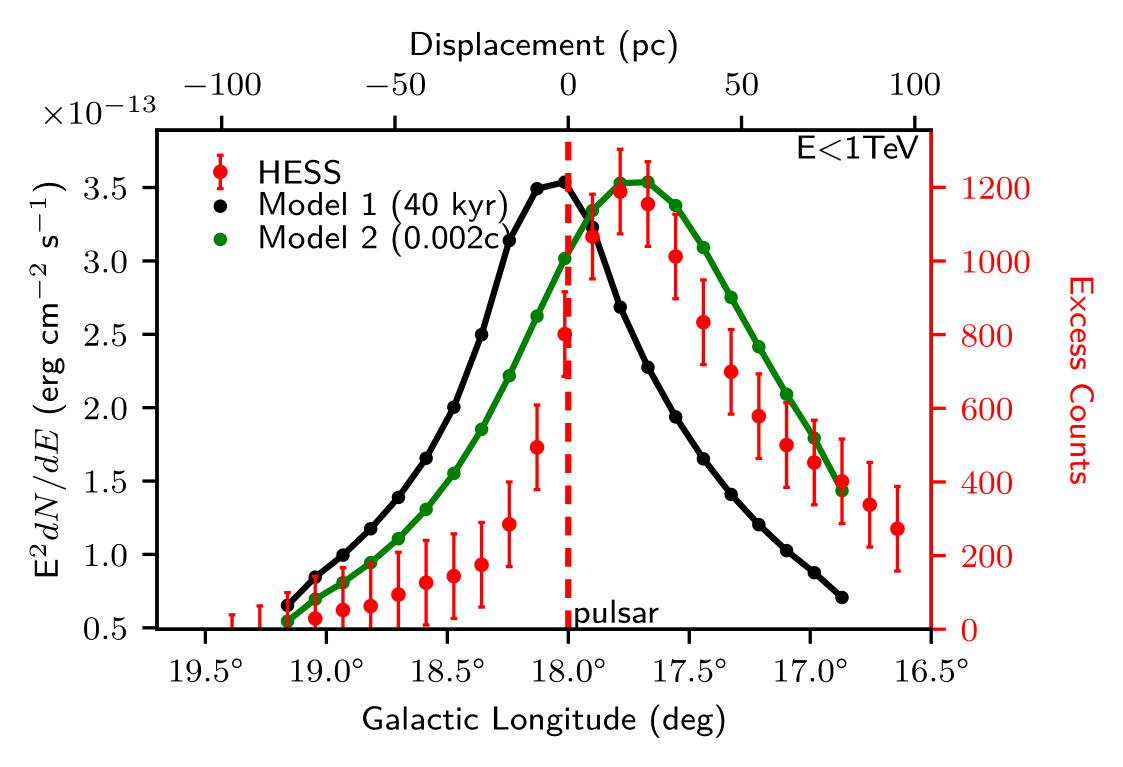}
    ~
   \includegraphics[width=\columnwidth]{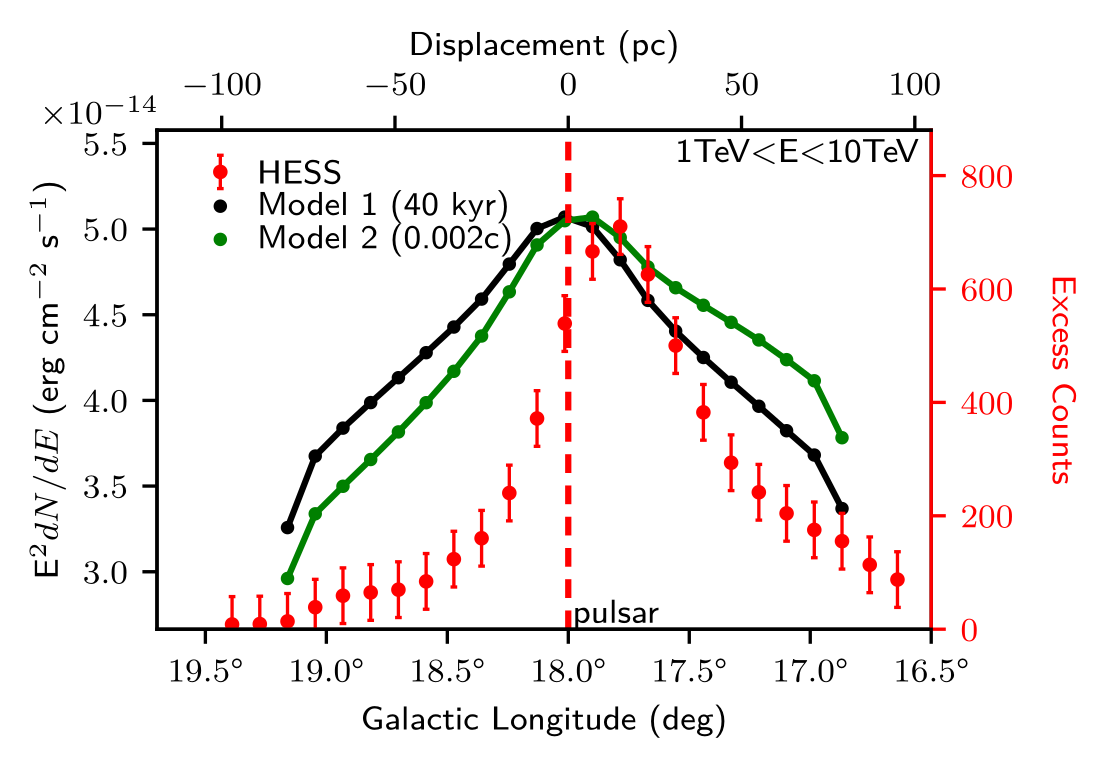}
    
    \centering
    \includegraphics[width=\columnwidth]{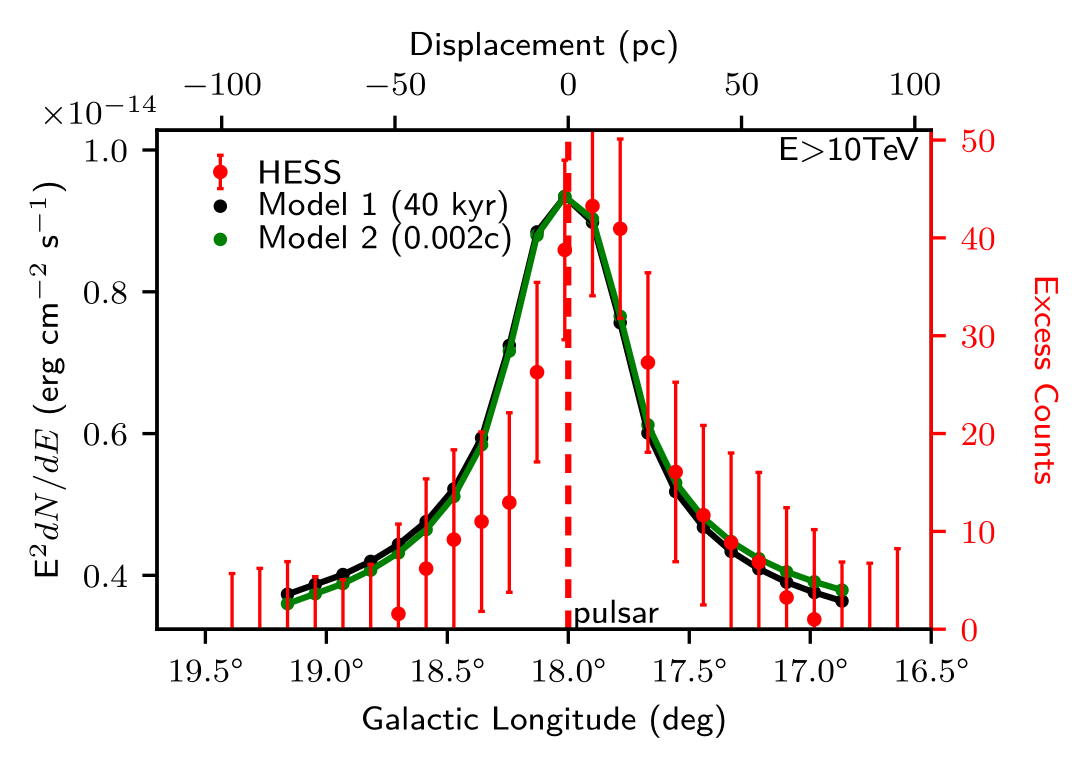}
    ~
    \includegraphics{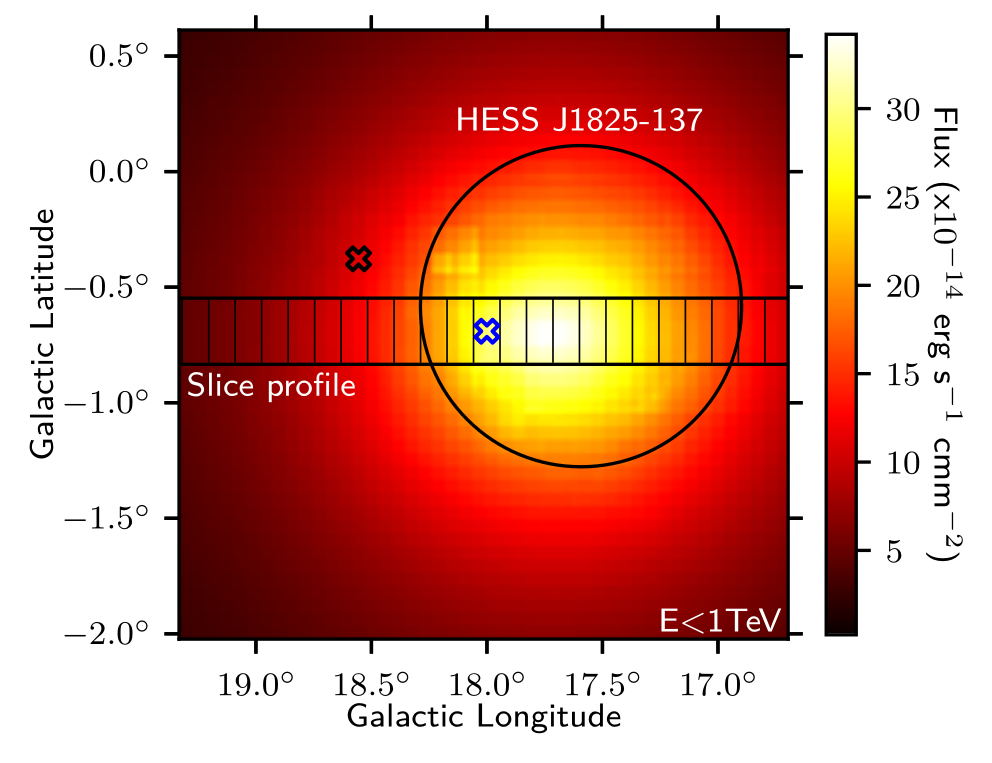}
    \caption{The energy flux along Galactic longitude for Model~1 ($40\,\kiloyear$, \textit{black}) and Model~2 ($0.002c$, \textit{green}) for energy bands $E<1\,\TeV$ (\textit{top-left}), $1 < E < 10\,\TeV$ (\textit{top-right}) and $E>10\TeV$ (\textit{bottom-left}) vs HESS excess counts \protect\citep{2019A&A...621A.116H}. (\textit{bottom-right}) Model~2 ($40\,\kiloyear$) gamma-ray morphology for energies  $<1\,\TeV$. The rectangular regions used to take the slice profile are indicated in black while the region used to extract the gamma-ray SED towards \mbox{HESS\,J1825-137} is shown by the black circle. The positions of \mbox{PSR\,J11826-1334} and \mbox{PSR\,J11826-1256} are depicted by the blue and black empty crosses respectively. See \autoref{tab:matched_parameters} for model parameters.}
  \label{fig:slices}
\end{figure*}
\par
For further comparison of the gamma-ray morphology towards \mbox{HESS\,J1825-137}, the energy flux was extracted from rectangular regions taken along Galactic longitude centred on \mbox{PSR\,J11826-1334} and are shown in \autoref{fig:slices}. An advective velocity of $0.002c$ was chosen so that the peak in the modelled gamma-ray morphology in energy range $E<1\,\TeV$ and $1\,\TeV<E<10\,\TeV$ corresponds to the HESS data (see the left-upper panel of \autoref{fig:multizone_40_advection_comparison}).
\par
While an additional advective flow of $0.002c$ lowered the gamma-ray SED towards \mbox{HESS\,J1826-130} for energies less than $2\,\TeV$, the emission still exceeds H.E.S.S. observations.

\subsubsection{Model~3 - Isotropic Diffusion + Advection + Magnetic Field towards HESS\,J1826-130} \label{sec:result_model3}

As discussed in \autoref{sec:morphology}, the turbulent molecular gas between \mbox{HESS\,J1825-137} and \mbox{HESS\,J1826-130} can act as a barrier for electrons escaping from the PWN. As clouds R1-R5 from \cite{2016MNRAS.458.2813V} are positioned in an approximate semi-circle around \mbox{PSR\,J11826-1256} (see \autoref{fig:nanten_data}), Model~3 expanded on Model~2 ($0.002c$) by including a shell of increased magnetic field strength, $B_{1826}$, centred on \mbox{HESS\,J1826-130} with inner and outer radii $0.17\,\si{\degree}$ and $0.33\,\si{\degree}$ respectively. Model~3$^*$ refers to Model~1 ($40\,\kiloyear$) with the shell of increased magnetic field strength with no advective component ($v_A=0$).
\par
\begin{figure*}

    \begin{subfigure}[c]{\columnwidth}
        \centering
       \includegraphics[width=\textwidth]{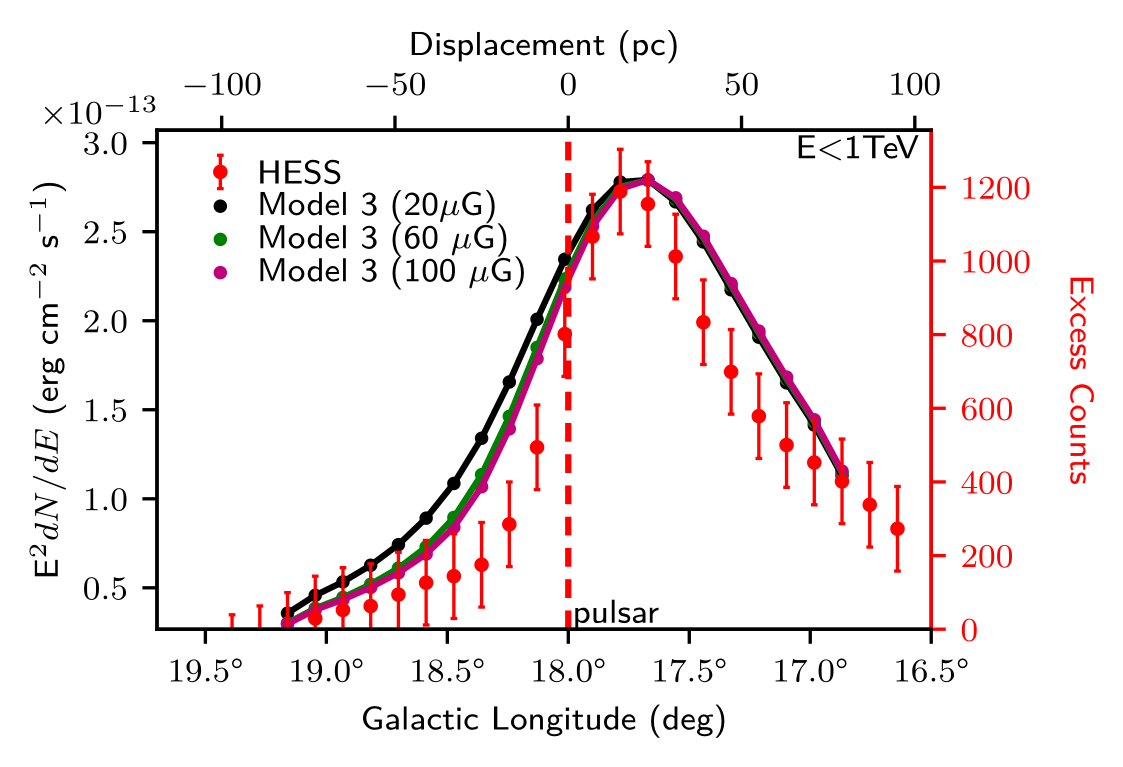}
    \end{subfigure}
    ~ 
    \begin{subfigure}[c]{\columnwidth}
        \centering
       \includegraphics[width=\textwidth]{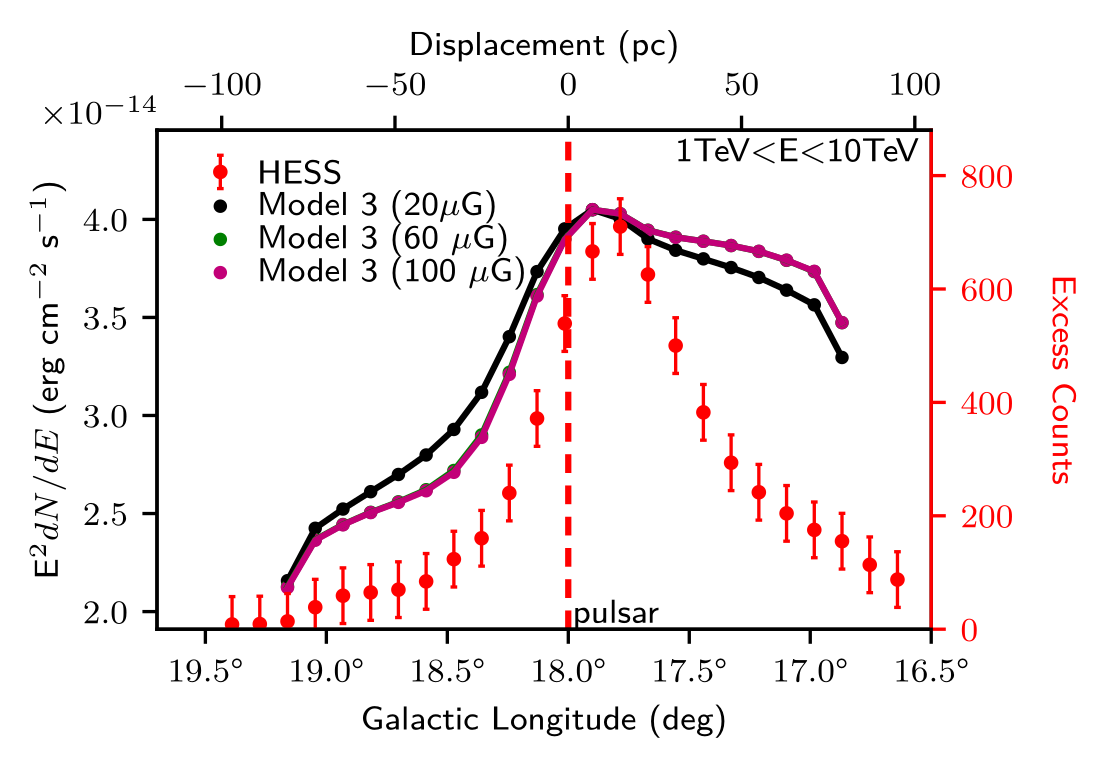}
    \end{subfigure}
    
   \begin{subfigure}[c]{\columnwidth}
        \centering
       \includegraphics[width=\textwidth]{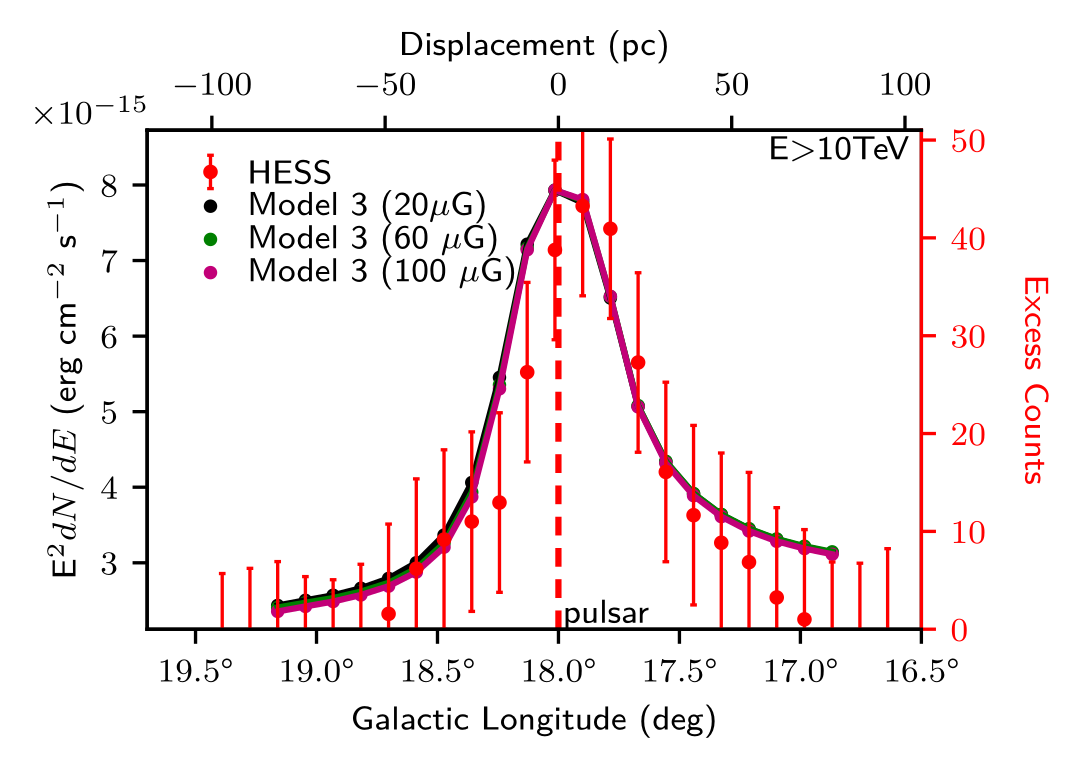}
    \end{subfigure}
    ~ 
    \begin{subfigure}[c]{\columnwidth}
        \centering
       \includegraphics[width=\textwidth]{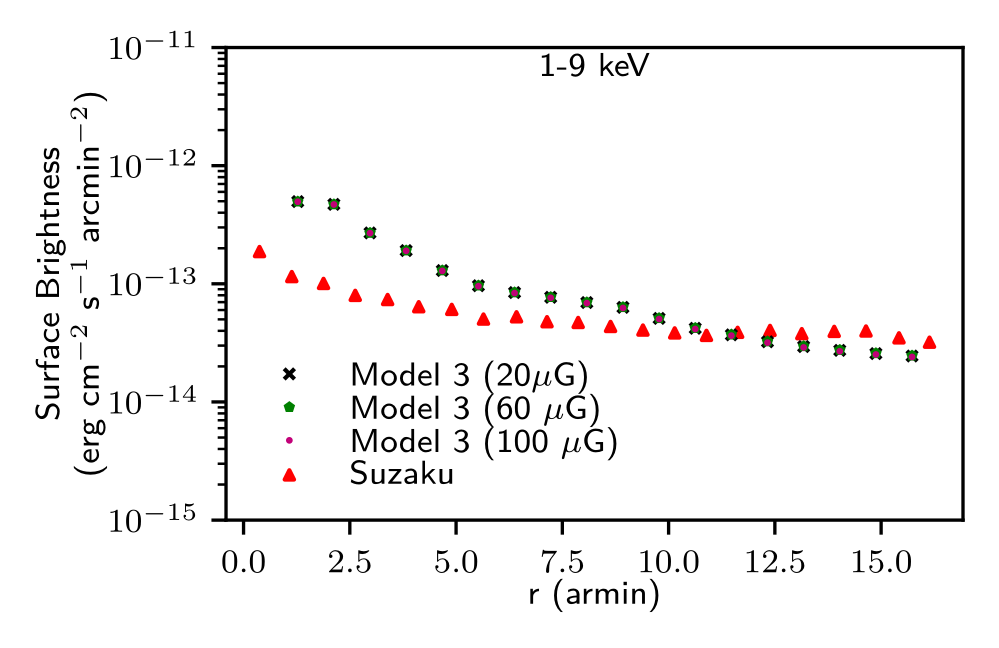}
    \end{subfigure}
    
    \begin{subfigure}[c]{\columnwidth}
        \centering
       \includegraphics[width=\textwidth]{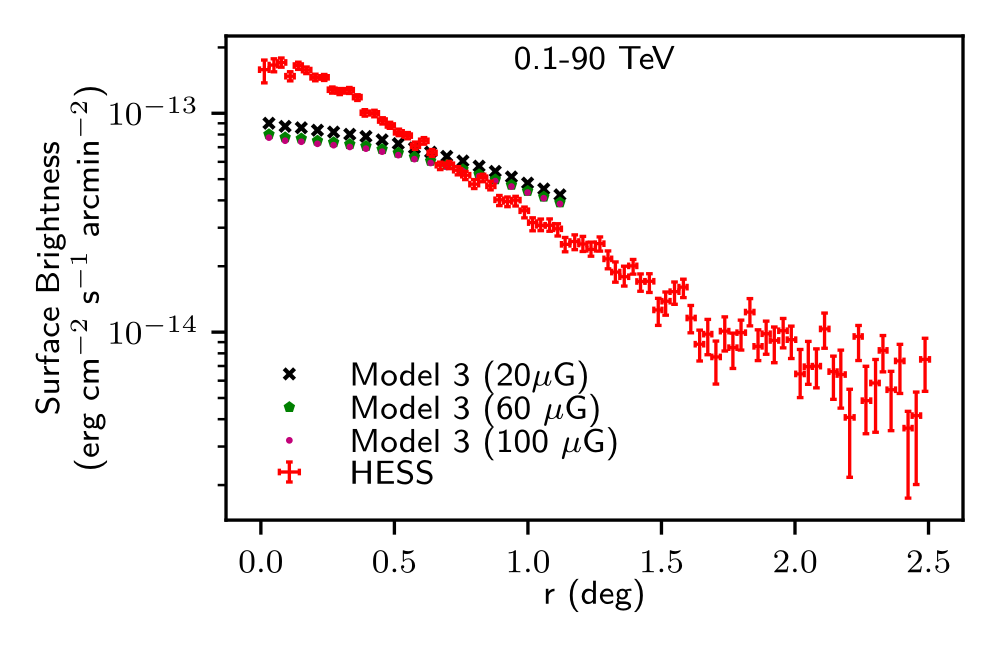}
    \end{subfigure}
    ~ 
    \begin{subfigure}[c]{\columnwidth}
        \centering
       \includegraphics[width=\textwidth]{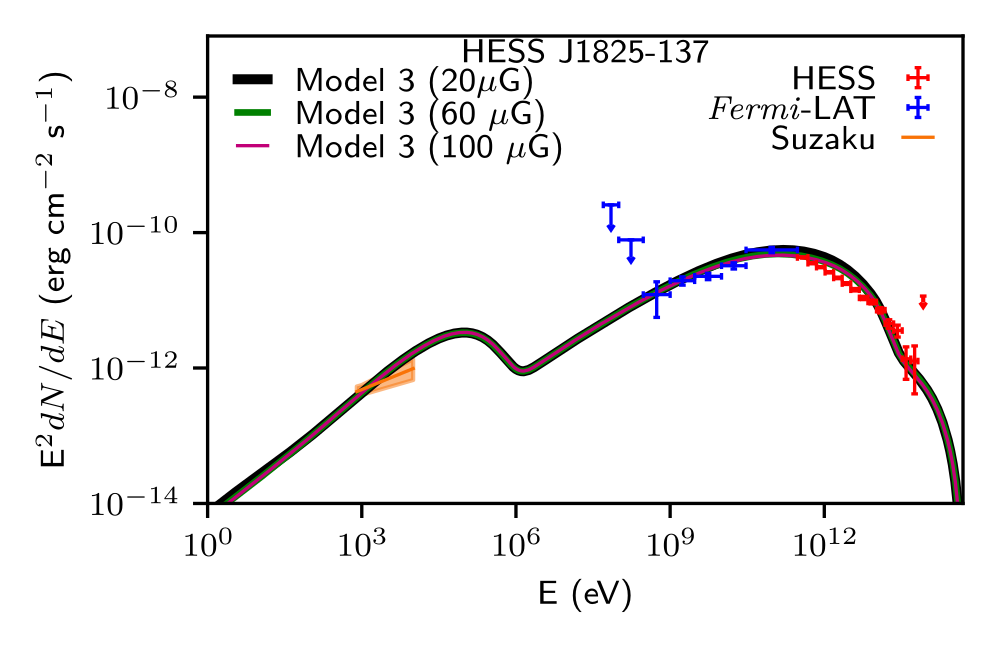}
    \end{subfigure}

    \begin{subfigure}[c]{\columnwidth}
        \centering
       \includegraphics[width=\textwidth]{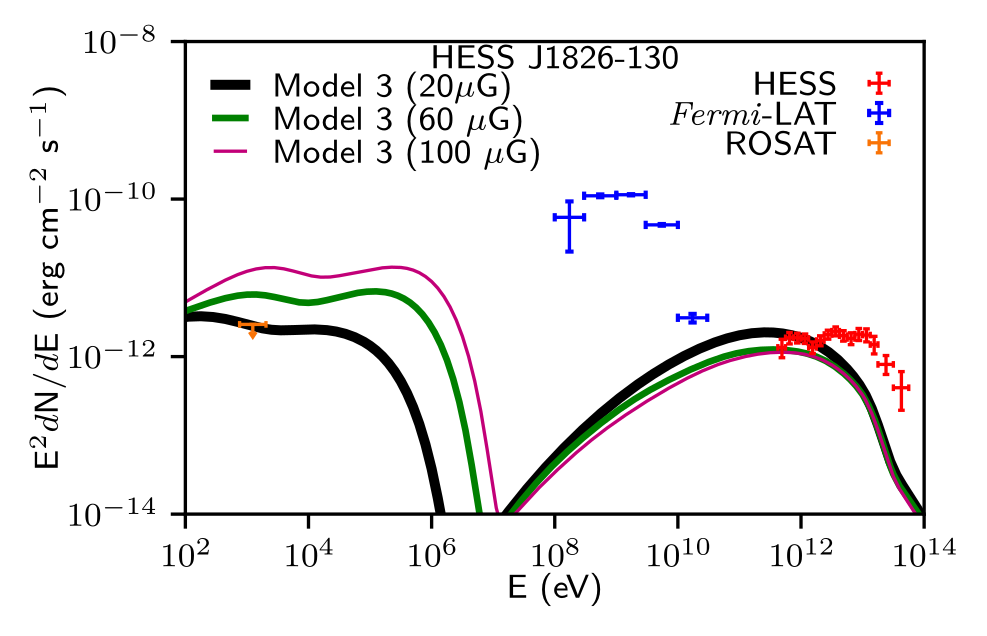}
    \end{subfigure}

    \caption{The energy flux along Galactic longitude profiles (\textit{top \& top-middle left}), surface brightness radial profiles (\textit{top-middle right \& bottom-middle left}) and SED towards \mbox{HESS\,J1825-137} (\textit{bottom-middle right}) and \mbox{HESS\,J1826-130} (\textit{bottom}) for Model~3 ($20\,\si{\micro G}$, \textit{black}), ($60\,\si{\micro G}$, \textit{green}) and ($100\,\si{\micro G}$, \textit{purple}) around HESS J1826-130. See \autoref{tab:matched_parameters} for model parameters.}
    \label{fig:1826_contam_multB}
\end{figure*}
\begin{figure*}
    \centering
    \includegraphics{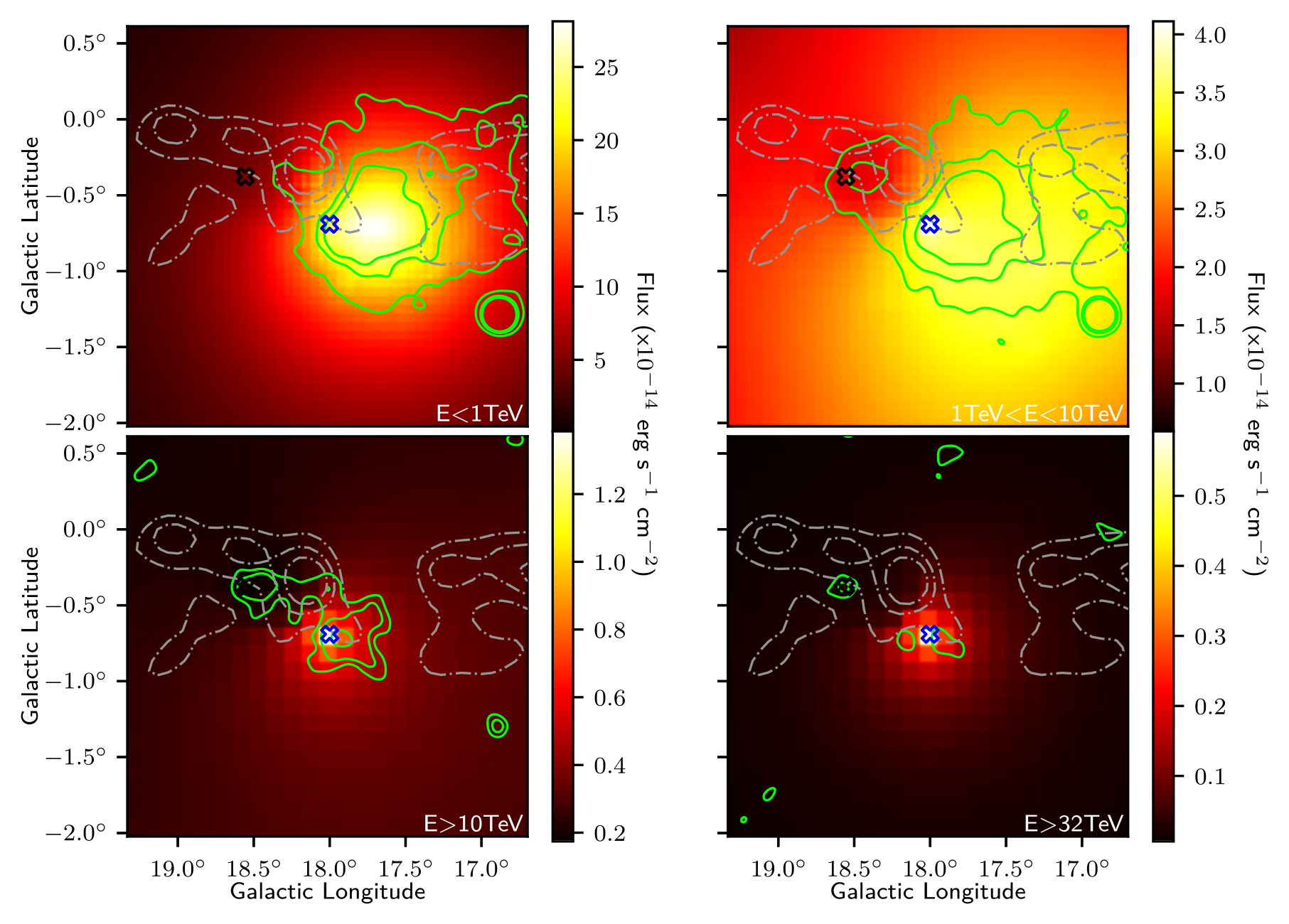}
    

    \begin{subfigure}[c]{\columnwidth}
       \includegraphics[width=\textwidth]{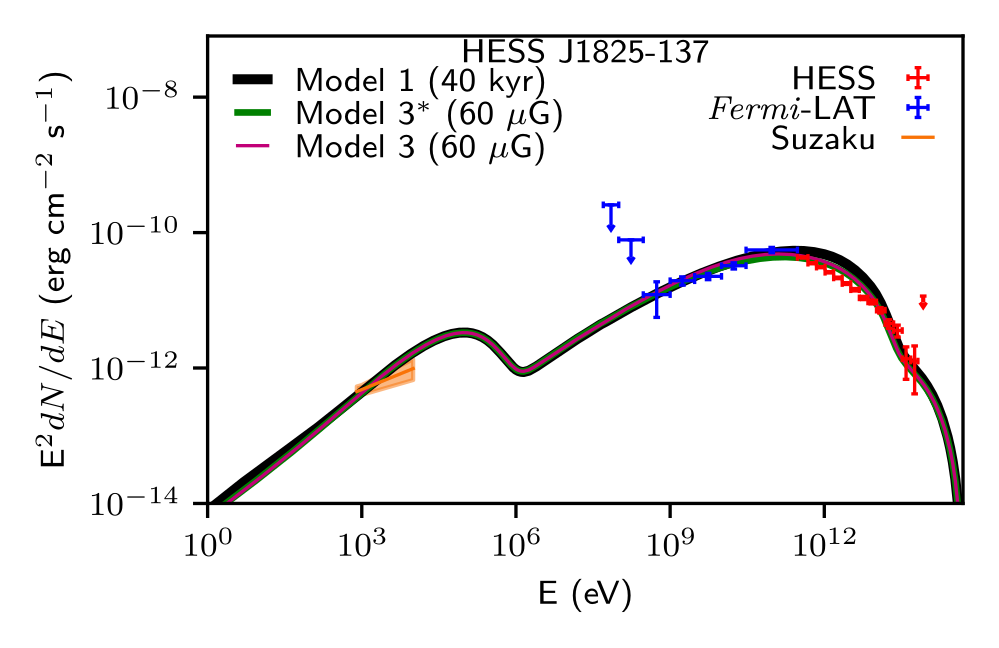}
    \end{subfigure}
    ~
    \begin{subfigure}[c]{\columnwidth}
       \includegraphics[width=\textwidth]{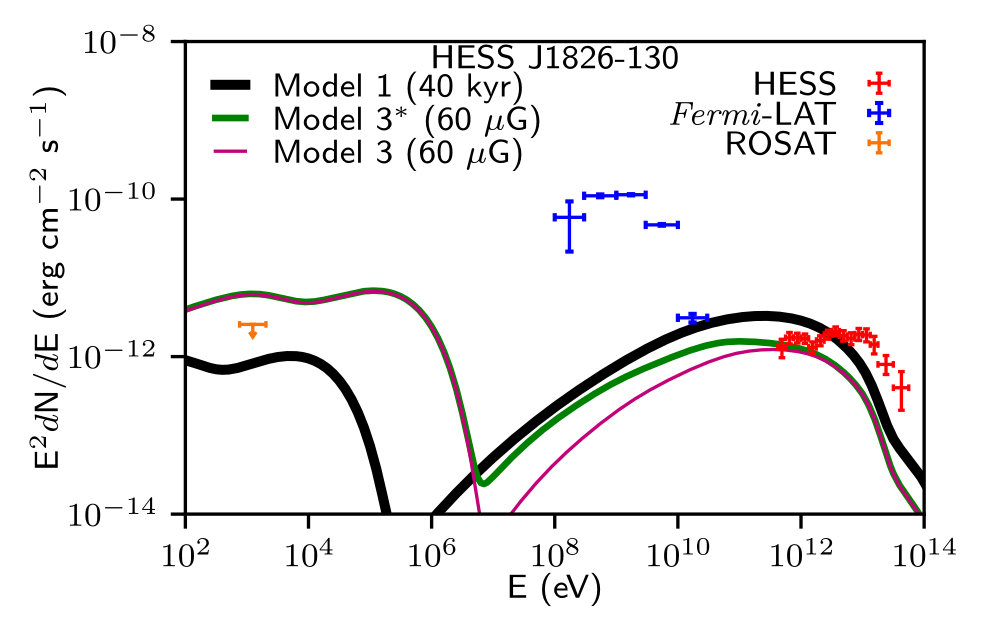}
    \end{subfigure}
    \caption{Comparison of the $40\,\kiloyear$ isotropic diffusion - Model~1 ($40\,\kiloyear$, \textit{black}), Model~3$^*$ (Model~1 + $60\,\si{\micro G}$, \textit{green}) and Model~3 ($60\,\si{\micro G}$, \textit{purple}). The morphology plots towards HESS\,J1825-137 for Model~3 ($60\,\si{\micro G}$) are shown in the top and top-middle panels. See \autoref{tab:matched_parameters} for model parameters.}
    \label{fig:1826_contam_60}
\end{figure*}

\autoref{fig:1826_contam_multB} shows the SED, surface brightness radial profiles and the gamma-ray flux along Galactic longitude of \mbox{HESS\,J1825-137} and \mbox{HESS\,J1826-130} for Model~3 with magnetic field strengths of $B=20$, $60$ and $100\,\si{\micro G}$. A comparison between Model~1 ($40\,\kiloyear$), Model~3$^*$ ($60\,\si{\micro G}$) and Model~3 $(60\,\si{\micro G})$ is shown in \autoref{fig:1826_contam_60} as well as the gamma-ray morphology for Model~3 ($60\,\si{\micro G}$). All models otherwise have the same parameters as Model~1 ($40\,\kiloyear$) (see \autoref{tab:matched_parameters}).

\subsection{Discussion}
\subsubsection{Model~1 - Isotropic Diffusion}
The $21\,\kiloyear$ and $40\,\kiloyear$ models were unable to reproduce both the X-ray and gamma-ray surface brightness radial profiles. For example, the diffusion suppression coefficient, $\chi$, could be increased to compensate for the steep X-ray surface brightness radial profile for the $21\,\kiloyear$ model. Electrons would then escape the PWN at a higher rate and the gamma-ray surface brightness radial profile will flatten. This can be seen in the $40\,\kiloyear$ model, which assumed a lower value of $\chi$ than the $21\,\kiloyear$ model. The shallow $40\,\kiloyear$ gamma-ray surface brightness radial profile indicates that lower energy electrons have started to accumulate near the pulsar, while high-energy electrons rapidly lose their energy through radiative cooling and do not escape far from the pulsar. This is demonstrated in the upper right panel in \autoref{fig:multizone_40} where the gamma-ray flux below $10\,\TeV$ is relatively constant over the grid while the flux above $10\,\TeV$ is constrained to the pulsar. This accumulation is not as apparent in the $40\,\kiloyear$ X-ray surface brightness radial profile and the SED as the regions used in extracting X-ray spectra and surface brightness radial profile are smaller than the regions used for the gamma-ray analysis (see \autoref{fig:nanten_data} and top-middle right panel of \autoref{fig:multizone_21}). The accumulation of lower-energy electrons is also a reflected as a bump in the $\TeV$ gamma-ray SED. The bump occurs when synchrotron losses start to dominate at electron energies $>9\,\TeV$, resulting in inverse Compton radiation $>6\,\TeV$, and radiative energy losses are balanced by the electron injection luminosity \citep{2007A&A...474..689M,2009ARA&A..47..523H}. This bump is not present for a slightly younger age of $36\,\kiloyear$ (with the same parameters as Model~1 ($40\,\kiloyear$), see \autoref{fig:multizone_time_comparison}), where the gamma-ray SED $\gtrsim 10\,\TeV$ at age $36\,\kiloyear$ matches Model~1 ($40\,\kiloyear$).
\par
The $21\,\kiloyear$ model required a spin-down conversion factor of $10.7$. To compensate, a braking index of $3$ would inject a greater quantity of electrons at earlier times (see \autoref{eq:Edot_evo}). However, this results in an accumulation of electrons at lower energies, consequently increasing the gamma-ray flux for photons with energies $<1\TeV$ (see \autoref{fig:multizone_n_comparison}) and the modelled SED no longer reproduces observations. This suggests that the age of \mbox{HESS\,J1825-137} lies between $21\,\kiloyear$ and $40\,\kiloyear$. The $40\,\kiloyear$ magnetic field profile takes values of $B_0=450\,\si{\micro G}$ and $\beta = -0.7$ (see \autoref{eq:PWN_Bfield}) in comparison to $B_0=400\,\si{\micro G}$ and $\beta=-0.69$ used in \citep{2011ApJ...742...62V}. \cite{2011ApJ...742...62V} considered an evolving magnetic field $B\propto \dot{E}\qty(t)$ where the magnetic field takes larger values at earlier times. This could explain the larger $B_0$ normalisation used in our modelling.

\subsubsection{Model~2 - Isotropic Diffusion + Advection} \label{sec:model_2_discussion}
An advective component of $0.002c$ towards lower Galactic longitudes was included into Model~1 ($40\,\kiloyear$). The SED and X-ray surface brightness radial profile with an advective transportation component remains unchanged to Model~1 ($40\,\kiloyear$). Electrons rapidly escape the small ($r=\ang{0.05}$) X-ray region, hence the subsequent X-ray SED and surface brightness radial profile depends more on the injected electron spectrum rather than the method of transport. On the other hand, for both Model~1 ($40\,\kiloyear$) and Model~2 ($0.002c$), the majority of electrons remain within the large ($\ang{0.7}$) HESS region leaving the gamma-ray SED unchanged. However, the electrons in Model~2 have migrated further from the pulsar while remaining within the HESS region. Subsequently, the gamma-ray profile for Model~2 ($0.002c$) is flatter than Model~1 ($40\,\kiloyear$).
\autoref{fig:cooling_time_distances} shows the distance that electrons are transported before losing their energy to radiation. It can be seen that advection is the dominant particle transport method for electron energies less than $25\,\TeV$, resulting in IC emission below $7\,\TeV$. Diffusion is dominant for electrons above $25\,\TeV$. However, these high-energy electrons do not travel far from their birthplace before losing their energy to radiative cooling.
\par 
At all energies, the gamma-ray energy flux along Galactic longitude for Model~1 (40\,\kiloyear) appears symmetric around the pulsar position and shows no preferential direction of transportation. However, the HESS uncorrelated excess data indicates that electrons are preferentially transported to lower Galactic longitudes. With the addition of an advective flow of $0.002c$, the peak in the $40\,\kiloyear$ gamma-ray slices for photons less than $1\,\TeV$ is now offset from the pulsar and follows the shape of the uncorrelated excess. For the $1\,\TeV<E<10\,\TeV$ energy band, both Model~1 and Model~2 show a flatter slice profile compared to HESS observations. Electrons resulting from this emission appear to be contained near their birthplace before escaping into the nebula. Our model assumed that the diffusion lies within the Kraichnan regime with the index being fixed at $\delta=0.5$. The top-right panel of \autoref{fig:slices} shows that the modelled gamma-ray slice morphology is broader than that observed by HESS, suggesting that electrons are constrained within the PWN. This suggests that the diffusion index inside the PWN may be somewhat less than the $\delta=0.5$ value we assumed.
\par
By assuming that diffusion was isotropic in \autoref{eq:Diffusion_Equation_numerical}, any preferential direction for particle transport was a result of magnetic field irregularities and/or advective flow. The highly asymmetric morphology towards HESS\,J1825-137 could be explained if diffusion is anisotropic with preferential diffusion towards lower Galactic longitudes. However, an anisotropic diffusion model can be approximated by an isotropic diffusion + advection model (i.e. Model~2).
\subsubsection{Model~3 - Isotropic Diffusion + Advection + Magnetic Field towards \mbox{HESS\,J1826-130}} \label{sec:discussion_model3}
Here, a spherical shell of increased magnetic field strength around \mbox{HESS\,J1826-130} was considered to replicate the turbulent gas towards cloud R1 from \cite{2016MNRAS.458.2813V}. 
\par
The bottom-right panel of \autoref{fig:slices} shows that cloud R1 lies within the area used to determine the gamma-ray SED of \mbox{HESS\,J1825-137}. The ratio of synchrotron to inverse Compton flux is proportional to the magnetic field \citep{1997MNRAS.291..162A}. Hence, as the magnetic field around \mbox{HESS\,J1826-130} increases, electrons lose more energy through synchrotron losses and the inverse Compton flux decreases at equivalent gamma-ray energies. This can be seen in the SED for \mbox{HESS\,J1825-137} in the bottom-middle right panel of \autoref{fig:1826_contam_multB}. This has the effect of improving the match to HESS observations between $1-10\,\TeV$ compared to Model~1 ($40\,\kiloyear$) as shown in the bottom-middle-left panel of \autoref{fig:1826_contam_60}.
\par 
The gamma-ray energy flux along Galactic longitudes are shown in the top and top-middle panels of \autoref{fig:1826_contam_multB}. As the magnetic field around \mbox{HESS\,J1826-130} increases, gamma-ray emission less than $1<\,\TeV$ and $>10\,\TeV$ remains unchanged at lower longitudes, with a decrease at higher longitudes. Between $1\,\TeV<E<10\,\TeV$, the gamma-ray slice profile drops of at a shallower rate compared to the HESS data at lower Galactic longitudes. However the gamma-ray emission at higher longitudes, representing the area towards \mbox{HESS\,J1826-130}, increases proportionally with the magnetic field.
\par
Additionally, the bottom-left panel of \autoref{fig:1826_contam_multB} indicates that increasing the magnetic field strength around \mbox{HESS\,J1826-130} lowers the contamination of \mbox{HESS\,J1826-130} by the PWN associated with \mbox{HESS\,J1825-137} for energies $<2\,\TeV$. Regions of high magnetic field strength experience a slower rate of diffusion (see \autoref{eq:diffusion}) and high synchrotron losses. Hence regions of high magnetic field tend to `block' cosmic rays from passing through. The model implies that a minimum magnetic field strength of $60\,\si{\micro G}$ is required to successfully lower the contamination of \mbox{HESS\,J1826-130} according to H.E.S.S. observations. The bottom-right panel of \autoref{fig:1826_contam_multB} shows the multi-wavelength SED towards \mbox{HESS\,J1826-130}. An upper-limit to the X-ray emission towards \mbox{HESS\,J1826-130} can be obtained using HEARSAC's X-Ray background tool utilising ROSAT data \citep{2019ascl.soft04001S}. The estimated synchrotron flux towards \mbox{HESS\,J1826-130} combined with the ROSAT X-ray upper limit (obtained from the same region used to extract the SED of \mbox{HESS\,J1826-130}) implies a maximum magnetic field strength of $\approx20\,\si{\micro G}$ around \mbox{HESS\,J1826-130}. This constraint violation suggests that the model is not fully encapsulating the transport of particles between PSR\,J1826-1334 and HESS\,1826-130.

\subsection{Future Work}

Presently, our model only considers isotropic diffusion and does not account for diffusion parallel and perpendicular to the magnetic field \citep{1983RPPh...46..973D,2023FrASS..1054760L}. The magnetic fields of PWNe are believed to be toroidal in nature \citep{1979ApJ...227..106S,2006ApJ...638..225K}, hence diffusion is expected to be suppressed perpendicular to the magnetic axis of the pulsar. Additionally, the recent detection of TeV halos \citep{2017Sci...358..911A} implies that the region surrounding the PWN experiences a higher diffusion suppression compared to the average Galactic diffusion coefficient \citep{2018PhRvD..98f3017E, 2020PhRvD.101j3035D, 2023PhRvD.107l3020S}.  While current models of particle transport suggest that advection dominates particle transport within the PWN and diffusion dominates at the edges, this could be described by two different regions of diffusion suppression. Model\,2 and 3 in this study considered spatially-independent advection towards lower Galactic longitudes to explain the asymmetric $\TeV$ gamma-ray morphology towards \mbox{HESS\,J1825-137}. As a result, losses due to adiabatic expansion were neglected. Future modelling of PWNe, in particular \mbox{HESS\,J1825-137}, could investigate the effects of inhomogeneous anisotropic diffusion and an azimuthal and surface brightness dependent advective velocity on the gamma-ray morphology and SED. This can then be applied to model the formation of the $\TeV$ halo around the PWN \citep{2020A&A...640A..76P}.
\par 
Our model assumed a time-independent magnetic field with decreasing strength from the distance to the pulsar (see \autoref{sec:B_field}). However, the average magnetic field of PWNe are expected to decrease over time from the conservation of magnetic energy density (e.g. \cite{2010ApJ...715.1248T}) and the normalisation, $B_0$, obtained from the modelling (see \autoref{tab:matched_parameters}) can be considered as the time-averaged normalisation. Any future predictions of the formation of the $\TeV$ halo around \mbox{HESS\,J1825-137} must consider time-dependency on the magnetic field.
\par
The implementation of a time-dependent source position will not affect the gamma-ray morphology $\gtrsim 13\,\TeV$ around \mbox{PSR\,J11826-1334} due to high synchrotron losses. However, lower energy photons will appear to originate at a position offset to the current position of the pulsar as seen in \cite{2020A&A...640A..76P}. This does not explain the extended $\TeV$ gamma-ray morphology towards lower Galactic longitudes as discussed in \autoref{sec:multizone}, but could affect the modelled formation of the TeV halo. Future work could investigate the effects of an evolving source position on the gamma-ray surface brightness radial profile and morphology towards PWNe and other sources. For example, the application of our model to SNRs would require cosmic rays to be injected by an expanding shell to model diffusive shock acceleration by the expanding SNR.
\par
The model presented in this study is not limited to \mbox{HESS\,J1825-137} and can be used to model the transport of cosmic-rays (electrons and protons) from other PWNe and cosmic ray sources.

\section{Summary}
By modelling the transport of electrons across a 3D Cartesian grid of {varying ISM density and magnetic field, we are able to reproduce the main characteristics of the multi-wavelength spectrum and morphology towards \mbox{HESS\,J1825-137}.
Three different models were considered. Model~1 assumed a simple case of isotropic diffusion and radiative losses for the characteristic age of $21\,\kiloyear$ and an older age of $40\,\kiloyear$ as suggested by \cite{2011ApJ...742...62V}. Model~2 included an additional advective component to Model~1 ($40\,\kiloyear$) and Model~3 introduced turbulent ISM towards \mbox{HESS\,J1826-130} to Model~2.
\par
The best fit $21\,\kiloyear$ and $40\,\kiloyear$ Model~1 consisted of a pulsar injecting electrons into the surrounding medium with a spin-down conversion factor of $10.7$ and $0.14\%$ respectively, indicating that the true age of the system is older than the characteristic age of \mbox{PSR\,J11826-1334}. While able to reproduce the multi-wavelength SED, neither model was able to sufficiently reproduce the gamma-ray flux along Galactic longitude described in \cite{2019A&A...621A.116H} for photons with energies $1\,\TeV<E<10\,\TeV$. However, the morphological profile could be matched for gamma-rays with energies $<1\,\TeV$ (with an offset of $\ang{0.3}$ towards higher Galactic longitudes compared to the HESS data) and energies $>10\,\TeV$.
\par
Applying an advective bulk flow (with velocity $v=0.002\,c$) of electrons towards lower Galactic longitudes did not alter the photon SED predicted by Model~1. By extracting the energy flux along Galactic longitude, we were able to compare the energy-dependent morphology towards \mbox{HESS\,J1825-137}. Model~2 ($0.002c$) was able to reproduce the energy flux for photons $<1\,\TeV$ and $>10\,\TeV$, however photons with energies $1<E<10\,\TeV$ experience a shallower drop-off compared to the uncorrelated HESS excess slices as revealed by \cite{2019A&A...621A.116H}. This suggests that the parent electrons are constrained within the PWN before escaping into the interstellar medium to form a $\TeV$ halo. The asymmetric gamma-ray morphology towards HESS J1825-137 cannot be predicted using a diffusion-only model and requires an advective component with bulk flow of $0.02c$.
\par
As the gamma-ray emission associated with \mbox{PSR\,J1826-1334} cannot exceed the observed emission towards \mbox{HESS\,J1826-130}, \mbox{HESS\,J1826-130} can be used to constrain the model. Model~1 and Model~2 were found to over-predict the SED of \mbox{HESS\,J1826-130} for photons $<1.5\,\TeV$. By placing a shell of increased magnetic field strength of at least $60\,\si{\micro G}$ around \mbox{HESS\,J1826-130}, representing the turbulent gas between the two HESS sources \citep{2016MNRAS.458.2813V}, the contamination was successfully lowered below the levels closer to those estimated by HESS. By combining the modelled synchrotron flux with the ROSAT X-ray upper limit towards \mbox{HESS\,J1826-130}, we were able to constrain the magnetic field shell to have maximum strength of $20\,\si{\micro G}$. This constraint violation suggests that further modelling of the turbulent gas is needed to fully disentangle the particle transport towards \mbox{HESS\,J1825-137}. 

\section*{Acknowledgements}
\mbox{T. Collins} acknowledges support through the provision of Australian Government Research Training Program Scholarship. The Nanten project is based on a mutual agreement between Nagoya University and the Carnegie Institution of Washington (CIW). This research has made use of the NASA’s Astrophysics Data System and the SIMBAD data base, operated at CDS, Strasbourg, France.

\section*{Data Availability}

No new data were generated or analysed in support of this research.




\bibliographystyle{mnras}
\bibliography{Main_Document} 

\begin{thebibliography}{}
\makeatletter
\relax
\def\mn@urlcharsother{\let\do\@makeother \do\$\do\&\do\#\do\^\do\_\do\%\do\~}
\def\mn@doi{\begingroup\mn@urlcharsother \@ifnextchar [ {\mn@doi@}
  {\mn@doi@[]}}
\def\mn@doi@[#1]#2{\def\@tempa{#1}\ifx\@tempa\@empty \href
  {http://dx.doi.org/#2} {doi:#2}\else \href {http://dx.doi.org/#2} {#1}\fi
  \endgroup}
\def\mn@eprint#1#2{\mn@eprint@#1:#2::\@nil}
\def\mn@eprint@arXiv#1{\href {http://arxiv.org/abs/#1} {{\tt arXiv:#1}}}
\def\mn@eprint@dblp#1{\href {http://dblp.uni-trier.de/rec/bibtex/#1.xml}
  {dblp:#1}}
\def\mn@eprint@#1:#2:#3:#4\@nil{\def\@tempa {#1}\def\@tempb {#2}\def\@tempc
  {#3}\ifx \@tempc \@empty \let \@tempc \@tempb \let \@tempb \@tempa \fi \ifx
  \@tempb \@empty \def\@tempb {arXiv}\fi \@ifundefined
  {mn@eprint@\@tempb}{\@tempb:\@tempc}{\expandafter \expandafter \csname
  mn@eprint@\@tempb\endcsname \expandafter{\@tempc}}}

\bibitem[\protect\citeauthoryear{{Abdollahi} et~al.,}{{Abdollahi}
  et~al.}{2020}]{2020ApJS..247...33A}
{Abdollahi} S.,  et~al., 2020, \mn@doi [\apjs] {10.3847/1538-4365/ab6bcb},
  \href {https://ui.adsabs.harvard.edu/abs/2020ApJS..247...33A} {247, 33}

\bibitem[\protect\citeauthoryear{{Abeysekara} et~al.,}{{Abeysekara}
  et~al.}{2017}]{2017Sci...358..911A}
{Abeysekara} A.~U.,  et~al., 2017, \mn@doi [Science] {10.1126/science.aan4880},
  \href {https://ui.adsabs.harvard.edu/abs/2017Sci...358..911A} {358, 911}

\bibitem[\protect\citeauthoryear{{Abeysekara} et~al.,}{{Abeysekara}
  et~al.}{2020}]{2020PhRvL.124b1102A}
{Abeysekara} A.~U.,  et~al., 2020, \mn@doi [\prl]
  {10.1103/PhysRevLett.124.021102}, \href
  {https://ui.adsabs.harvard.edu/abs/2020PhRvL.124b1102A} {124, 021102}

\bibitem[\protect\citeauthoryear{{Aharonian} \& {Atoyan}}{{Aharonian} \&
  {Atoyan}}{1996}]{1996A&A...309..917A}
{Aharonian} F.~A.,  {Atoyan} A.~M.,  1996, \aap, \href
  {https://ui.adsabs.harvard.edu/abs/1996A&A...309..917A} {309, 917}

\bibitem[\protect\citeauthoryear{{Aharonian}, {Atoyan}  \&
  {Kifune}}{{Aharonian} et~al.}{1997}]{1997MNRAS.291..162A}
{Aharonian} F.~A.,  {Atoyan} A.~M.,   {Kifune} T.,  1997, \mn@doi [\mnras]
  {10.1093/mnras/291.1.162}, \href
  {https://ui.adsabs.harvard.edu/abs/1997MNRAS.291..162A} {291, 162}

\bibitem[\protect\citeauthoryear{{Aharonian} et~al.,}{{Aharonian}
  et~al.}{2005}]{}
{Aharonian} F.,  et~al., 2005, \mn@doi [Science] {10.1126/science.1108643},
  \href {https://ui.adsabs.harvard.edu/abs/2005Sci...307.1938A} {307, 1938}

\bibitem[\protect\citeauthoryear{{Araya}, {Mitchell}  \& {Parsons}}{{Araya}
  et~al.}{2019}]{2019MNRAS.485.1001A}
{Araya} M.,  {Mitchell} A.~M.~W.,   {Parsons} R.~D.,  2019, \mn@doi [\mnras]
  {10.1093/mnras/stz462}, \href
  {https://ui.adsabs.harvard.edu/abs/2019MNRAS.485.1001A} {485, 1001}

\bibitem[\protect\citeauthoryear{{Atoyan}, {Aharonian}  \& {V{\"o}lk}}{{Atoyan}
  et~al.}{1995}]{1995PhRvD..52.3265A}
{Atoyan} A.~M.,  {Aharonian} F.~A.,   {V{\"o}lk} H.~J.,  1995, \mn@doi [\prd]
  {10.1103/PhysRevD.52.3265}, \href
  {https://ui.adsabs.harvard.edu/abs/1995PhRvD..52.3265A} {52, 3265}

\bibitem[\protect\citeauthoryear{{Benbow}}{{Benbow}}{2005}]{2005AIPC..745..611B}
{Benbow} W.,  2005, in {Aharonian} F.~A.,  {V{\"o}lk} H.~J.,   {Horns} D.,
  eds,  American Institute of Physics Conference Series Vol. 745, High Energy
  Gamma-Ray Astronomy. pp 611--616, \mn@doi{10.1063/1.1878471}

\bibitem[\protect\citeauthoryear{{Berezinskii}, {Bulanov}, {Dogiel}  \&
  {Ptuskin}}{{Berezinskii} et~al.}{1990}]{1990acr..book.....B}
{Berezinskii} V.~S.,  {Bulanov} S.~V.,  {Dogiel} V.~A.,   {Ptuskin} V.~S.,
  1990, {Astrophysics of cosmic rays}.
John Wiley \& Sons, Inc.

\bibitem[\protect\citeauthoryear{{Blumenthal} \& {Gould}}{{Blumenthal} \&
  {Gould}}{1970}]{1970RvMP...42..237B}
{Blumenthal} G.~R.,  {Gould} R.~J.,  1970, \mn@doi [Reviews of Modern Physics]
  {10.1103/RevModPhys.42.237}, \href
  {https://ui.adsabs.harvard.edu/abs/1970RvMP...42..237B} {42, 237}

\bibitem[\protect\citeauthoryear{{Brand} \& {Blitz}}{{Brand} \&
  {Blitz}}{1993}]{1993A&A...275...67B}
{Brand} J.,  {Blitz} L.,  1993, \aap, \href
  {https://ui.adsabs.harvard.edu/abs/1993A&A...275...67B} {275, 67}

\bibitem[\protect\citeauthoryear{{Brogan}, {Gelfand}, {Gaensler}, {Kassim}  \&
  {Lazio}}{{Brogan} et~al.}{2006}]{2006ApJ...639L..25B}
{Brogan} C.~L.,  {Gelfand} J.~D.,  {Gaensler} B.~M.,  {Kassim} N.~E.,   {Lazio}
  T.~J.~W.,  2006, \mn@doi [\apjl] {10.1086/501500}, \href
  {https://ui.adsabs.harvard.edu/abs/2006ApJ...639L..25B} {639, L25}

\bibitem[\protect\citeauthoryear{{Cao} et~al.,}{{Cao}
  et~al.}{2021}]{2021Natur.594...33C}
{Cao} Z.,  et~al., 2021, \mn@doi [\nat] {10.1038/s41586-021-03498-z}, \href
  {https://ui.adsabs.harvard.edu/abs/2021Natur.594...33C} {594, 33}

\bibitem[\protect\citeauthoryear{{Castor}, {McCray}  \& {Weaver}}{{Castor}
  et~al.}{1975}]{1975ApJ...200L.107C}
{Castor} J.,  {McCray} R.,   {Weaver} R.,  1975, \mn@doi [\apjl]
  {10.1086/181908}, \href
  {https://ui.adsabs.harvard.edu/abs/1975ApJ...200L.107C} {200, L107}

\bibitem[\protect\citeauthoryear{{Cesarsky} \& {Volk}}{{Cesarsky} \&
  {Volk}}{1978}]{1978A&A....70..367C}
{Cesarsky} C.~J.,  {Volk} H.~J.,  1978, \aap, \href
  {https://ui.adsabs.harvard.edu/abs/1978A&A....70..367C} {70, 367}

\bibitem[\protect\citeauthoryear{{Collins}, {Rowell}, {Mitchell}, {Voisin},
  {Fukui}, {Sano}, {Alsulami}  \& {Einecke}}{{Collins}
  et~al.}{2021}]{2021MNRAS.tmp..976C}
{Collins} T.,  {Rowell} G.,  {Mitchell} A.~M.~W.,  {Voisin} F.,  {Fukui} Y.,
  {Sano} H.,  {Alsulami} R.,   {Einecke} S.,  2021, \mn@doi [\mnras]
  {10.1093/mnras/stab983}, \href
  {https://ui.adsabs.harvard.edu/abs/2021MNRAS.tmp..976C} {}

\bibitem[\protect\citeauthoryear{{Cordes} \& {Lazio}}{{Cordes} \&
  {Lazio}}{2002}]{2002astro.ph..7156C}
{Cordes} J.~M.,  {Lazio} T.~J.~W.,  2002, \mn@doi [arXiv e-prints]
  {10.48550/arXiv.astro-ph/0207156}, \href
  {https://ui.adsabs.harvard.edu/abs/2002astro.ph..7156C} {pp
  astro--ph/0207156}

\bibitem[\protect\citeauthoryear{{Crutcher}, {Wandelt}, {Heiles}, {Falgarone}
  \& {Troland}}{{Crutcher} et~al.}{2010}]{2010ApJ...725..466C}
{Crutcher} R.~M.,  {Wandelt} B.,  {Heiles} C.,  {Falgarone} E.,   {Troland}
  T.~H.,  2010, \mn@doi [\apj] {10.1088/0004-637X/725/1/466}, \href
  {https://ui.adsabs.harvard.edu/abs/2010ApJ...725..466C} {725, 466}

\bibitem[\protect\citeauthoryear{{Di Mauro}, {Manconi}  \& {Donato}}{{Di Mauro}
  et~al.}{2020}]{2020PhRvD.101j3035D}
{Di Mauro} M.,  {Manconi} S.,   {Donato} F.,  2020, \mn@doi [\prd]
  {10.1103/PhysRevD.101.103035}, \href
  {https://ui.adsabs.harvard.edu/abs/2020PhRvD.101j3035D} {101, 103035}

\bibitem[\protect\citeauthoryear{{Drury}}{{Drury}}{1983}]{1983RPPh...46..973D}
{Drury} L.~O.,  1983, \mn@doi [Reports on Progress in Physics]
  {10.1088/0034-4885/46/8/002}, \href
  {https://ui.adsabs.harvard.edu/abs/1983RPPh...46..973D} {46, 973}

\bibitem[\protect\citeauthoryear{{Evoli}, {Linden}  \& {Morlino}}{{Evoli}
  et~al.}{2018}]{2018PhRvD..98f3017E}
{Evoli} C.,  {Linden} T.,   {Morlino} G.,  2018, \mn@doi [\prd]
  {10.1103/PhysRevD.98.063017}, \href
  {https://ui.adsabs.harvard.edu/abs/2018PhRvD..98f3017E} {98, 063017}

\bibitem[\protect\citeauthoryear{{Gabici}, {Aharonian}  \& {Blasi}}{{Gabici}
  et~al.}{2007}]{2007Ap&SS.309..365G}
{Gabici} S.,  {Aharonian} F.~A.,   {Blasi} P.,  2007, \mn@doi [\apss]
  {10.1007/s10509-007-9427-6}, \href
  {https://ui.adsabs.harvard.edu/abs/2007Ap&SS.309..365G} {309, 365}

\bibitem[\protect\citeauthoryear{{Gabici}, {Casanova}, {Aharonian}  \&
  {Rowell}}{{Gabici} et~al.}{2010}]{2010sf2a.conf..313G}
{Gabici} S.,  {Casanova} S.,  {Aharonian} F.~A.,   {Rowell} G.,  2010, in
  {Boissier} S.,  {Heydari-Malayeri} M.,  {Samadi} R.,   {Valls-Gabaud} D.,
  eds, SF2A-2010: Proceedings of the Annual meeting of the French Society of
  Astronomy and Astrophysics. p.~313 (\mn@eprint {arXiv} {1009.5291})

\bibitem[\protect\citeauthoryear{{Gaensler} \& {Slane}}{{Gaensler} \&
  {Slane}}{2006}]{2006ARA&A..44...17G}
{Gaensler} B.~M.,  {Slane} P.~O.,  2006, \mn@doi [\araa]
  {10.1146/annurev.astro.44.051905.092528}, \href
  {https://ui.adsabs.harvard.edu/abs/2006ARA&A..44...17G} {44, 17}

\bibitem[\protect\citeauthoryear{{Giacinti}, {Mitchell}, {L{\'o}pez-Coto},
  {Joshi}, {Parsons}  \& {Hinton}}{{Giacinti}
  et~al.}{2020}]{2020A&A...636A.113G}
{Giacinti} G.,  {Mitchell} A.~M.~W.,  {L{\'o}pez-Coto} R.,  {Joshi} V.,
  {Parsons} R.~D.,   {Hinton} J.~A.,  2020, \mn@doi [\aap]
  {10.1051/0004-6361/201936505}, \href
  {https://ui.adsabs.harvard.edu/abs/2020A&A...636A.113G} {636, A113}

\bibitem[\protect\citeauthoryear{{Giuliani} et~al.,}{{Giuliani}
  et~al.}{2010}]{2010A&A...516L..11G}
{Giuliani} A.,  et~al., 2010, \mn@doi [\aap] {10.1051/0004-6361/201014256},
  \href {https://ui.adsabs.harvard.edu/abs/2010A&A...516L..11G} {516, L11}

\bibitem[\protect\citeauthoryear{{H.E.S.S. Collaboration} et~al.,}{{H.E.S.S.
  Collaboration} et~al.}{2018a}]{2018A&A...612A...1H}
{H.E.S.S. Collaboration} et~al., 2018a, \mn@doi [\aap]
  {10.1051/0004-6361/201732098}, \href
  {https://ui.adsabs.harvard.edu/abs/2018A&A...612A...1H} {612, A1}

\bibitem[\protect\citeauthoryear{{H.E.S.S. Collaboration} et~al.,}{{H.E.S.S.
  Collaboration} et~al.}{2018b}]{2018A&A...612A...2H}
{H.E.S.S. Collaboration} et~al., 2018b, \mn@doi [\aap]
  {10.1051/0004-6361/201629377}, \href
  {https://ui.adsabs.harvard.edu/abs/2018A&A...612A...2H} {612, A2}

\bibitem[\protect\citeauthoryear{{H.E.S.S. Collaboration} et~al.,}{{H.E.S.S.
  Collaboration} et~al.}{2019}]{2019A&A...621A.116H}
{H.E.S.S. Collaboration} et~al., 2019, \mn@doi [\aap]
  {10.1051/0004-6361/201834335}, \href
  {https://ui.adsabs.harvard.edu/abs/2019A&A...621A.116H} {621, A116}

\bibitem[\protect\citeauthoryear{{H.E.S.S. Collaboration} et~al.,}{{H.E.S.S.
  Collaboration} et~al.}{2020}]{2020A&A...644A.112H}
{H.E.S.S. Collaboration} et~al., 2020, \mn@doi [\aap]
  {10.1051/0004-6361/202038851}, \href
  {https://ui.adsabs.harvard.edu/abs/2020A&A...644A.112H} {644, A112}

\bibitem[\protect\citeauthoryear{{Haensel}, {Potekhin}  \&
  {Yakovlev}}{{Haensel} et~al.}{2007}]{neutron_star}
{Haensel} P.,  {Potekhin} A.~Y.,   {Yakovlev} D.~G.,  2007, Neutron Stars 1 :
  Equation of State and Structure.
Springer, New York

\bibitem[\protect\citeauthoryear{{Hinton} \& {Hofmann}}{{Hinton} \&
  {Hofmann}}{2009}]{2009ARA&A..47..523H}
{Hinton} J.~A.,  {Hofmann} W.,  2009, \mn@doi [\araa]
  {10.1146/annurev-astro-082708-101816}, \href
  {https://ui.adsabs.harvard.edu/abs/2009ARA&A..47..523H} {47, 523}

\bibitem[\protect\citeauthoryear{Isaacson}{Isaacson}{1966}]{alma9929637001811}
Isaacson E.,  1966, Analysis of numerical methods.
Wiley, New York

\bibitem[\protect\citeauthoryear{{Kennel} \& {Coroniti}}{{Kennel} \&
  {Coroniti}}{1984a}]{1984ApJ...283..694K}
{Kennel} C.~F.,  {Coroniti} F.~V.,  1984a, \mn@doi [\apj] {10.1086/162356},
  \href {https://ui.adsabs.harvard.edu/abs/1984ApJ...283..694K} {283, 694}

\bibitem[\protect\citeauthoryear{{Kennel} \& {Coroniti}}{{Kennel} \&
  {Coroniti}}{1984b}]{1984ApJ...283..710K}
{Kennel} C.~F.,  {Coroniti} F.~V.,  1984b, \mn@doi [\apj] {10.1086/162357},
  \href {https://ui.adsabs.harvard.edu/abs/1984ApJ...283..710K} {283, 710}

\bibitem[\protect\citeauthoryear{{Khangulyan}, {Koldoba}, {Ustyugova},
  {Bogovalov}  \& {Aharonian}}{{Khangulyan} et~al.}{2018}]{2018ApJ...860...59K}
{Khangulyan} D.,  {Koldoba} A.~V.,  {Ustyugova} G.~V.,  {Bogovalov} S.~V.,
  {Aharonian} F.,  2018, \mn@doi [\apj] {10.3847/1538-4357/aac20f}, \href
  {https://ui.adsabs.harvard.edu/abs/2018ApJ...860...59K} {860, 59}

\bibitem[\protect\citeauthoryear{{Kothes}, {Reich}  \& {Uyan{\i}ker}}{{Kothes}
  et~al.}{2006}]{2006ApJ...638..225K}
{Kothes} R.,  {Reich} W.,   {Uyan{\i}ker} B.,  2006, \mn@doi [\apj]
  {10.1086/498666}, \href
  {https://ui.adsabs.harvard.edu/abs/2006ApJ...638..225K} {638, 225}

\bibitem[\protect\citeauthoryear{{Lazarian}, {Xu}  \& {Hu}}{{Lazarian}
  et~al.}{2023}]{2023FrASS..1054760L}
{Lazarian} A.,  {Xu} S.,   {Hu} Y.,  2023, \mn@doi [Frontiers in Astronomy and
  Space Sciences] {10.3389/fspas.2023.1154760}, \href
  {https://ui.adsabs.harvard.edu/abs/2023FrASS..1054760L} {10, 1154760}

\bibitem[\protect\citeauthoryear{{Lemoine} \& {Pelletier}}{{Lemoine} \&
  {Pelletier}}{2010}]{2010MNRAS.402..321L}
{Lemoine} M.,  {Pelletier} G.,  2010, \mn@doi [\mnras]
  {10.1111/j.1365-2966.2009.15869.x}, \href
  {https://ui.adsabs.harvard.edu/abs/2010MNRAS.402..321L} {402, 321}

\bibitem[\protect\citeauthoryear{{Li} \& {Chen}}{{Li} \&
  {Chen}}{2010}]{2010MNRAS.409L..35L}
{Li} H.,  {Chen} Y.,  2010, \mn@doi [\mnras]
  {10.1111/j.1745-3933.2010.00944.x}, \href
  {https://ui.adsabs.harvard.edu/abs/2010MNRAS.409L..35L} {409, L35}

\bibitem[\protect\citeauthoryear{{Lu}, {Zhu}, {Hu}  \& {Zhang}}{{Lu}
  et~al.}{2023}]{2023MNRAS.518.3949L}
{Lu} F.-W.,  {Zhu} B.-T.,  {Hu} W.,   {Zhang} L.,  2023, \mn@doi [\mnras]
  {10.1093/mnras/stac3298}, \href
  {https://ui.adsabs.harvard.edu/abs/2023MNRAS.518.3949L} {518, 3949}

\bibitem[\protect\citeauthoryear{{Manchester}, {Hobbs}, {Teoh}  \&
  {Hobbs}}{{Manchester} et~al.}{2005}]{2005AJ....129.1993M}
{Manchester} R.~N.,  {Hobbs} G.~B.,  {Teoh} A.,   {Hobbs} M.,  2005, \mn@doi
  [\aj] {10.1086/428488}, \href
  {https://ui.adsabs.harvard.edu/abs/2005AJ....129.1993M} {129, 1993}

\bibitem[\protect\citeauthoryear{{Manolakou}, {Horns}  \& {Kirk}}{{Manolakou}
  et~al.}{2007}]{2007A&A...474..689M}
{Manolakou} K.,  {Horns} D.,   {Kirk} J.~G.,  2007, \mn@doi [\aap]
  {10.1051/0004-6361:20078298}, \href
  {https://ui.adsabs.harvard.edu/abs/2007A&A...474..689M} {474, 689}

\bibitem[\protect\citeauthoryear{{Mizuno} \& {Fukui}}{{Mizuno} \&
  {Fukui}}{2004}]{2004ASPC..317...59M}
{Mizuno} A.,  {Fukui} Y.,  2004, in {Clemens} D.,  {Shah} R.,   {Brainerd} T.,
  eds,  Astronomical Society of the Pacific Conference Series Vol. 317, Milky
  Way Surveys: The Structure and Evolution of our Galaxy. p.~59

\bibitem[\protect\citeauthoryear{{Odegard}}{{Odegard}}{1986}]{1986AJ.....92.1372O}
{Odegard} N.,  1986, \mn@doi [\aj] {10.1086/114270}, \href
  {https://ui.adsabs.harvard.edu/abs/1986AJ.....92.1372O} {92, 1372}

\bibitem[\protect\citeauthoryear{{Popescu}, {Yang}, {Tuffs}, {Natale},
  {Rushton}  \& {Aharonian}}{{Popescu} et~al.}{2017}]{2017MNRAS.470.2539P}
{Popescu} C.~C.,  {Yang} R.,  {Tuffs} R.~J.,  {Natale} G.,  {Rushton} M.,
  {Aharonian} F.,  2017, \mn@doi [\mnras] {10.1093/mnras/stx1282}, \href
  {https://ui.adsabs.harvard.edu/abs/2017MNRAS.470.2539P} {470, 2539}

\bibitem[\protect\citeauthoryear{{Porth}, {Vorster}, {Lyutikov}  \&
  {Engelbrecht}}{{Porth} et~al.}{2016}]{2016MNRAS.460.4135P}
{Porth} O.,  {Vorster} M.~J.,  {Lyutikov} M.,   {Engelbrecht} N.~E.,  2016,
  \mn@doi [\mnras] {10.1093/mnras/stw1152}, \href
  {https://ui.adsabs.harvard.edu/abs/2016MNRAS.460.4135P} {460, 4135}

\bibitem[\protect\citeauthoryear{{Principe}, {Mitchell}, {Caroff}, {Hinton},
  {Parsons}  \& {Funk}}{{Principe} et~al.}{2020}]{2020A&A...640A..76P}
{Principe} G.,  {Mitchell} A.~M.~W.,  {Caroff} S.,  {Hinton} J.~A.,  {Parsons}
  R.~D.,   {Funk} S.,  2020, \mn@doi [\aap] {10.1051/0004-6361/202038375},
  \href {https://ui.adsabs.harvard.edu/abs/2020A&A...640A..76P} {640, A76}

\bibitem[\protect\citeauthoryear{{Prosekin}, {Kelner}  \&
  {Aharonian}}{{Prosekin} et~al.}{2015}]{2015PhRvD..92h3003P}
{Prosekin} A.~Y.,  {Kelner} S.~R.,   {Aharonian} F.~A.,  2015, \mn@doi [\prd]
  {10.1103/PhysRevD.92.083003}, \href
  {https://ui.adsabs.harvard.edu/abs/2015PhRvD..92h3003P} {92, 083003}

\bibitem[\protect\citeauthoryear{{Protheroe}, {Ott}, {Ekers}, {Jones}  \&
  {Crocker}}{{Protheroe} et~al.}{2008}]{2008MNRAS.390..683P}
{Protheroe} R.~J.,  {Ott} J.,  {Ekers} R.~D.,  {Jones} D.~I.,   {Crocker}
  R.~M.,  2008, \mn@doi [\mnras] {10.1111/j.1365-2966.2008.13752.x}, \href
  {https://ui.adsabs.harvard.edu/abs/2008MNRAS.390..683P} {390, 683}

\bibitem[\protect\citeauthoryear{{Recchia}, {Di Mauro}, {Aharonian}, {Orusa},
  {Donato}, {Gabici}  \& {Manconi}}{{Recchia}
  et~al.}{2021}]{2021PhRvD.104l3017R}
{Recchia} S.,  {Di Mauro} M.,  {Aharonian} F.~A.,  {Orusa} L.,  {Donato} F.,
  {Gabici} S.,   {Manconi} S.,  2021, \mn@doi [\prd]
  {10.1103/PhysRevD.104.123017}, \href
  {https://ui.adsabs.harvard.edu/abs/2021PhRvD.104l3017R} {104, 123017}

\bibitem[\protect\citeauthoryear{{Rees} \& {Gunn}}{{Rees} \&
  {Gunn}}{1974}]{1974MNRAS.167....1R}
{Rees} M.~J.,  {Gunn} J.~E.,  1974, \mn@doi [\mnras] {10.1093/mnras/167.1.1},
  \href {https://ui.adsabs.harvard.edu/abs/1974MNRAS.167....1R} {167, 1}

\bibitem[\protect\citeauthoryear{{Sabol} \& {Snowden}}{{Sabol} \&
  {Snowden}}{2019}]{2019ascl.soft04001S}
{Sabol} E.~J.,  {Snowden} S.~L.,  2019, {sxrbg: ROSAT X-Ray Background Tool},
  Astrophysics Source Code Library, record ascl:1904.001 (\mn@eprint {ascl}
  {1904.001})

\bibitem[\protect\citeauthoryear{{Sano} et~al.,}{{Sano}
  et~al.}{2017}]{2017ApJ...843...61S}
{Sano} H.,  et~al., 2017, \mn@doi [\apj] {10.3847/1538-4357/aa73e0}, \href
  {https://ui.adsabs.harvard.edu/abs/2017ApJ...843...61S} {843, 61}

\bibitem[\protect\citeauthoryear{{Schmidt}, {Angel}  \& {Beaver}}{{Schmidt}
  et~al.}{1979}]{1979ApJ...227..106S}
{Schmidt} G.~D.,  {Angel} J.~R.~P.,   {Beaver} E.~A.,  1979, \mn@doi [\apj]
  {10.1086/156708}, \href
  {https://ui.adsabs.harvard.edu/abs/1979ApJ...227..106S} {227, 106}

\bibitem[\protect\citeauthoryear{{Schroer}, {Evoli}  \& {Blasi}}{{Schroer}
  et~al.}{2023}]{2023PhRvD.107l3020S}
{Schroer} B.,  {Evoli} C.,   {Blasi} P.,  2023, \mn@doi [\prd]
  {10.1103/PhysRevD.107.123020}, \href
  {https://ui.adsabs.harvard.edu/abs/2023PhRvD.107l3020S} {107, 123020}

\bibitem[\protect\citeauthoryear{{Sironi}, {Keshet}  \& {Lemoine}}{{Sironi}
  et~al.}{2015}]{2015SSRv..191..519S}
{Sironi} L.,  {Keshet} U.,   {Lemoine} M.,  2015, \mn@doi [\ssr]
  {10.1007/s11214-015-0181-8}, \href
  {https://ui.adsabs.harvard.edu/abs/2015SSRv..191..519S} {191, 519}

\bibitem[\protect\citeauthoryear{{Skilling}}{{Skilling}}{1975}]{1975MNRAS.172..557S}
{Skilling} J.,  1975, \mn@doi [\mnras] {10.1093/mnras/172.3.557}, \href
  {https://ui.adsabs.harvard.edu/abs/1975MNRAS.172..557S} {172, 557}

\bibitem[\protect\citeauthoryear{{Strong}, {Moskalenko}, {Reimer}, {Digel}  \&
  {Diehl}}{{Strong} et~al.}{2004}]{2004A&A...422L..47S}
{Strong} A.~W.,  {Moskalenko} I.~V.,  {Reimer} O.,  {Digel} S.,   {Diehl} R.,
  2004, \mn@doi [\aap] {10.1051/0004-6361:20040172}, \href
  {https://ui.adsabs.harvard.edu/abs/2004A&A...422L..47S} {422, L47}

\bibitem[\protect\citeauthoryear{{Strong}, {Moskalenko}  \& {Ptuskin}}{{Strong}
  et~al.}{2007}]{2007ARNPS..57..285S}
{Strong} A.~W.,  {Moskalenko} I.~V.,   {Ptuskin} V.~S.,  2007, \mn@doi [Annual
  Review of Nuclear and Particle Science]
  {10.1146/annurev.nucl.57.090506.123011}, \href
  {https://ui.adsabs.harvard.edu/abs/2007ARNPS..57..285S} {57, 285}

\bibitem[\protect\citeauthoryear{{Stupar}, {Parker}  \&
  {Filipovi{\'c}}}{{Stupar} et~al.}{2008}]{2008MNRAS.390.1037S}
{Stupar} M.,  {Parker} Q.~A.,   {Filipovi{\'c}} M.~D.,  2008, \mn@doi [\mnras]
  {10.1111/j.1365-2966.2008.13761.x}, \href
  {https://ui.adsabs.harvard.edu/abs/2008MNRAS.390.1037S} {390, 1037}

\bibitem[\protect\citeauthoryear{{Tanaka} \& {Takahara}}{{Tanaka} \&
  {Takahara}}{2010}]{2010ApJ...715.1248T}
{Tanaka} S.~J.,  {Takahara} F.,  2010, \mn@doi [\apj]
  {10.1088/0004-637X/715/2/1248}, \href
  {https://ui.adsabs.harvard.edu/abs/2010ApJ...715.1248T} {715, 1248}

\bibitem[\protect\citeauthoryear{{Tang} \& {Chevalier}}{{Tang} \&
  {Chevalier}}{2012}]{2012ApJ...752...83T}
{Tang} X.,  {Chevalier} R.~A.,  2012, \mn@doi [\apj]
  {10.1088/0004-637X/752/2/83}, \href
  {https://ui.adsabs.harvard.edu/abs/2012ApJ...752...83T} {752, 83}

\bibitem[\protect\citeauthoryear{{Taylor} \& {Cordes}}{{Taylor} \&
  {Cordes}}{1993}]{1993ApJ...411..674T}
{Taylor} J.~H.,  {Cordes} J.~M.,  1993, \mn@doi [\apj] {10.1086/172870}, \href
  {https://ui.adsabs.harvard.edu/abs/1993ApJ...411..674T} {411, 674}

\bibitem[\protect\citeauthoryear{{Uchiyama}, {Matsumoto}, {Tsuru}, {Koyama}  \&
  {Bamba}}{{Uchiyama} et~al.}{2009}]{2009PASJ...61S.189U}
{Uchiyama} H.,  {Matsumoto} H.,  {Tsuru} T.~G.,  {Koyama} K.,   {Bamba} A.,
  2009, \mn@doi [\pasj] {10.1093/pasj/61.sp1.S189}, \href
  {https://ui.adsabs.harvard.edu/abs/2009PASJ...61S.189U} {61, S189}

\bibitem[\protect\citeauthoryear{{Van Etten} \& {Romani}}{{Van Etten} \&
  {Romani}}{2011}]{2011ApJ...742...62V}
{Van Etten} A.,  {Romani} R.~W.,  2011, \mn@doi [\apj]
  {10.1088/0004-637X/742/2/62}, \href
  {https://ui.adsabs.harvard.edu/abs/2011ApJ...742...62V} {742, 62}

\bibitem[\protect\citeauthoryear{{Voisin}, {Rowell}, {Burton}, {Walsh}, {Fukui}
   \& {Aharonian}}{{Voisin} et~al.}{2016}]{2016MNRAS.458.2813V}
{Voisin} F.,  {Rowell} G.,  {Burton} M.~G.,  {Walsh} A.,  {Fukui} Y.,
  {Aharonian} F.,  2016, \mn@doi [\mnras] {10.1093/mnras/stw473}, \href
  {https://ui.adsabs.harvard.edu/abs/2016MNRAS.458.2813V} {458, 2813}

\bibitem[\protect\citeauthoryear{{Weaver}, {McCray}, {Castor}, {Shapiro}  \&
  {Moore}}{{Weaver} et~al.}{1977}]{1977ApJ...218..377W}
{Weaver} R.,  {McCray} R.,  {Castor} J.,  {Shapiro} P.,   {Moore} R.,  1977,
  \mn@doi [\apj] {10.1086/155692}, \href
  {https://ui.adsabs.harvard.edu/abs/1977ApJ...218..377W} {218, 377}

\bibitem[\protect\citeauthoryear{{de Jager} \& {Djannati-Ata{\"\i}}}{{de Jager}
  \& {Djannati-Ata{\"\i}}}{2009}]{2009ASSL..357..451D}
{de Jager} O.~C.,  {Djannati-Ata{\"\i}} A.,  2009, in {Becker} W.,  ed.,
  Astrophysics and Space Science Library Vol. 357, Astrophysics and Space
  Science Library. p.~451 (\mn@eprint {arXiv} {0803.0116}),
  \mn@doi{10.1007/978-3-540-76965-1\_17}

\makeatother
\end{thebibliography}



\appendix

\section{Non-thermal emission} \label{sec:non_thermal_em}
This section will provide an overview of leptonic interactions and the subsequent photon emission via synchrotron, Bremsstrahlung and inverse Compton Processes.
\par~\par 
\noindent Synchrotron radiation occurs when an electron interacts with background magnetic fields. The resulting photon emission from a single electron with Lorentz factor $\gamma$ with pitch angle $\alpha$ to the magnetic field $B$ is given by:
	
\begin{equation}
    \begin{aligned}
        \dv{N}{E}&=\frac{\sqrt{3}e^3B}{mc^2} \frac{\nu}{\nu_c}\int_{\frac{\nu}{\nu_c}}^\infty K_\frac{5}{3}(x) \dd{x}\text{ ,}
    \end{aligned} \label{eq:synchrotron}
\end{equation}
\noindent where $e$ and $m$ are the charge and mass of an electron respectively, $K_\frac{5}{3}$ is the modified Bessel Function, $\nu$ is the frequency of the gamma ray and $\nu_c$ is the critical frequency of the emission:

\begin{equation}
    \begin{aligned}
        \nu_c&=\gamma^2\frac{3eB\sin\alpha}{4\pi mc}\text{ .}
    \end{aligned}
\end{equation}
\noindent The inverse Compton gamma-ray emission from an electron with energy $E_e$ scattering off a target photon with energy in range ($\epsilon+\dd{\epsilon}$) and number density $n\qty(\epsilon)$ can be found using:

\begin{subequations}
    \begin{alignat}{1}
        \frac{\dd{N}}{\dd{E_\gamma}}&=\frac{3\sigma_T mc^3}{4\gamma}\int_{E_\gamma/4\gamma^2}^{E_\gamma}\frac{n\qty(\epsilon)\dd{\epsilon}}{\epsilon}f\qty(q,\Gamma) \label{eq:IC_emission} \\ 
        f\qty(q,\Gamma)&=2q\ln q +\qty(1+2q)\qty(1-q)+\frac{1}{2}\frac{\qty(\Gamma q)^2}{1+\Gamma q}\qty(1-q) \\
        q&=\frac{E_\gamma}{\Gamma\qty(E_e-E_\gamma)}\text{,}\quad \Gamma=\frac{4\epsilon\gamma}{m_ec^2}
    \end{alignat} 
\end{subequations}
\noindent where $\sigma_T=(3/8\pi)r_0^2$ is the Thompson cross section, $r_0$ is the classical electron radius and $F_{\text{KN}}$ takes account the full Klein-Nishina cross section for inverse Compton scattering \citep{2007A&A...474..689M}:

\begin{equation}
	\begin{split}
		F_\text{KN}&=\frac{1}{u_0}\int_{0}^\infty \mathscr{f}\qty(\gamma,\epsilon)u_{\epsilon}\dd{\epsilon}\text{,}\quad \mathscr{f}\qty(\gamma,\epsilon)=\qty(1+4\gamma\epsilon)^{-\frac{3}{2}}\text{ .}
	\end{split} \label{eq:klein_nishina_cross_section}
\end{equation}
\noindent For a Planck distribution of photon energies, $F_\text{KN}$ can be approximated by:

\begin{equation}
    \begin{aligned}
    F_\text{KN}&=\qty(1+4\gamma \epsilon_\text{eff})^{-3/2}\text{,}\quad\epsilon_\text{eff}=\frac{2.8kT}{m_ec^2}\text{ .}
    \end{aligned}
\end{equation}

\noindent Finally, the photon emission from Bremsstrahlung interactions is given by:

\begin{equation}
    \begin{aligned}
        \dv{N}{E_\gamma}= nc \int \dd{\sigma}\qty(E_e, E_\gamma, Z)\dd{E_e} \text{ ,} 
    \end{aligned} \label{eq:bremsstrahlung_flux}
\end{equation}

\noindent where $Z$ is the atomic number of the target material and $\dd{\sigma}$ is the Bremsstrahlung differential cross section as defined in \cite{1970RvMP...42..237B}. 
\par~\par 
\noindent The coefficients for leptonic losses in \autoref{eq:cooling_rate} are:
\setlist[itemize]{wide=0pt,leftmargin=10pt}
\begin{itemize}
    \itemsep0em
    \renewcommand\labelitemi{--}
    \item $b_s\equiv 1.292\times 10^{-15}\qty(B/10^3\si{\micro G})^2\,\si{\per\second}$ is the synchrotron loss coefficient
    \item $b_c\equiv 1.491\times 10^{-14} \qty(n_H/1\centimeterminusthree)\,\si{\per\second}$ is the Coulomb loss coefficient
    \item $b_b\equiv 1.37\times 10^{-16}\qty(n_H/1\centimeterminusthree)\,\si{\per\second}$ is the Bremsstrahlung loss coefficient
    \item $b_\text{IC}\equiv 5.204\times 10^{-20}\qty(u_0/\si{\electronvolt\per\centi\meter\cubed})\,\si{\per\second}$ is the IC loss coefficient with the energy density of photons given by $u_0$
    \item $n_H$ is the density of the ambient hydrogen gas
\end{itemize}

\par 
\noindent The diffusion length for electrons \citep{1995PhRvD..52.3265A}:

\begin{equation}
    \begin{aligned}
        R_\text{diff}&=\sqrt{\frac{4D\qty(\gamma)}{b_s\gamma\qty(1-\delta)}\qty[1-\qty(1-\gamma b_st)^{1-\delta}]}
    \end{aligned} \text{ .} \label{eq:diffusion_radius}
\end{equation}

\section{Magnetic field due to turbulent ISM gas}
The magnetic field due to the ISM gas with number density $n$ is given through Crutcher's relation \citep{2010ApJ...725..466C}:

\begin{equation}
    \begin{aligned}
    B_\text{gas}\qty(n) &=
    \begin{cases}
        B_{0,\text{gas}}\text{,} &  n < 300\,\centimeterminusthree \\
        B_{0,\text{gas}}\qty({n}/{300\,\centimeterminusthree})^\alpha\text{,} & n > 300\,\centimeterminusthree
    \end{cases}
    \end{aligned} \text{ ,} \label{eq:crutchers_relation}
\end{equation}
\noindent where $B_{0,\text{gas}}=10\,\si{\micro G}$ and $\alpha=0.65$.
\section{Single-Zone Modelling} \label{sec:singlezone_mod}

Here we considered a `single-zone' model, where electrons are injected into a spherical region of constant number density and magnetic field \citep{2017ApJ...843...61S, 2021MNRAS.tmp..976C}. The final electron number density is calculated by solving Eq.\,C7 from \cite{2021MNRAS.tmp..976C} over the age of the system where electrons escape the region at a rate dependent on diffusion. The multi-wavelength SED from this region is then calculated. While unable to encapsulate the complexity towards \mbox{HESS\,J1825-137}, a general insight of the system was gained before more detailed modelling of the morphology and time evolution.

\subsection{Method} \label{sec:1825_single}

\begin{figure}
    \centering
    \includegraphics[width=\columnwidth]{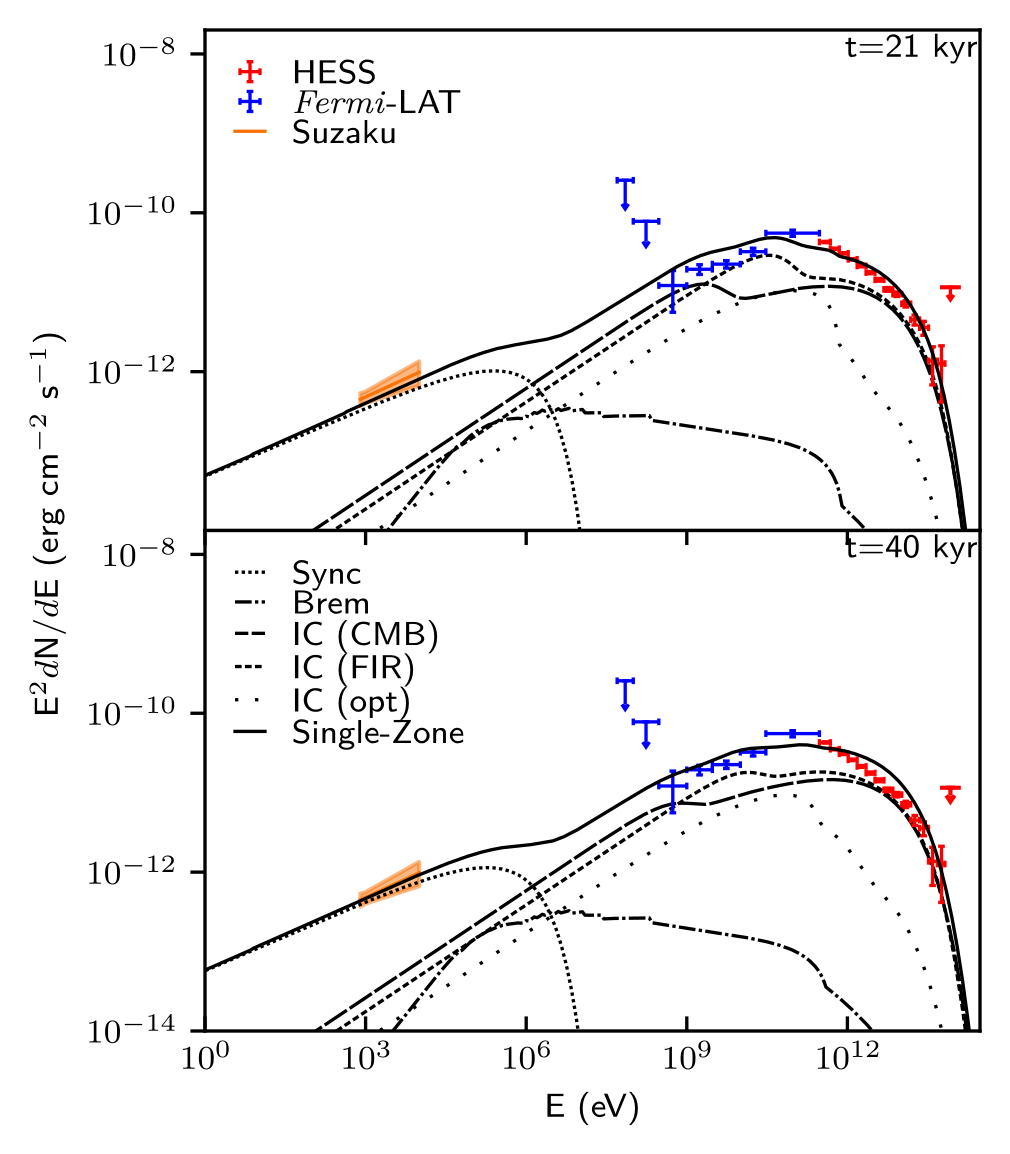}
    \begin{minipage}[t]{\columnwidth}
        \begin{tabular}{ccccc}
    		\toprule
    		& \multicolumn{2}{c}{${t=21\,\kiloyear}$} & \multicolumn{2}{c}{${t=40\,\kiloyear}$} \\
            Parameter & HESS & Suzaku & HESS & Suzaku \\
    		\midrule
    		$\dot{E}$ ($\ergspersecond$) & $2\times 10^{38}$ & $8\times 10^{35}$& $1\times 10^{38}$ & $4\times 10^{35}$ \\
            $r$ ($\si{\degree}$) & $0.70$ & $0.025$ & $0.70$ & $0.025$ \\
            $n$ ($\centimeterminusthree$) & $0.5$ & $0.5$ & $0.5$ & $0.5$ \\
            $B$ ($\si{\micro G}$) & $5$  & $40$ & $5$  & $40$ \\
            $\chi$ & $0.25$ & $0.25$ & $0.1$ & $0.1$ \\
            $\Gamma$ & $2.1$ & $1.9$ & $2.1$ & $1.9$ \\
            $E_c$ ($\TeV$) & $40$ & $1000$ & $50$ & $1000$\\
    		\bottomrule
    	\end{tabular}
    \end{minipage}
    \caption{SED for leptonic interactions towards \HESSmain using single-zone modelling for ages $21\,\kiloyear$ (\textit{top}) and $40\,\kiloyear$ (\textit{bottom}). The X-ray and gamma-ray spectra are fitted separately due to the different coverage areas of HESS and Suzaku. The orange line shows the Suzaku observations of X-rays between $1-9\,\keV$ towards the pulsar associated with \HESSmain \citep{2009PASJ...61S.189U}. Blue data points represent the spectrum from the \fermi-LAT 4FGL source catalogue towards \HESSmain while the red data shows the H.E.S.S. energy flux towards \HESSmain \citep{2019A&A...621A.116H}. The corresponding model parameters are shown in the table.}
    \label{fig:nsp_1825}
\end{figure}

The X-ray and gamma-ray emission was modelled separately using two spheres with radii $r_\text{X-ray}$ and $r_\text{gamma}$ (see \autoref{fig:nanten_data}) following the extraction regions used by \cite{2009PASJ...61S.189U} and \cite{2019A&A...621A.116H}. Electrons were injected into the spherical region at a constant rate $\dot{E}$ and followed a power-law spectrum with an exponential cutoff: $\dv{N}{E}\propto E^{-\Gamma}\cdot\exp(E/E_c)$, where $\Gamma$ is the spectral index and $E_c$ is the cutoff energy. Two different ages were modelled, $21\,\kiloyear$ based on the characteristic age of the pulsar and $40\,\kiloyear$ based on modelling conductede by \citep{2011ApJ...742...62V}.
\par
The HESS region adopted a uniform magnetic field of $5\,\si{\micro G}$ as suggested by \cite{2020A&A...640A..76P} from comparing the estimated synchrotron emission to the Suzaku X-ray emission. Subsequently, it was assumed that the smaller X-ray region has a higher magnetic field strength than the HESS region due to the proximity of the pulsar and was left as a free parameter. Both the gamma-ray and X-ray region assumed a constant ISM density of $0.5\,\centimeterminusthree$
\par 
The fits to the SED towards \HESSmain as well as the modelled parameters can be seen in \autoref{fig:nsp_1825}.

\subsection{Discussion}

\autoref{fig:nsp_1825} shows the modelled SED with corresponding parameters to the gamma and X-ray spectra towards \HESSmain. The majority of gamma rays in this model are predicted to come from inverse Compton interactions from the infrared and CMB photon fields. An electron injection luminosity of $2\times 10^{38}\,\ergspersecond$ and $1\times 10^{38}\,\ergspersecond$ is needed to match the gamma-ray spectra at ages $21$ and $40\,\kiloyear$ respectively. This is a factor ten times greater than the spin-down power of \mbox{PSR\,J1826-1334} ($\dot{E}=2.8\times 10^{36}\,\ergspersecond$). The single-zone model assumes a time-independent injection luminosity, whereas the spin-down power of the pulsar decreases over time. The spin-down power of \mbox{PSR\,J1826-1334} could have been as high as $10^{39}\,\ergspersecond$ at a pulsar age of $1\,\kiloyear$ (see \autoref{ref:particle_spectra_evolution}. Therefore the modeled injection luminosities represents the average electron injection luminosity over the age of the pulsar.
\par 
The X-ray emission towards \mbox{PSR\,J1826-1334} can be predicted with an injection luminosity of $8\times 10^{35}\,\ergspersecond$ and $4\times 10^{35}\,\ergspersecond$ for the $21$ and $40\,\kiloyear$ model respectively. The single-zone model can reasonably predict both the X-ray and gamma-ray SED, yet the X-ray and gamma-ray photon models require different injection spectra for both ages of the system. The single-zone model assumes constant density and magnetic field strength across the region of interest. However, the magnetic field structures towards PWNe have been suggested to be toroidal in nature but the viewing angle results in magnetic fields appearing radially dependent or tangled \citep{2006ApJ...638..225K}. If the dense clouds towards \mbox{HESS\,J1826-130}, as seen in \autoref{fig:nanten_data}, lie at the same distance as the pulsar, diffusion will be suppressed towards this region with electrons losing their energy to bremsstrahlung losses. As previously mentioned, the spin-down power of the pulsar decreases over time which has an effect on the injection luminosity of electrons in the ISM. While the single-zone model is able to predict the X-ray and gamma-ray SED towards \mbox{HESS\,J1825-137}, it is unable to encapsulate the complexity of the PWN.

\section{Systematic Variation of Multizone Parameters} \label{sec:systematic_variation}

\begin{figure*}
    \includegraphics{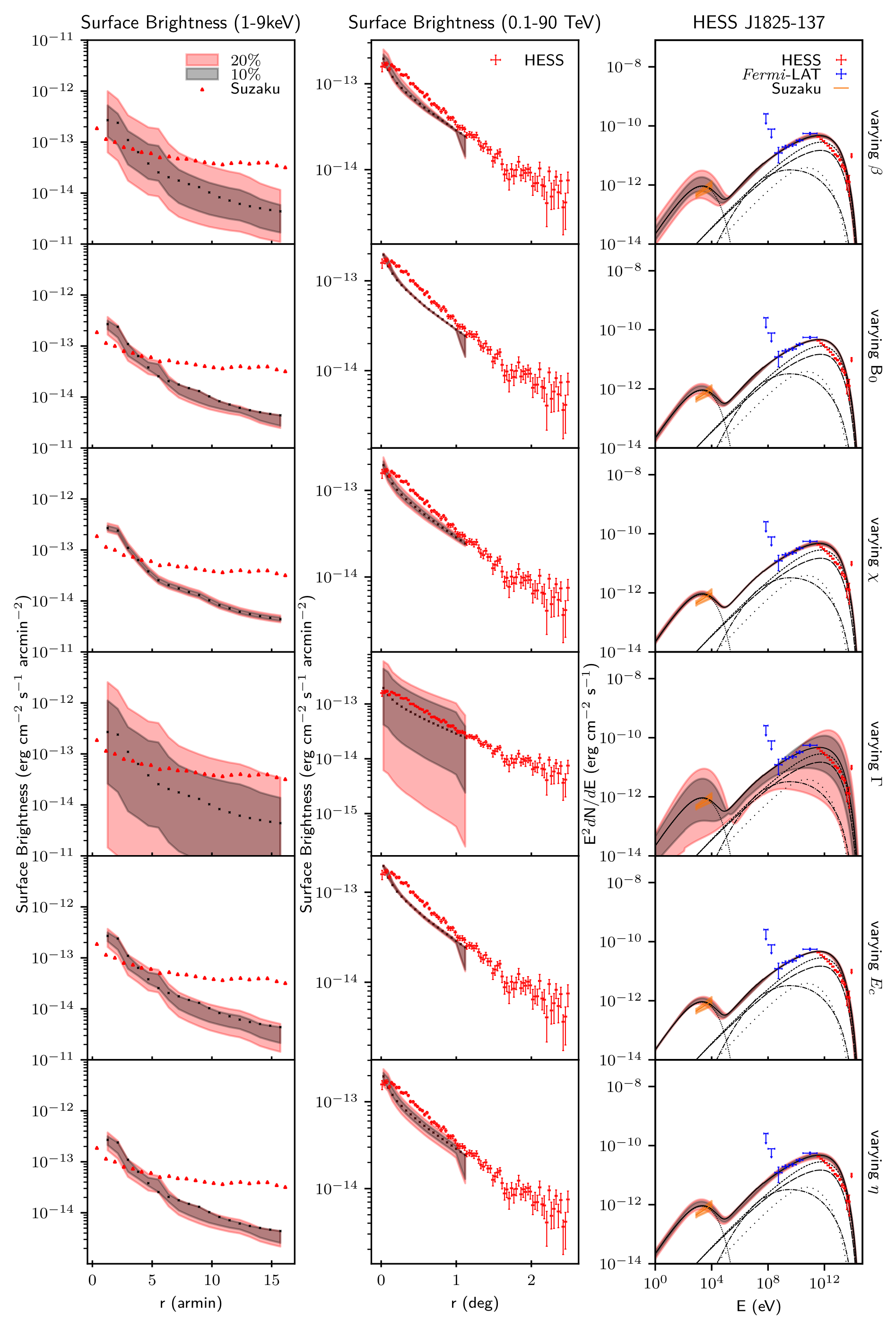}
    \caption{Model~1 ($21\,\kiloyear$) as in \autoref{fig:multizone_21} but with $10\%$ (grey shaded band) and $20\%$ (pink shaded band) variation in parameters.}
    \label{fig:multizone_21_changing_paramaters}
\end{figure*}

\begin{figure*}
    \includegraphics{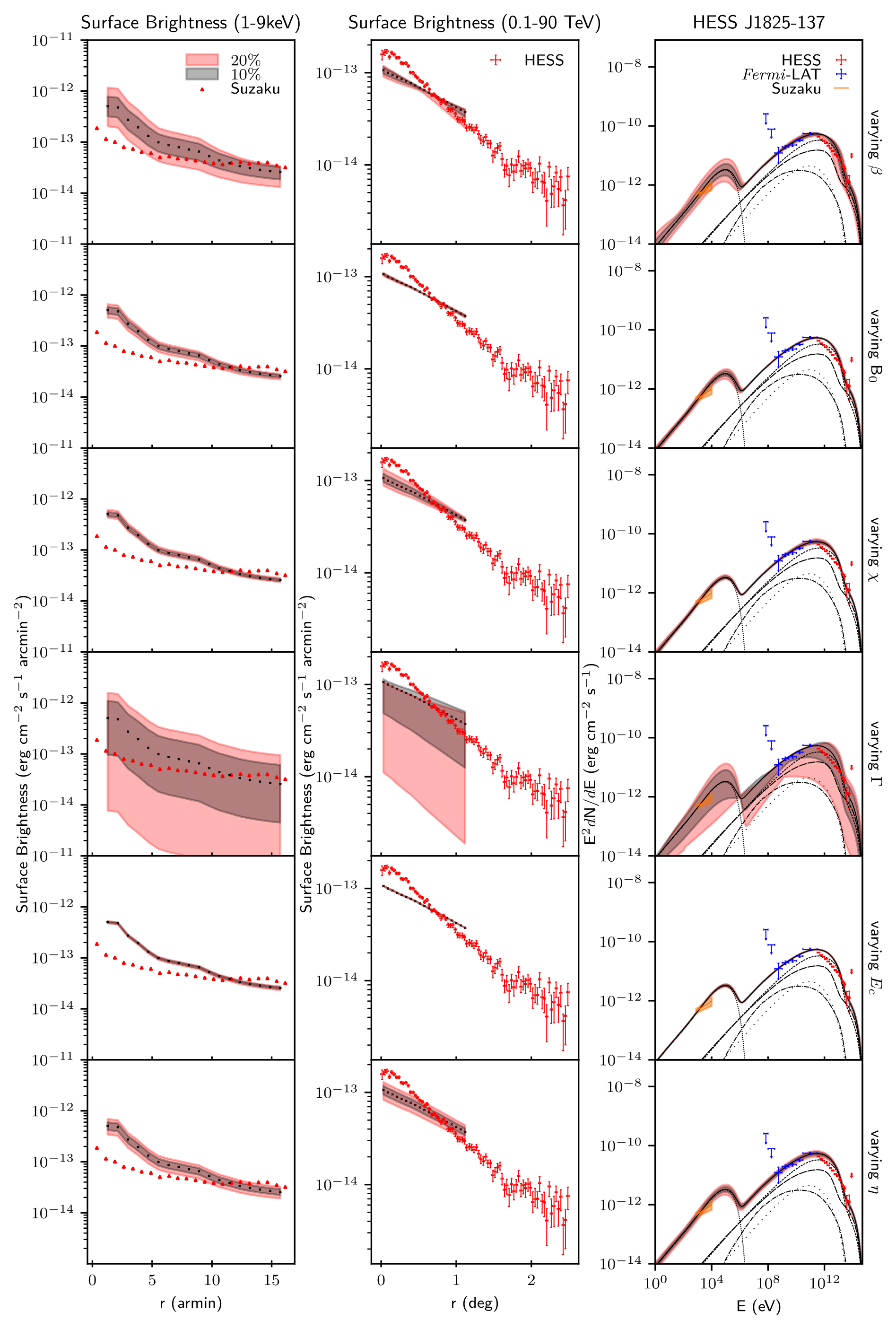}   
    \caption{Model~2 ($40\,\kiloyear$) as in \autoref{fig:multizone_40} but with $10\%$ (grey shaded band) and $20\%$ (pink shaded band) variation in parameters.}
    \label{fig:multizone_40_changing_paramaters}
\end{figure*}

\autoref{fig:multizone_21_changing_paramaters} and \autoref{fig:multizone_40_changing_paramaters} shows the $10\%$ and $20\%$ systematic variation of the free parameters $\beta$, $B_0$, $\chi$, $\Gamma$, $E_c$ and $\eta$ for the $21\,\kiloyear$ and $40\,\kiloyear$ models. These figures show that the spectral index of injection electrons, $\Gamma$, has the largest systematic variation, where the X-ray SED and surface brightness radial profiles show more sensitivity than the gamma-ray emission. This is a result of the smaller region used to extract the X-ray and SED (see \autoref{fig:nanten_data} and \autoref{fig:multizone_21}).
\par
The modelled X-ray surface brightness radial profiles for the $21$ and $40\,\kiloyear$ models are steeper than observations, indicating that the model over-predicts the synchrotron emission closer to PSR\,J1826-1334. This may be corrected by decreasing the rate at which the magnetic field drops off with distance from the pulsar ($\beta$), allowing electrons to escape the PWN at a faster rate. The outer edges of the PWN experiences greater synchrotron losses at the cost of $\TeV$ gamma-ray emission from IC interactions, flattening out the gamma-ray surface brightness radial profile. This is demonstrated in the $10$ and $20\%$ variation of $\beta$ shown in the top row of \autoref{fig:multizone_21_changing_paramaters} and \autoref{fig:multizone_40_changing_paramaters}. Alternatively, decreasing the overall magnetic field strength, $B_0$, decreases synchrotron losses towards \mbox{HESS\,J1825-137} at the cost of increasing the gamma-ray to X-ray flux ratio. With flux being dependent on the observational area, any changes to the gamma-ray and X-ray ratio will be more prominent in the X-ray SED as shown in \autoref{fig:multizone_21_changing_paramaters} and \autoref{fig:multizone_40_changing_paramaters}.
\par
To better fit the X-ray surface brightness radial profile, the diffusion suppression coefficient, $\chi$, towards \mbox{HESS\,J1825-137} could be increased to allow electrons to escape further from the pulsar before losing their energy to synchrotron radiation. High-energy electrons rapidly lose their energy to radiative losses and remain close to the pulsar, resulting in a shallower gamma-ray surface brightness radial profile as shown in \autoref{fig:multizone_21_changing_paramaters} and \autoref{fig:multizone_40_changing_paramaters}. As the region used to extract the X-ray data is small ($<2pc$) compared to the HESS region ($\approx 49\,\pc$), electrons quickly escape the X-ray region while remaining in the HESS region. Thus the X-ray SED far more sensitive to the value of $\chi$ than the gamma-ray SED.
\par
Both the surface brightness radial profiles and SED are very sensitive to the injected electron spectral index, $\Gamma$, as seen in \autoref{fig:multizone_21_changing_paramaters} and \autoref{fig:multizone_40_changing_paramaters}. If $\beta$, $B_0$ or $\chi$ was altered to fit the observed Suzaku X-ray surface brightness radial profile, the predicted SED from the model will no longer fit to the data. In turn, the spectral index can be modified to refit the modelled SED. Consequently, the X-ray surface brightness radial profile will no longer match the Suzaku observations.
\par
The cutoff energies for the $21\,\kiloyear$ and $40\,\kiloyear$ models are $45\,\TeV$ and $500\,\TeV$ respectively. As the cutoff energy for an exponential cutoff power-law increases, the energy spectra starts to follow a power-law. Hence, the systematic variation of $E_C$ is less apparent for $40\,\kiloyear$ than $21\,\kiloyear$ as seen in the fourth row of \autoref{fig:multizone_21_changing_paramaters} and \autoref{fig:multizone_40_changing_paramaters} respectively.

\section{Additional Figures}

 \begin{center}
    \includegraphics[width=\columnwidth]{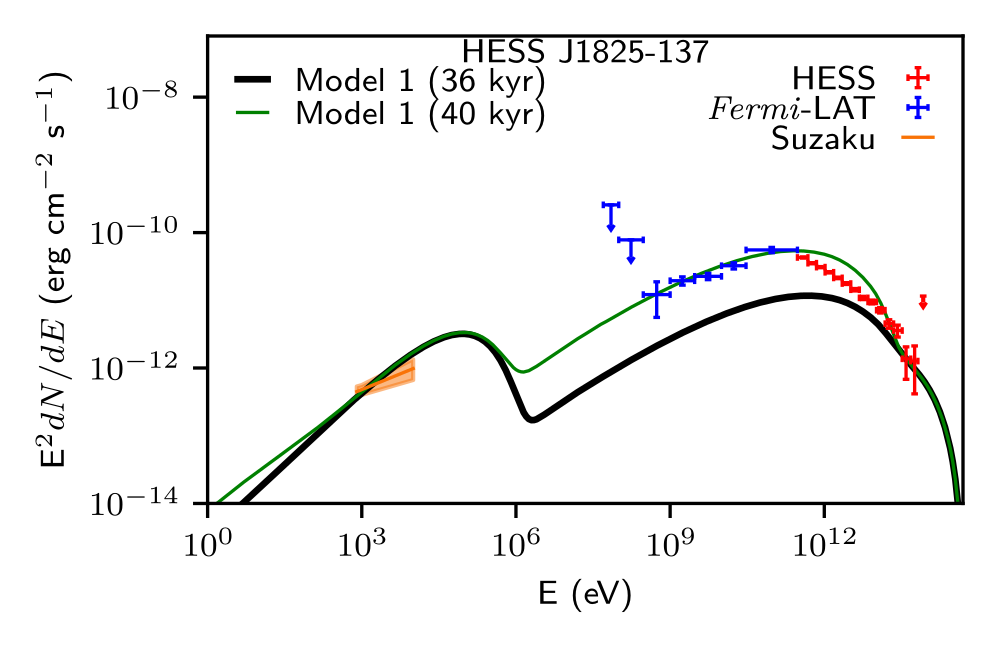}
    \captionof{figure}{SED towards \mbox{HESS\,J1825-137} for Model~1 ($36\,\kiloyear$, \textit{green}) shown vs Model~1 ($40\,\kiloyear$). A `bump' is present in the SED above $10\,\TeV$ for Model~1 ($40\,\kiloyear$) where radiative losses are balanced by the electron injection luminosity. The $36\,\kiloyear$ model has the same parameters as the $40\,\kiloyear$ model (see \autoref{tab:matched_parameters}).}
    \label{fig:multizone_time_comparison}
\end{center}

\begin{center}
    \includegraphics[width=\columnwidth]{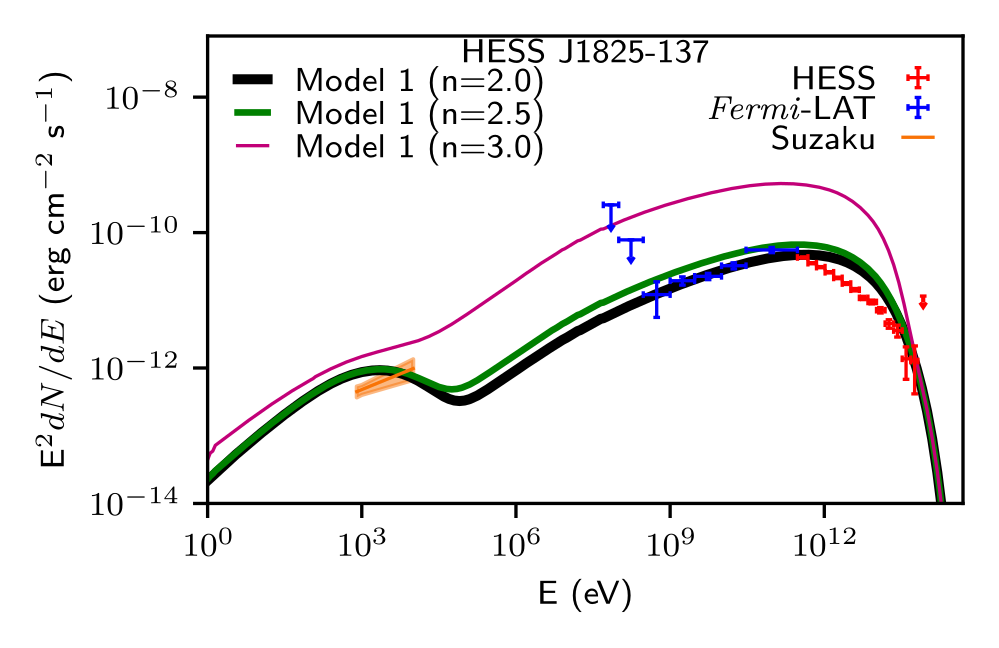}
    \captionof{figure}{SED towards \mbox{HESS\,J1825-137} for Model~1 ($21\,\kiloyear$, $n=2$, \textit{black}, see \autoref{tab:matched_parameters}) vs Model~1 ($21\,\kiloyear$, $n=2.5$, \textit{green}) and Model~1 ($21\,\kiloyear$, $n=3$, \textit{purple}).}
    \label{fig:multizone_n_comparison}
\end{center}

\begin{center}
    \includegraphics[width=\columnwidth]{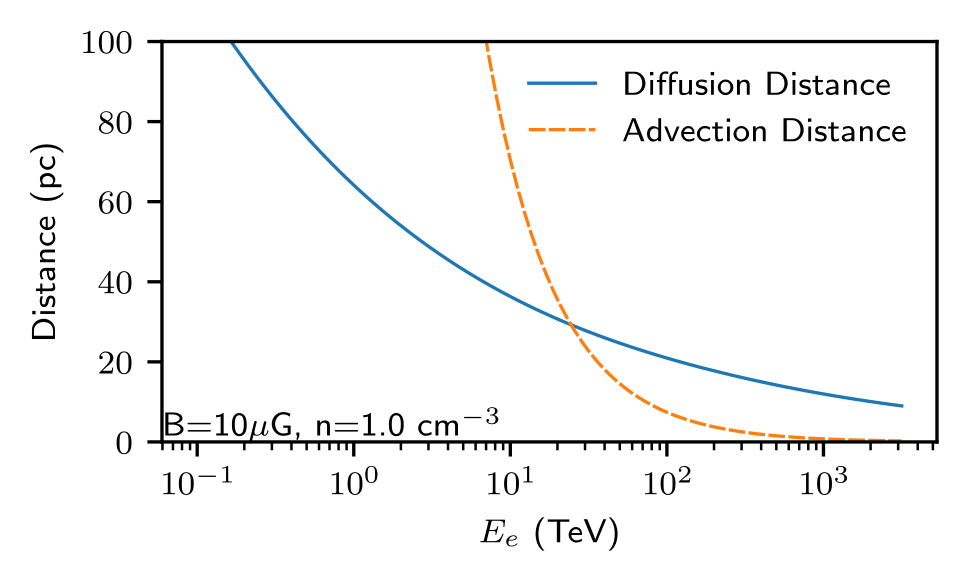}
    \captionof{figure}{The distance that electrons are transported before losing their energy through radiative cooling (synchrotron, inverse Compton and Bremsstrahlung) assuming purely diffusive (solid line, $\chi=0.1$) or advective transport (dashed line, $v=0.02c$).}
    \label{fig:cooling_time_distances}
\end{center}

\begin{figure*}
    \centering
    \begin{subfigure}[c]{\columnwidth}
        \centering
       \includegraphics[width=\textwidth]{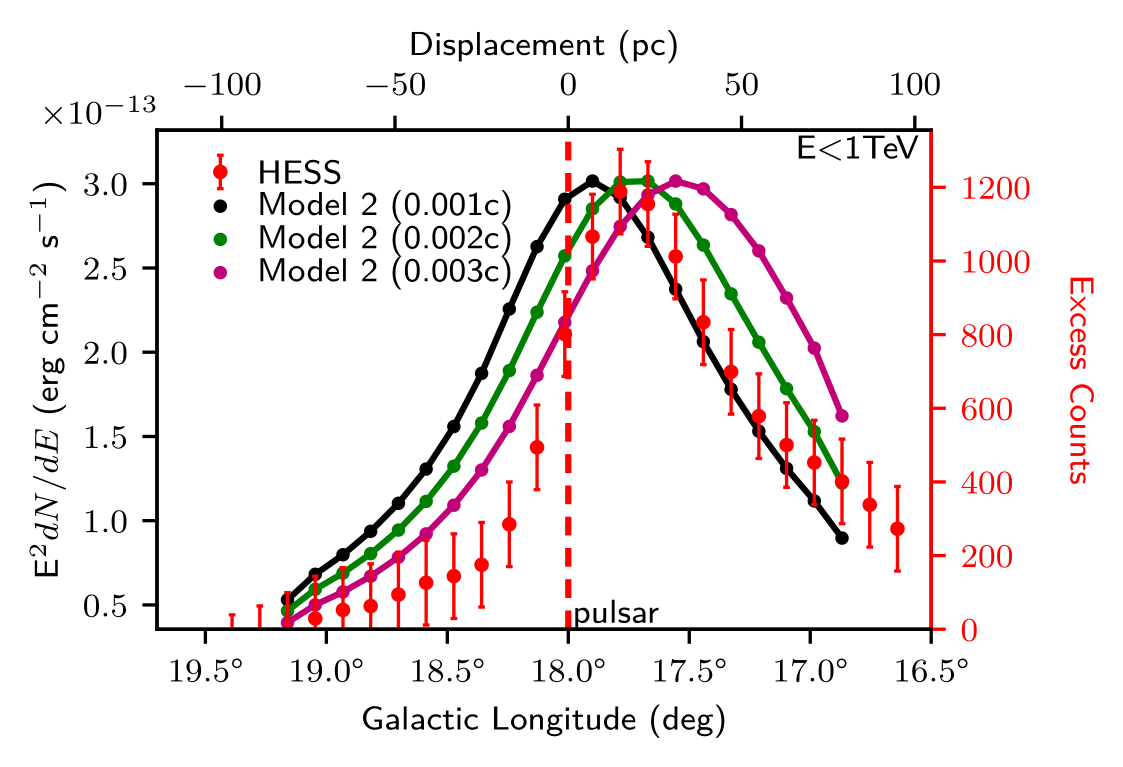}
    \end{subfigure}
    ~ 
    \begin{subfigure}[c]{\columnwidth}
        \centering
       \includegraphics[width=\textwidth]{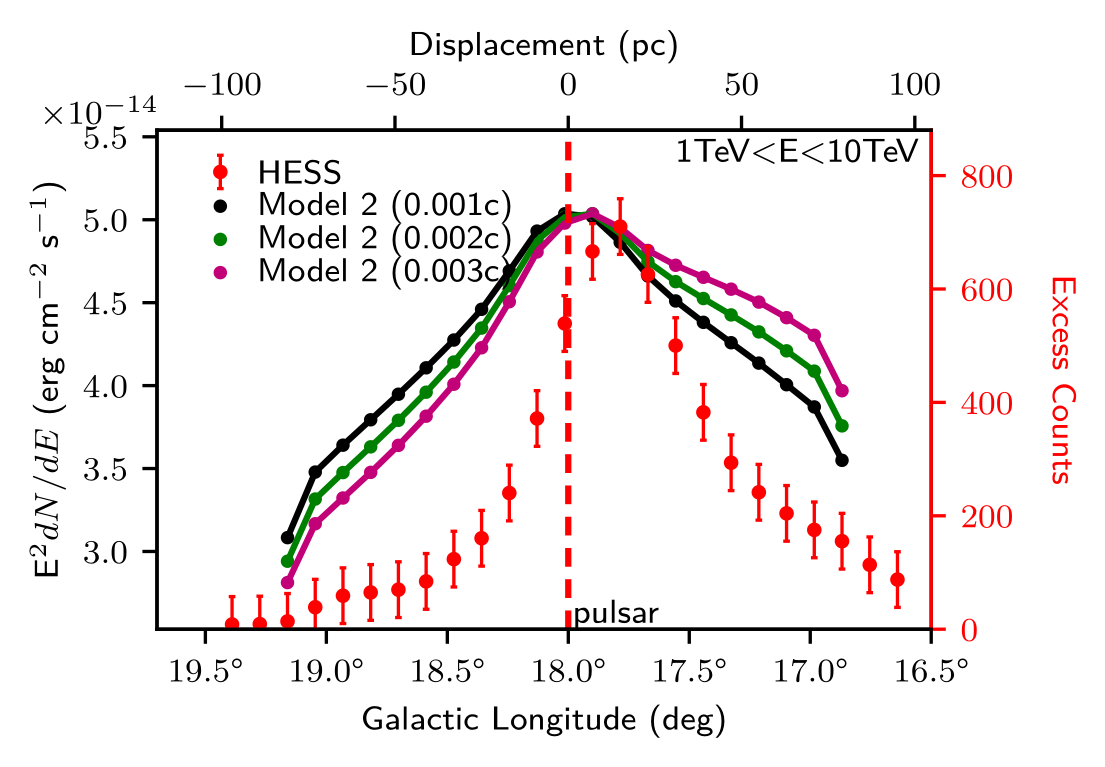}
    \end{subfigure}
    
   \begin{subfigure}[c]{\columnwidth}
        \centering
       \includegraphics[width=\textwidth]{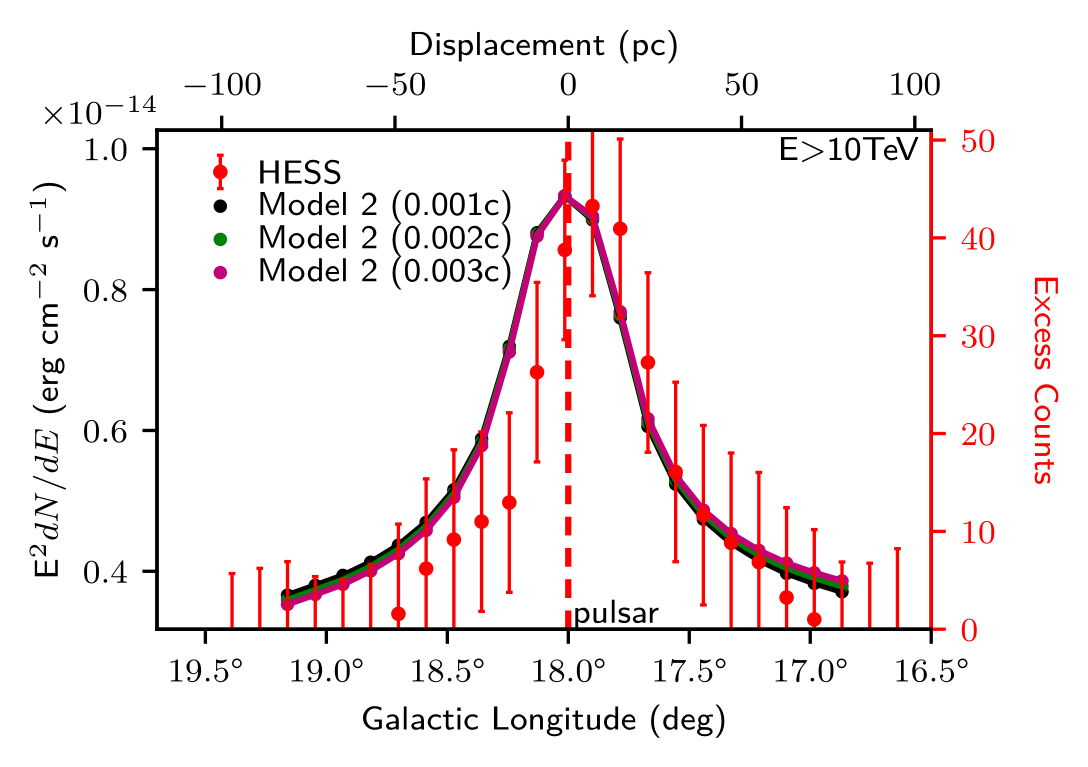}
    \end{subfigure}
    ~ 
    \begin{subfigure}[c]{\columnwidth}
        \centering
       \includegraphics[width=\textwidth]{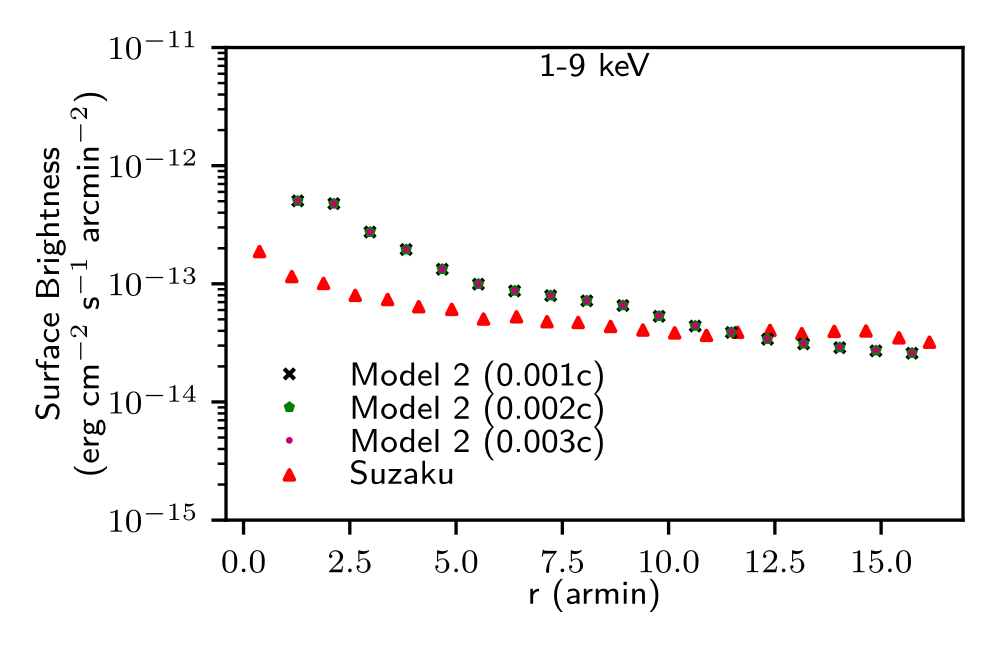}
    \end{subfigure}
    
    \begin{subfigure}[c]{\columnwidth}
       \includegraphics[width=\textwidth]{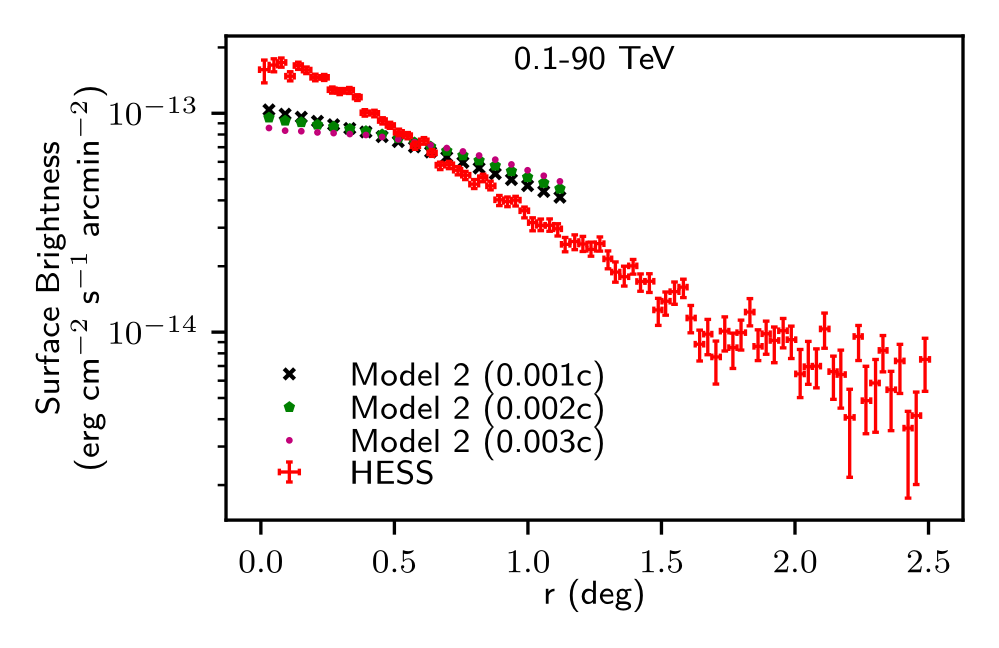}
    \end{subfigure}
    ~ 
    \begin{subfigure}[c]{\columnwidth}
       \includegraphics[width=\textwidth]{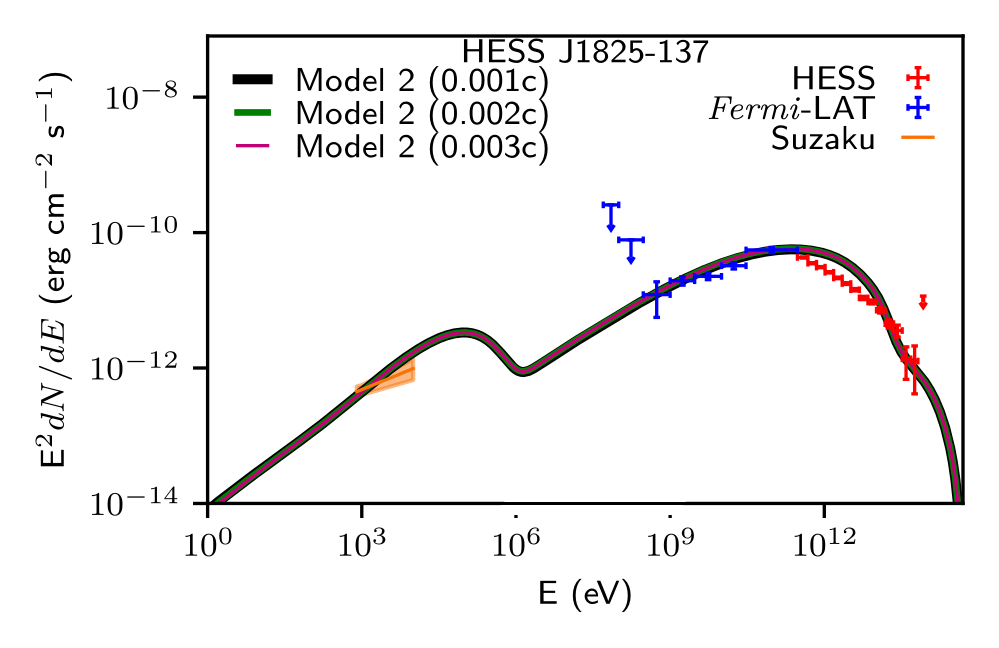}
    \end{subfigure}

    \begin{subfigure}[c]{\columnwidth}
       \includegraphics[width=\textwidth]{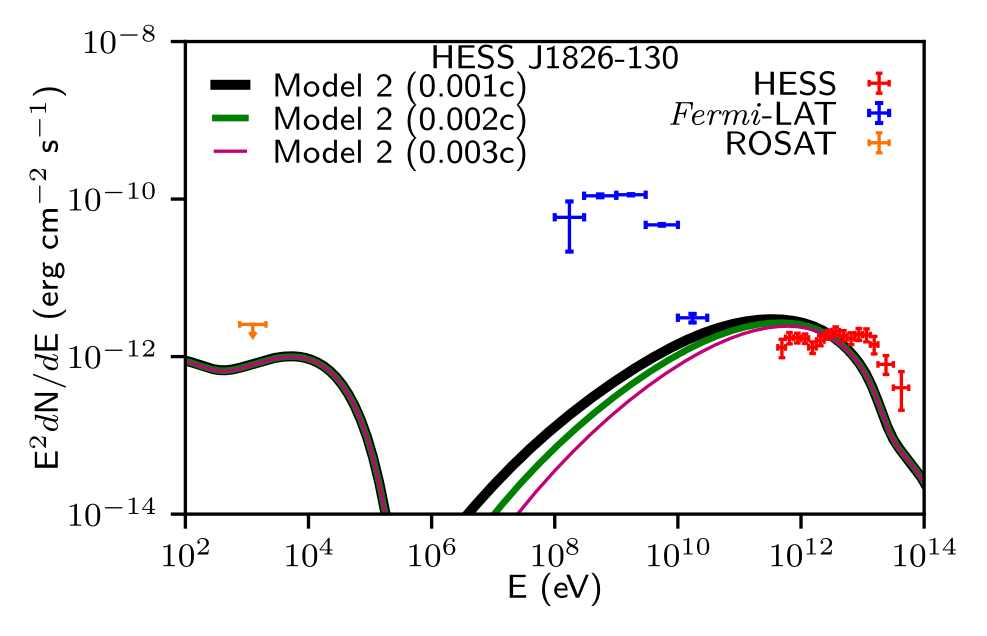}
    \end{subfigure}

    \caption{The slice profiles (\textit{top \& top-middle left}), surface brightness radial profiles (\textit{top-middle right \& bottom-middle left}) and SED towards \mbox{HESS\,J1825-137} (\textit{bottom-middle right}) and \mbox{HESS\,J1826-130} (\textit{bottom}) for Model~2 $0.001c$ (\textit{black}), $0.002c$ (\textit{green}) and $0.003c$ (\textit{purple}). See \autoref{tab:matched_parameters} for model parameters.}
    \label{fig:multizone_40_advection_comparison}
\end{figure*}


\bsp	
\label{lastpage}
\end{document}